\documentclass[a4paper,11pt]{article}
\pdfoutput=1 

\usepackage{jcappub} 

\usepackage[T1]{fontenc} 
\usepackage{graphicx,caption}\graphicspath{{figs/}}
\usepackage{subcaption}
\usepackage{bm}
\usepackage{lineno}
\usepackage{float}
\usepackage{caption}

\title{Probing for anisotropies at different angular scales in the arrival directions of the cosmic rays detected at the Pierre Auger Observatory}
\author[13]{A.~Abdul Halim,}
\author[67]{P.~Abreu,}
\author[50,49]{M.~Aglietta,}
\author[31]{M.~Ahmed,}
\author[1]{I.~Allekotte,}
\author[74,73]{K.~Almeida Cheminant,}
\author[42,43]{R.~Aloisio,}
\author[72]{J.~Alvarez-Mu\~niz,}
\author[42,43]{A.~Ambrosone,}
\author[73]{J.~Ammerman Yebra,}
% ...

\affiliation[1]{Centro At\'omico Bariloche and Instituto Balseiro (CNEA-UNCuyo-CONICET), San Carlos de Bariloche, Argentina}
% ...
\affiliation[13]{Adelaide University, Adelaide, S.A., Australia}
% ...
\affiliation[31]{CNRS/IN2P3, IJCLab, Universit\'e Paris-Saclay, Orsay, France}
% ...
\affiliation[42]{Gran Sasso Science Institute, L'Aquila, Italy}
\affiliation[43]{INFN Laboratori Nazionali del Gran Sasso, Assergi (L'Aquila), Italy}
% ...
\affiliation[49]{INFN, Sezione di Torino, Torino, Italy}
\affiliation[50]{Osservatorio Astrofisico di Torino (INAF), Torino, Italy}
% ...
\affiliation[67]{Laborat\'orio de Instrumenta\c{c}\~ao e F\'\i{}sica Experimental de Part\'\i{}culas -- LIP and Instituto Superior T\'ecnico -- IST, Universidade de Lisboa -- UL, Lisboa, Portugal}
% ...
\affiliation[72]{Instituto Galego de F\'\i{}sica de Altas Enerx\'\i{}as (IGFAE), Universidade de Santiago de Compostela, Santiago de Compostela, Spain}
\affiliation[73]{IMAPP, Radboud University Nijmegen, Nijmegen, The Netherlands}
\affiliation[74]{Nationaal Instituut voor Kernfysica en Hoge Energie Fysica (NIKHEF), Science Park, Amsterdam, The Netherlands}
% ...
\affiliation[c]{School of Physics and Astronomy, University of Leeds, Leeds, United Kingdom}
% ...

\emailAdd{spokespersons@auger.org}

\abstract{
    Studies of the arrival-direction distribution of ultra-high-energy cosmic rays are crucial for revealing their origin.
    In this work, we investigate the distribution of the arrival directions of events detected with the Pierre Auger Observatory over different angular scales.
    To do this, we adopt a novel method where we use a von Mises--Fisher distribution to model the excesses with respect to isotropic expectations.
    This method allows us to study the distribution of arrival directions over different angular scales consistently and simultaneously.
    We consider the events in different energy bins above 4\,EeV detected from 1$^\text{st}$ January 2004 to 31$^\text{st}$ December 2022, i.e.\ during the entire Phase~I of the Pierre Auger Observatory, with a total exposure of $123{,}000$\,km$^2$\,sr\,yr.
    The most significant excess in the bins considered is a dipolar modulation in the energy range (8--16)\,EeV, which is compatible with previous findings of the Pierre Auger Collaboration. 
    A dipolar modulation appears to be present for the higher energies considered as well, and at the highest energies, i.e.\ above 32\,EeV, we find indications of its coexistence with an overdensity of intermediate angular scale, located in the Centaurus region.
}

\begin{document}
\maketitle
\flushbottom

\section{Introduction}
\label{sec:intro}

The study of the arrival-direction distribution of ultra-high-energy cosmic rays (UHECRs) is central to understanding their origin.
These analyses are classified into two regimes based on the angular scale: large-scale anisotropies and small- to intermediate-scale anisotropies.

At large angular scales, a significant dipolar modulation in right ascension (R.A.) was observed above 8\,EeV\footnote{$1\,\mathrm{EeV}=10^{18}$\,eV} by the Pierre Auger Collaboration.
The modulation was first reported in~\cite{ref:lsa_science}, and has now reached 6.8$\sigma$ significance~\cite{ref:lsa}.
Its direction lies around 115$^\circ$ away from the Galactic Center, suggesting that UHECRs at these energies are mostly of extragalactic origin.
Moreover, the dipole amplitude has an approximately linear growth with energy~\cite{ref:lsa}. 

At small to intermediate angular scales, no significant anisotropy has been observed in the data collected with the two largest detector arrays for UHECRs: the Pierre Auger Observatory~\cite{ref:auger} and the Telescope Array~\cite{ref:ta}.
Nonetheless, both collaborations reported the presence of several hints of anisotropies above a few tens of EeV, see for example~\cite{ref:apj2022, ref:small_icrc23, ref:ta_hot, ref:ta_icrc25}.
The most significant one is an excess observed above 38\,EeV within a top-hat circle of $27^\circ$ centered on the radiogalaxy Centaurus~A with a  post-trial significance of 4.0$\sigma$~\cite{ref:apj2022,ref:small_icrc23}.
Interestingly, the Pierre Auger Collaboration also observed overdensities in the Centaurus region above 20\,EeV,  by scanning circular regions crossing the Supergalactic plane with a fixed angular radius of 27$^\circ$~\cite{ref:sgp}.
The scan was repeated for different energy thresholds, and notably, an excess present in the Centaurus region is observed above each energy threshold.
Moreover, the central position of the excess region is stable within few degrees as a function of the energy threshold.

The above results were obtained by treating the two angular-scale regimes separately.
In this work, we introduce a novel approach that enables a search for overdensities across all angular scales.
This analysis is complementary to the power-spectrum analysis which is sensitive to the combination of excesses at a given angular scale all over the sky \cite{ref:multipole, ref:lsa}.
With the approach presented here, we look instead for the most significant excess of events and provide a more comprehensive characterization thereof.
As a result, we get new insight into the transition from large-scale anisotropies at lower energies to intermediate-scale excesses at higher energies.

We structure this paper as follows.
In Section~\ref{sec:dataset}, we describe the dataset used for the analysis.
In Section~\ref{sec:fish}, we give a description of the likelihood and the test statistics used in the analysis.
In Section~\ref{sec:res}, we report the results obtained with the analysis focusing first, in Section~\ref{sec:res_st}, on a set of energy bins already introduced in previous analyses of the Pierre Auger Collaboration, then, in Section~\ref{sec:res_add}, we perform the analysis introducing additional energy bins.
Finally, a discussion of the results and the conclusion are presented in Section~\ref{sec:concl}.

\section{The dataset}
\label{sec:dataset}
The Pierre Auger Observatory is the largest facility in the world to detect UHECRs.
It is located in Argentina, near the city of Malargüe (35.2$^\circ$ S, 69.5$^\circ$ W) at an average altitude of 1{,}400\,m a.s.l.\ and it covers an area of 3{,}000\,km$^2$.
The Observatory is a multi-hybrid system, composed of a surface detector (SD) and a fluorescence detector (FD).
The SD is an array of water-Cherenkov detectors located in a triangular grid.
The baseline array for the study of the highest-energy cosmic rays has 1600 detectors with a spacing of 1{,}500\,m (SD-1500).
The SD has a duty cycle of $\sim\!100\%$.
The FD consists of 24 fluorescence telescopes distributed between 4 sites along the perimeter of the SD-1500 and overlooking the atmosphere above the array.
The FD operates only in clear moonless nights, with a duty cycle of $\sim\!15\%$.

In this work, we adopt the same data selection as it was defined in~\cite{ref:lsa}.
We consider only data registered with the SD-1500 between 1$^\text{st}$ January 2004 and 31$^\text{st}$ December 2022.
For the years 2021 and 2022, when the installation of the \textit{AugerPrime} upgrade~\cite{ref:auger_prime} was underway, we consider only data registered with stations with non-upgraded electronics.

The SD-1500 is fully efficient above 4\,EeV for events entering the atmosphere with zenith angle $\theta < 80^\circ$ (achieving 85$\%$ of sky coverage).
According to $\theta$, physical events are reconstructed using different algorithms.
The \textit{vertical} reconstruction is applied to events entering the atmosphere with $\theta<60^\circ$, while the \textit{inclined} reconstruction is applied to events with $60^\circ\leq \theta < 80^\circ$. 
Vertical events are included in the analysis if the SD station with the largest signal is surrounded by at least five functioning stations.
Inclined events are included if the SD station closest to the reconstructed core is surrounded by at least five functioning stations.
Additionally, we require that the reconstructed core falls in an  isosceles triangle of nearby functioning stations. 
Events detected during periods of unreliable data acquisition are removed from the analysis.
The total exposure over the whole period is 123{,}000\,km$^2$\,sr\,yr.

The arrival directions of vertical events are derived by fitting a spherical model to the arrival times of the particles contained in the front of the showers at the SD detectors.
Given that the shower front of the inclined events is mostly formed by muons, the arrival direction is obtained by fitting the arrival times with a model considering the muon propagation from the production site.
A detailed description of the reconstruction procedure for the vertical events can be found in~\cite{ref:vert_rec}, while a description for the inclined events is present in~\cite{ref:incl_rec}.
In both cases, the angular resolution is better than 1$^\circ$.

The energy estimation is performed differently for vertical and inclined events.
For vertical showers, the proxy for the primary energy is the signal of water-Cherenkov detectors estimated at a reference distance of 1000\,m, $S(1000)$. 
For inclined showers, the proxy is the so-called $N_{19}$: the scaling factor by which the 2D muon density at ground simulated for proton-induced showers at $E=10^{19}\,\mathrm{eV}$ needs to be multiplied to fit the signal of the water-Cherenkov detectors.
Since the particle density at ground depends on the zenith angle due to the attenuation of shower particles in the atmosphere, both $S(1000)$ and $N_{19}$ need to be converted into zenith-independent energy estimators using the constant intensity cut method~\cite{ref:prd2020}, by which $S(1000)$ ($N_{19}$) is converted to the value expected for a shower arriving with a zenith angle of $38^\circ$ ($68^\circ$). 
The energy estimator $S_{38}$ ($N_{68}$) is calibrated using a subset of high-quality \textit{hybrid} events (i.e.\ events observed simultaneously by SD and FD).
Using the hybrid data, we perform a quasi-calorimetric measurement of the shower energy.
Therefore, the calibration procedure  enables a good energy estimation for all the events detected with the SD.
The statistical uncertainty on the energy calibration is $\sim\!15\%$ for events with energy above 4\,EeV and reduces to $\sim\!7\%$ at the highest energies, above 20\,EeV. 
The systematic uncertainty is $\sim\!14\%$~\cite{ref:incl_rec, ref:prl2020}.

Atmospheric~\cite{ref:atmo} and geomagnetic corrections~\cite{ref:geom} are applied to the energies of vertical events, to avoid spurious effects in the distribution in R.A.\ and azimuth angle.
Inclined events do not require atmospheric corrections, since atmospheric effects are negligible for muons, while geomagnetic effects are included in the reconstruction procedure.

\section{Description of the analysis}
\label{sec:fish}

In this work, we characterize the anisotropies in the arrival-direction distribution of the events using a likelihood-ratio test.
To construct it, we start by considering a localized source signal distribution superposed over an isotropic background, described by
\begin{equation}
    \label{eq:fish_distr}
    \frac{\text{d}N}{\text{d}\cos{\Theta}} = \frac{N}{2}\biggl[(1-\alpha) + \alpha\frac{\kappa}{\sinh{\kappa}}e^{\kappa\cos{\Theta}}\biggr].
\end{equation}
The first term, $(1-\alpha)$, represents the isotropic fraction, where $\alpha$ is the signal fraction.
The second term follows a von Mises--Fisher distribution, which is the generalization of a Gaussian distribution on a sphere.
The angular distance measured from the central position of the excess is denoted by $\Theta$.
The concentration parameter $\kappa$ controls the steepness of the profile around the reference direction.

The distribution in equation~\eqref{eq:fish_distr} exhibits two distinct asymptotical regimes depending on the value of $\kappa$.
In the limit $\kappa \ll 1$, the distribution is nearly isotropic and can be expanded in dipole and quadrupole moments
\begin{equation}
    \frac{\text{d}N}{\text{d}\cos{\Theta}}\approx\frac{N}{2}\biggl[ 1 + d\cos{\Theta} + q\biggl(\cos^2\Theta - \frac{1}{3}\biggr) \biggr],
\end{equation}
with dipole and quadrupole amplitudes  $d = \alpha \kappa$ and $q = \alpha\kappa^2/2$, respectively.
We note that $d$ and $q$ represent the dipole and quadrupole amplitudes associated to a modulation centered in the excess position.
Since the development is dominated by the dipolar component, $d$ can be compared with previous findings of the Pierre Auger Collaboration \cite{ref:lsa_science, ref:lsa}.
However, the quadrupole term $q$ describes the axisymmetric component ($a_{20}$) with axis in the excess direction, and this is not directly comparable with the amplitudes derived in previous analyses that tested the presence of a quadrupolar modulation \cite{ref:lsa, ref:lsa_2018}.

As $\kappa$ increases, the steepness of the von Mises--Fisher term of the distribution becomes more pronounced.
For $\kappa\gg1$ it can be approximated, in a small angular region around the center of the excess, by a Gaussian profile with width  $\sigma=1/\sqrt{\kappa}$.

In this regime, the distribution can be used to test whether a fraction $\alpha$ of events is clustered within an angular scale $1/\sqrt{\kappa}$ around the reference direction.

Equation~\eqref{eq:fish_distr} provides an accurate description of the expected arrival-direction distribution of the events emitted by a single source located in the reference direction, and propagating through a turbulent magnetic field~\cite{ref:fisher}.
Large values of $\kappa$ correspond to quasi–ballistic propagation, whereas small values describe magnetic turbulent diffusion.
With this model, we can simultaneously probe dipolar patterns and the transition between large-angular scale anisotropies and small- to intermediate angular scale anisotropies. 

For a dataset of $N$ events detected with the Pierre Auger Observatory, the likelihood of the events being distributed according to equation~\eqref{eq:fish_distr} is given by
\begin{equation}
    \label{eq:likelihood}
    \mathcal{L}(\alpha, \kappa, \bm{u_0})=\prod_{j=1}^N \biggl[ (1-\alpha) + \alpha \frac{\kappa}{\sinh{\kappa}}e^{\kappa \bm{u_j}\cdot\bm{u_0}} \biggr] \frac{\omega(\bm{u_j})}{\mathcal{N}(\alpha,\kappa,\bm{u_0})},
\end{equation}
where $\bm{u_0}$ denotes the reference direction, $\bm{u_j}$ is the arrival direction of the $j$-th event, and $w(\bm{u_j})$ accounts for the directional exposure of the array.
The geometrical exposures for vertical events and inclined events are evaluated separately.
To keep the analysis as data-driven as possible, $w(\bm{u_j})$ is evaluated combining the vertical and inclined geometrical exposures using the ratio of inclined to vertical events observed in each energy bins as the expected exposure ratio for that bin.
$\mathcal{N}(\alpha,\kappa,\bm{u_0})$ is a normalization factor depending on the signal fraction, the concentration parameter of the Von Mises--Fisher term and the reference direction
\begin{equation}
    \mathcal{N}(\alpha,\kappa,\bm{u_0}) = \int \bigg[ (1-\alpha) + \alpha \frac{\kappa}{\sinh{\kappa}}e^{\kappa \bm{u}\cdot\bm{u_0}}\bigg] \omega(\bm{u}) \text{d}\Omega.
\end{equation}
Under the isotropic hypothesis ($\alpha=0$), the likelihood reduces to
\begin{equation}
    \mathcal{L}(\alpha=0) = \prod_{j=1}^N\frac{\omega(\bm{u_j})}{\int \omega(\bm{u}) \text{d}\Omega}.
\end{equation}
The likelihood-ratio between $\mathcal{L}(\alpha,\kappa,\bm{u_0})$ and its value obtained for the isotropic hypothesis $\mathcal{L}(\alpha=0)$ defines the test statistics ($TS$) that we employ in the analysis
\begin{equation}
    \label{eq:ts}
    \mathrm{TS} = -2\ln \bigg( \frac{\mathcal{L}(\alpha=0)}{\mathcal{L}(\alpha, \kappa, \bm{u_0})} \bigg).
\end{equation}

As outlined above, the concentration parameter $\kappa$ controls the angular width of the tested distribution.
In this work, we compute a profile likelihood as a function of ($\kappa$), i.e., for each value of $\kappa = 10^{-1.0}, 10^{-0.9}, 10^{-0.8}, \dots, 10^{+3.5}$, we find the $\alpha$ and $\bm{u}_0$ that minimize $-2\ln(\mathcal{L}(\alpha,\bm{u}_0;\kappa))$.  
Since the equivalent dipole amplitude is $d=\alpha\kappa$ and the quadrupole amplitude $q=\alpha \kappa^2/2$, with the selected lower limit $k=10^{-1}$, we are able to probe a purely dipolar flux (with a quadrupolar term a factor 20 times smaller than the dipolar one).
On the other hand, the upper limit of the $\kappa$ range is selected to match the angular resolution of the Observatory, since $\kappa=10^{3.5}$ corresponds to the equivalent Gaussian width $\sigma=1.0^\circ$.

The minimization is performed using the differential evolution algorithm present in the $\texttt{python}$ library $\texttt{scipy}$~\cite{ref:scipy}.
In the minimization, the declination of $\bm{u}_0$ is constrained to be below $+44.8^\circ$, requiring the center of the von Mises--Fisher distribution to lie within the field of view of the Observatory. 
Although a von Mises--Fisher distribution centered outside the field of view can still produce a non-negligible contribution to the observed event distribution, restricting the center to the field of view avoids border effects, whereby a small number of clustered events close to the edge of the field of view could result in an artificially large signal.
Since we profile over $\kappa$, the likelihood described in equation~\eqref{eq:likelihood} has three degrees of freedom, given by the signal fraction and the celestial coordinates of the reference position ($\mathcal{L}(\alpha, \bm{u_0}(\text{R.A.}, \text{dec}); \kappa)$).
At any given $\kappa$, for all parameters, the uncertainties are derived from the confidence intervals obtained from the likelihood fit, in which all parameters are varied simultaneously during the minimization. 

For every step in the $\kappa$-parameter space, we evaluate the isotropic chance probability $p_\mathrm{iso}$ of obtaining the maximum $\mathrm{TS}$.
$p_\mathrm{iso}$ is evaluated performing 10{,}000 isotropic Monte-Carlo  simulations, and it is defined as the fraction of isotropic simulations yielding a maximum $\mathrm{TS}$ at a fixed $\kappa$ equal or larger than the $\mathrm{TS}$ obtained on data for the same $\kappa$.
Whenever we do not obtain a maximum $\mathrm{TS}$ in simulations larger than the $\mathrm{TS}$ obtained on data, we evaluate $p_\mathrm{iso}$ by doing an extrapolation using a $\chi^2$ distribution with three degrees of freedom.
This procedure allows us to probe the most significant excess in the observed arrival-direction distribution across different angular scales, providing a more detailed characterization of how these excesses evolve.

For each energy interval that we consider, we report either $d$ (if $\kappa<1$) or $\sigma$ (if $\kappa\geq1$) of the excess with the lowest $p_\mathrm{iso}$.
The values of d are calculated from using the combination of $\alpha$ and $\kappa$ that maximizes the $\mathrm{TS}$, with the center of the excess fixed to the position corresponding to the minimum $p_\mathrm{iso}$.
This procedure allows us to probe the dipolar modulation centered in the excess position that best describes the data.
The uncertainties on $d$ are derived from the profile likelihood obtained by profiling the $\mathrm{TS}$ over $\alpha$
\begin{equation}
    \mathrm{TS}_\text{prof}(d)= \max_{\alpha} \mathrm{TS} (\alpha, \kappa=d/\alpha).
    \label{eq:prof_ts}
\end{equation}
where the minimization is performed within the allowed parameter ranges $\alpha\in[0,1]$ and $\kappa\in[0.1,1]$. 
The confidence interval on $d$ is then determined from the values for which $\Delta\mathrm{TS}_\text{prof}(d)=1$.

In this work, we do not account for the small time-dependent non-uniformities in the array exposure, arising from the gradual deployment of the detector and the occasional downtime of individual stations, nor for the effect of the tilt of the array, both of which were included in some previous analyses of large-scale anisotropies by  the Pierre Auger Collaboration (see e.g.\ ~\cite{ref:lsa}).
Both effects have a minimal impact on the analysis performed here ($\le 0.1\%$ in the evaluation of the dipole amplitude), therefore, including them does not change the results.

\section{Results}
\label{sec:res}
\subsection{Results for standard energy bins}
\label{sec:res_st}

We present the results of the analysis described in section~\ref{sec:fish}.
In this section, we show the results obtained by dividing the dataset into four energy intervals: (4--8, 8--16, 16--32, >32)\,EeV.
These intervals follow the binning adopted in previous anisotropy studies performed by the Pierre Auger Collaboration for large angular scales~\cite{ref:lsa_science,ref:lsa}.
\begin{table}[tbp]
    \begin{center}
    \renewcommand{\arraystretch}{1.3}
        \begin{tabular}{|c|c|c|c|c|c|c|c|c|}
        \hline
        $\bm{E/}$\,\textbf{EeV} & $\bm{\kappa}$ & $\bm{\alpha\,(\%)}$ &  $\bm{\textbf{R.A.}/{}^\circ}$ & $\bm{{\textbf{dec}/{}^\circ}}$ & $\bm{\mathrm{TS}}$ & $\bm{d\,(\%)}$ & $\bm{\sigma/{}^\circ}$ & $\bm{p_\textbf{iso}}$\\
        \hline
        \phantom{0}4--8\phantom{0} & $2.5\phantom{0}$ & $\phantom{0}0.7^{+0.2}_{-0.2}$ & $112^{+26}_{-33}$ & $-65^{+16}_{-13}$ & $10.1$ & --- & $36.2$ & $0.023$ \\
        \phantom{0}8--16 & $0.10$ & $\phantom{0.}67^{+10\phantom{.}}_{-10\phantom{.}}$ & $\phantom{0}94^{+9\phantom{0}}_{-9\phantom{0}}$ & $-28^{+10}_{-10}$ & $49.2$ & $6.7^{+1.0}_{-1.0}$ & --- & $1.1\times10^{-10}{}^*$ \\
        16--32 & $0.10$ & $\phantom{.}100_{-17\phantom{.}}$ & $\phantom{0}95^{+17}_{-16}$ & $-53^{+11}_{-10}$ & $25.2$ & $\phantom{.}10^{+2\phantom{.0}}_{-2\phantom{.0}}$ & --- & $1.3\times10^{-5}{}^*\phantom{0}$  \\
        32--$\infty$ & $0.16$ & $\phantom{0.}90^{+10\phantom{.}}_{-25\phantom{.}}$ & $141^{+19}_{-19}$ & $-44^{+18}_{-15}$ & $13.8$ & $\phantom{.}14^{+4\phantom{.0}}_{-4\phantom{.0}}$ & --- & $3.0\times10^{-3}\phantom{00}$ \\
        \hline
    \end{tabular}
    \caption{\label{tab:st} Results obtained maximizing $\mathrm{TS}$ while profiling over $\kappa$. 
    For each energy interval, the $\kappa$ associated to the lowest $p_\mathrm{iso}$, the estimators for $\alpha$, R.A.\ and dec, the $\mathrm{TS}$, the equivalent dipole amplitude $d$ or the equivalent Gaussian $\sigma$, and $p_\mathrm{iso}$ are reported.
    $d$ is reported only if $\kappa<1$, while $\sigma$ is reported otherwise.
    The values of $p_\mathrm{iso}$ marked with a * are obtained with the $\chi^2$ extrapolation.
    }
    \end{center}
\end{table}
\begin{figure*}[t]
    \centering
    \begin{subfigure}{0.47\textwidth}
        \includegraphics[width=\linewidth]{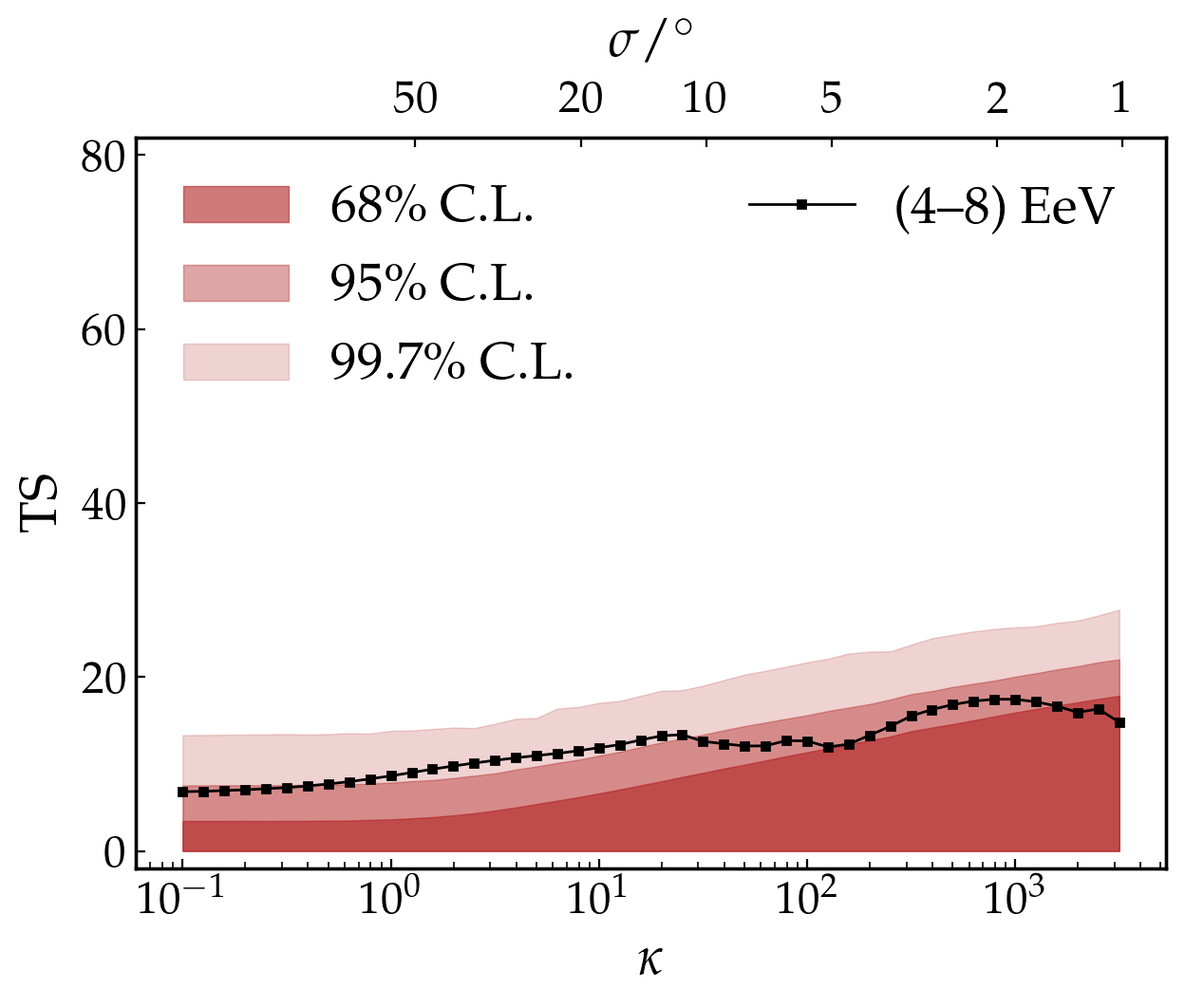}
        \caption{}
    \end{subfigure}\hspace{0.1cm}
    \begin{subfigure}{0.47\textwidth}
        \includegraphics[width=\linewidth]{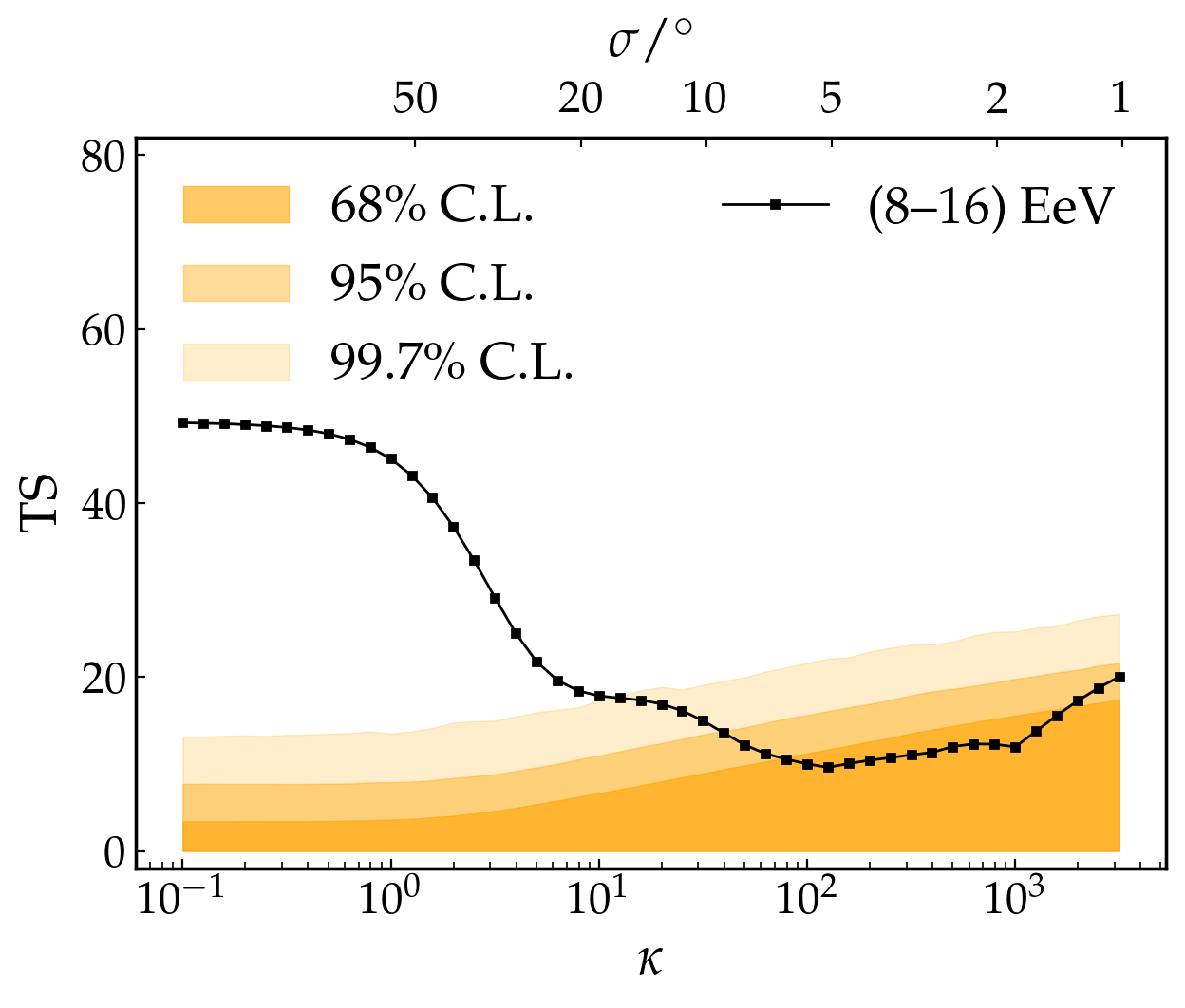}
        \caption{}
    \end{subfigure}
    \\[0.1cm]
    \begin{subfigure}{0.47\textwidth}
        \includegraphics[width=\linewidth]{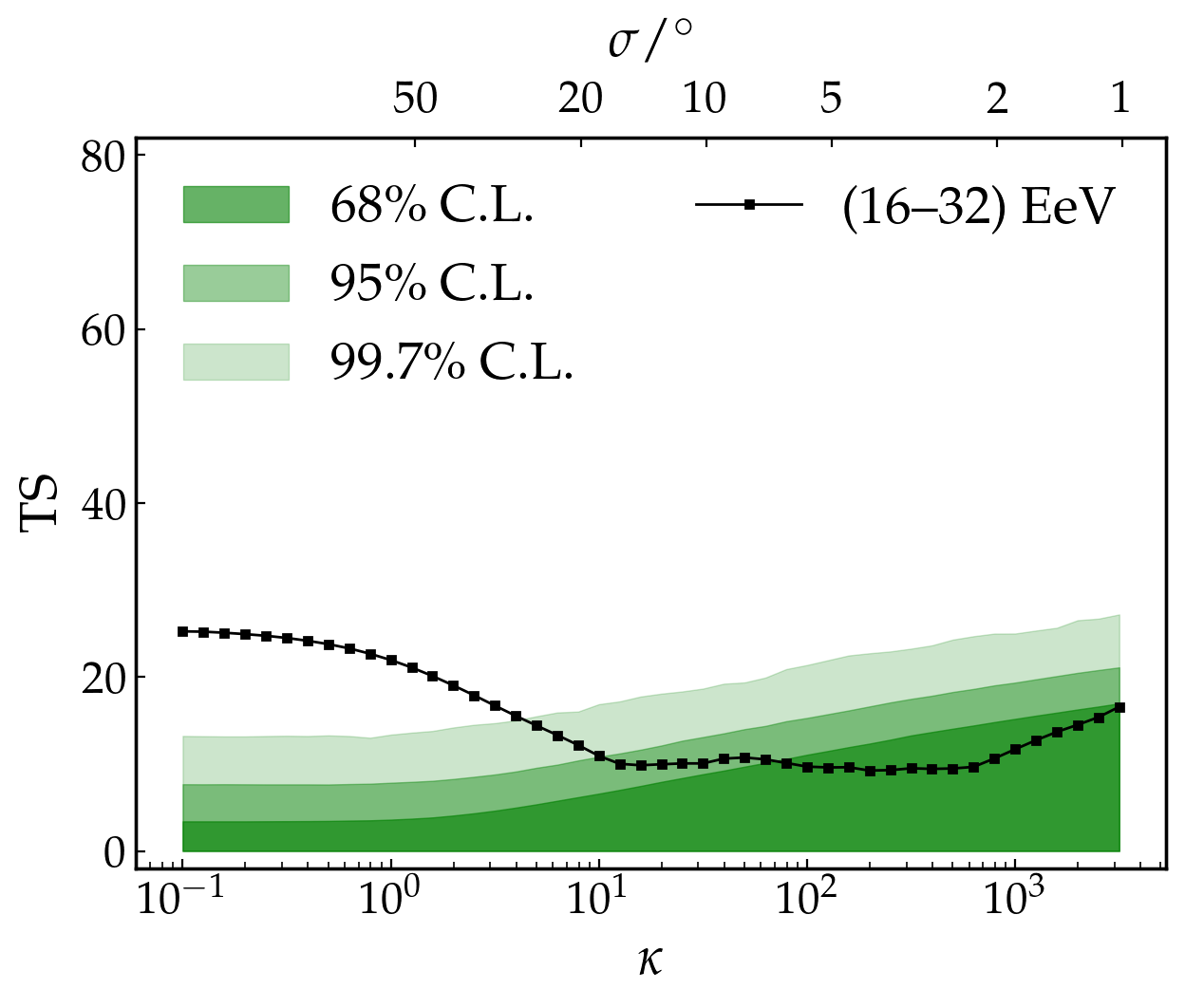}
        \caption{}
    \end{subfigure}\hspace{0.1cm}
    \begin{subfigure}{0.47\textwidth}
        \includegraphics[width=\linewidth]{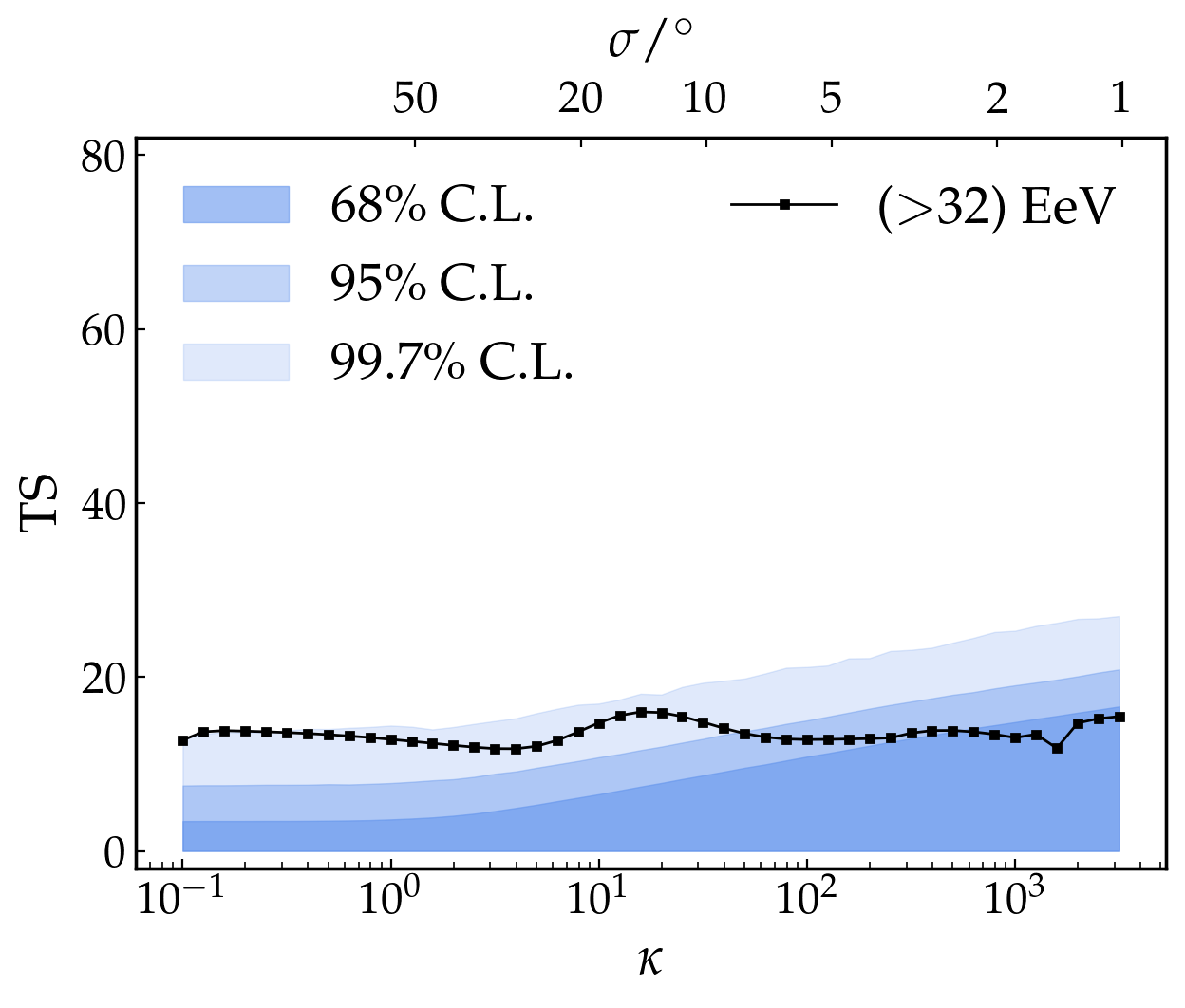}
        \caption{}
    \end{subfigure}
    \caption{
    $\mathrm{TS}$ derived by profiling over $\kappa$ for the energy intervals (a) (4--8)\,EeV, (b) (8--16)\,EeV, (c) (16--32)\,EeV, and (d) (>32)\,EeV.
    For each energy interval, the shadow bands correspond to the 68$\%$, 95$\%$, and 99.7$\%$ C.L. intervals derived with 10{,}000 isotropic simulations, using the same number of events present in the energy interval.}
    \label{fig:plots_st}
\end{figure*}
The results corresponding to the lowest $p_\mathrm{iso}$, obtained by profiling over $\kappa$ in each energy bin, are listed in table~\ref{tab:st}.
The values of $p_\mathrm{iso}$ marked with an asterisk (*) are derived with the $\chi^2$ extrapolation.

In table~\ref{tab:st}, we quote either the equivalent dipole amplitude $d=\alpha\kappa$ (if $\kappa < 1$) or the equivalent Gaussian width $\sigma=1/\sqrt{\kappa}$ (if $\kappa \geq 1$).
Dipole amplitudes are reported for the (8--16, 16--32, >32)\,EeV intervals.
The resulting dipole amplitudes are compatible with those obtained from the 3D dipole reconstruction  reported in~\cite{ref:lsa}, and the lowest $p_\mathrm{iso}$ is found in the (8–16)\,EeV interval.

In figure~\ref{fig:plots_st}, the full results obtained by profiling over $\kappa$ are shown.
In the plots, the maximum $\mathrm{TS}$ derived for each scanned $\kappa$ is shown together with the $68\%$, $95\%$, and $99.7\%$ confidence level (C.L.) bands derived from 10{,}000 isotropic simulations.
We note that the C.L.\ bands get broader with increasing $\kappa$.
Indeed, at larger $\kappa$ (smaller angular widths of the distribution), higher $\mathrm{TS}$ can be obtained from isotropic simulations than at smaller $\kappa$, given that more independent regions are probed as a function of increasing $\kappa$, making it more likely to find larger values of $\mathrm{TS}$ by chance.
Moreover, few clustered events can lead to large $\mathrm{TS}$ for small angular scales.
With the introduction of $p_\mathrm{iso}$, we are able to compute the significance of the excesses observed at different values of $\kappa$.
In the (4--8)\,EeV interval, no $\mathrm{TS}$ exceeding the $99.7\%$ C.L.\ band is observed.
In contrast, for (8--16, 16--32, >32)\,EeV, values of $\mathrm{TS}$ that exceed the $99.7\%$ C.L.\ interval are observed.
These values correspond to the results reported in table~\ref{tab:st}.
A secondary excess of the $\mathrm{TS}$ outside the $95\%$ C.L.\ band is observed in (>32)\,EeV at $\kappa = 10^{1.1}\simeq13$.
The equivalent Gaussian $\sigma$ is 16$^\circ$.
The reference direction associated to this excess is $(\text{R.A.},\text{dec})=(209^\circ, -50^\circ)$.
The equivalent Gaussian width derived with the minimization can be compared to the top-hat radius $\Psi$ of the Centaurus~A targeted analysis through the relation $\Psi=1.59\times\sigma$.
The Gaussian width obtained is compatible with a top-hat excess with radius $\Psi=25^\circ$.
Therefore, the excess is compatible with the previous finding of the Pierre Auger Collaboration of an overdensity of intermediate angular scale within the $27^\circ$ circular region around Centaurus~A, $(\text{R.A.},\text{dec}) = (201.4^\circ, -43.0^\circ)$~\cite{ref:apj2022, ref:small_icrc23}.
No excess with a $\mathrm{TS}$ above $95\%$ C.L.\ is found at concentrations larger than 100 in any of the four energy intervals considered.

\begin{figure*}[!ht]
    \centering
    \begin{subfigure}{0.47\textwidth}
        \includegraphics[width=\linewidth]{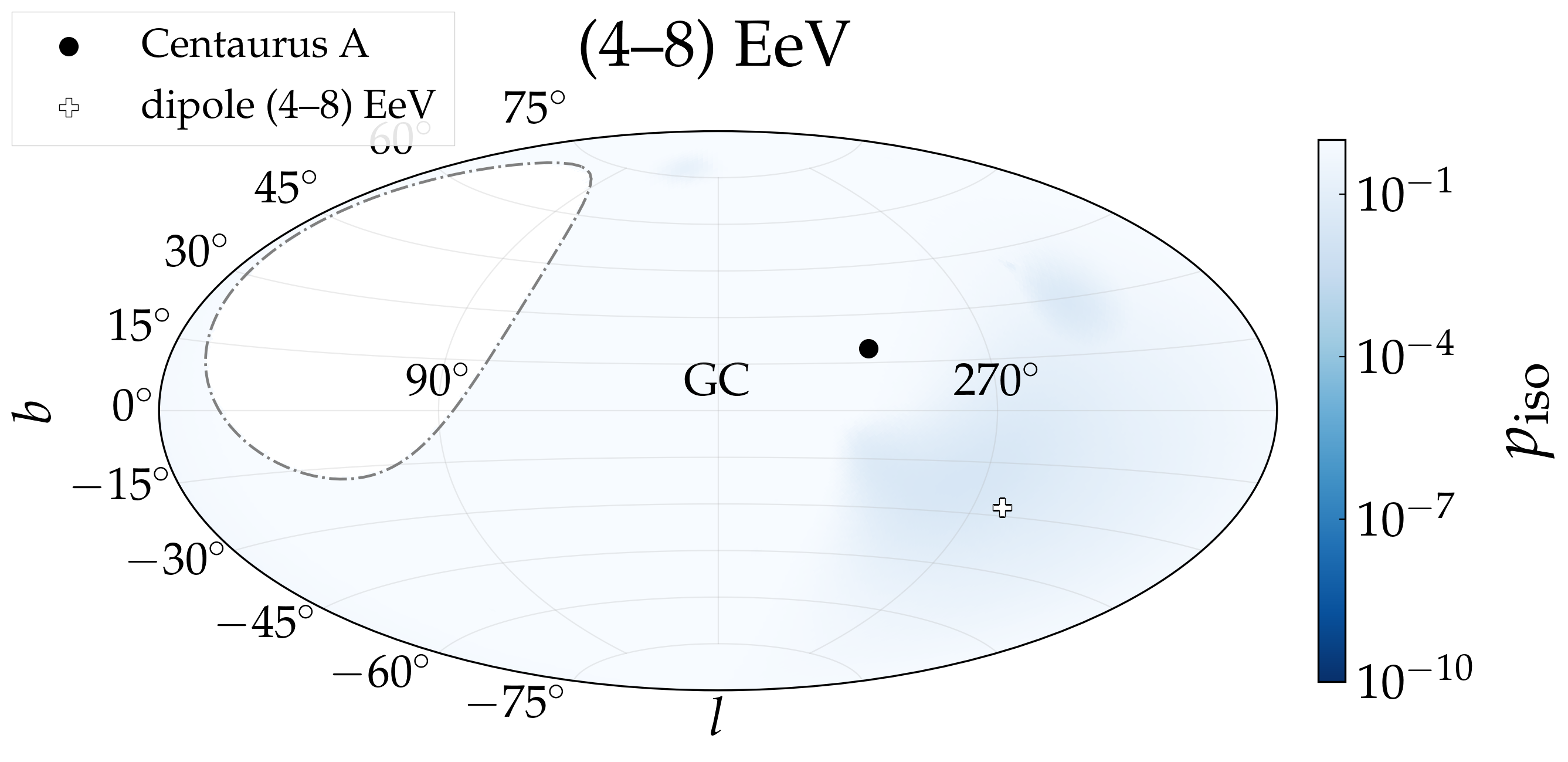}
        \caption{}
    \end{subfigure}
    \begin{subfigure}{0.47\textwidth}
        \includegraphics[width=\linewidth]{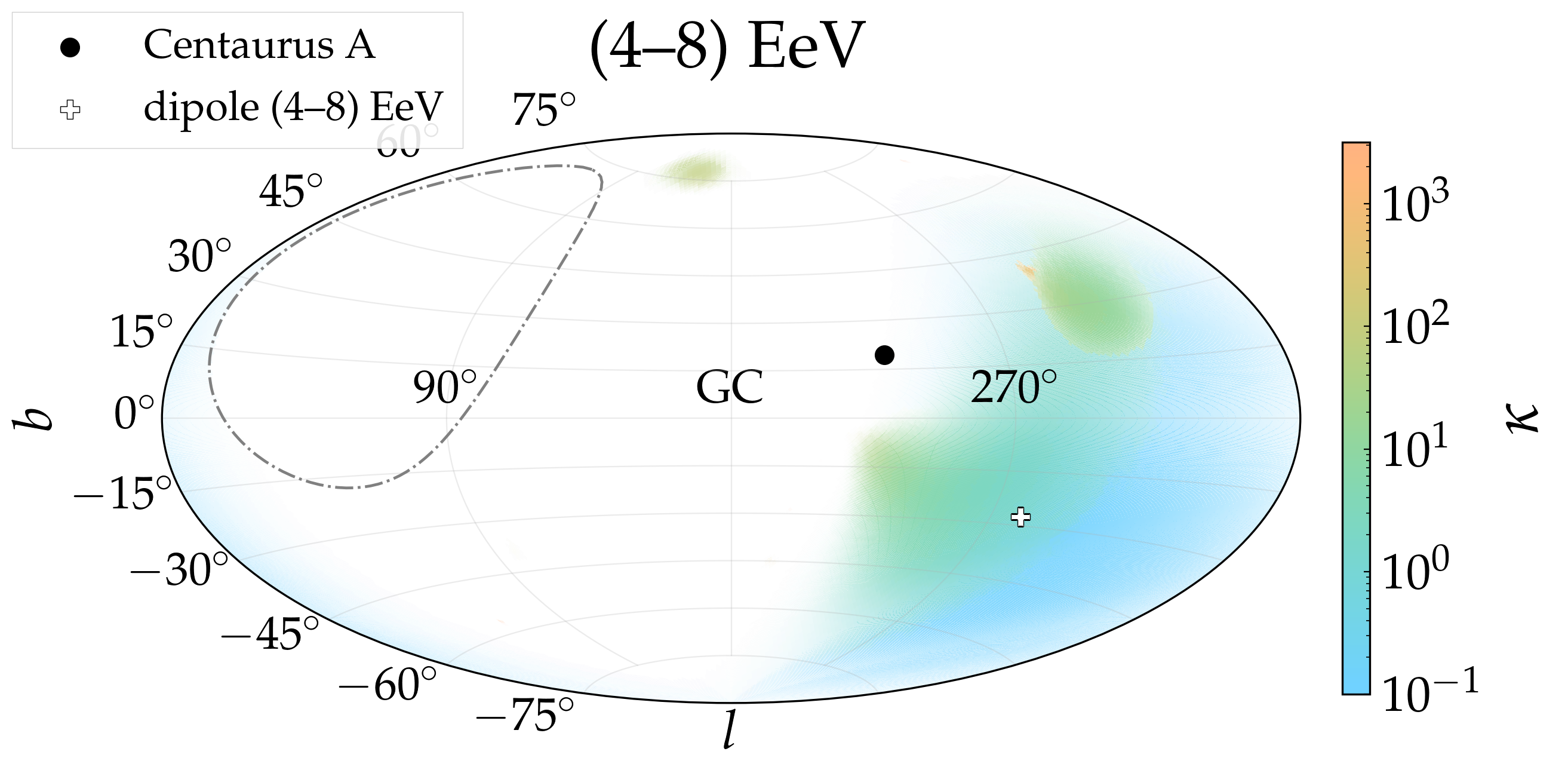}
        \caption{}
    \end{subfigure}
    \\[0.01cm]
    \begin{subfigure}{0.47\textwidth}
        \includegraphics[width=\linewidth]{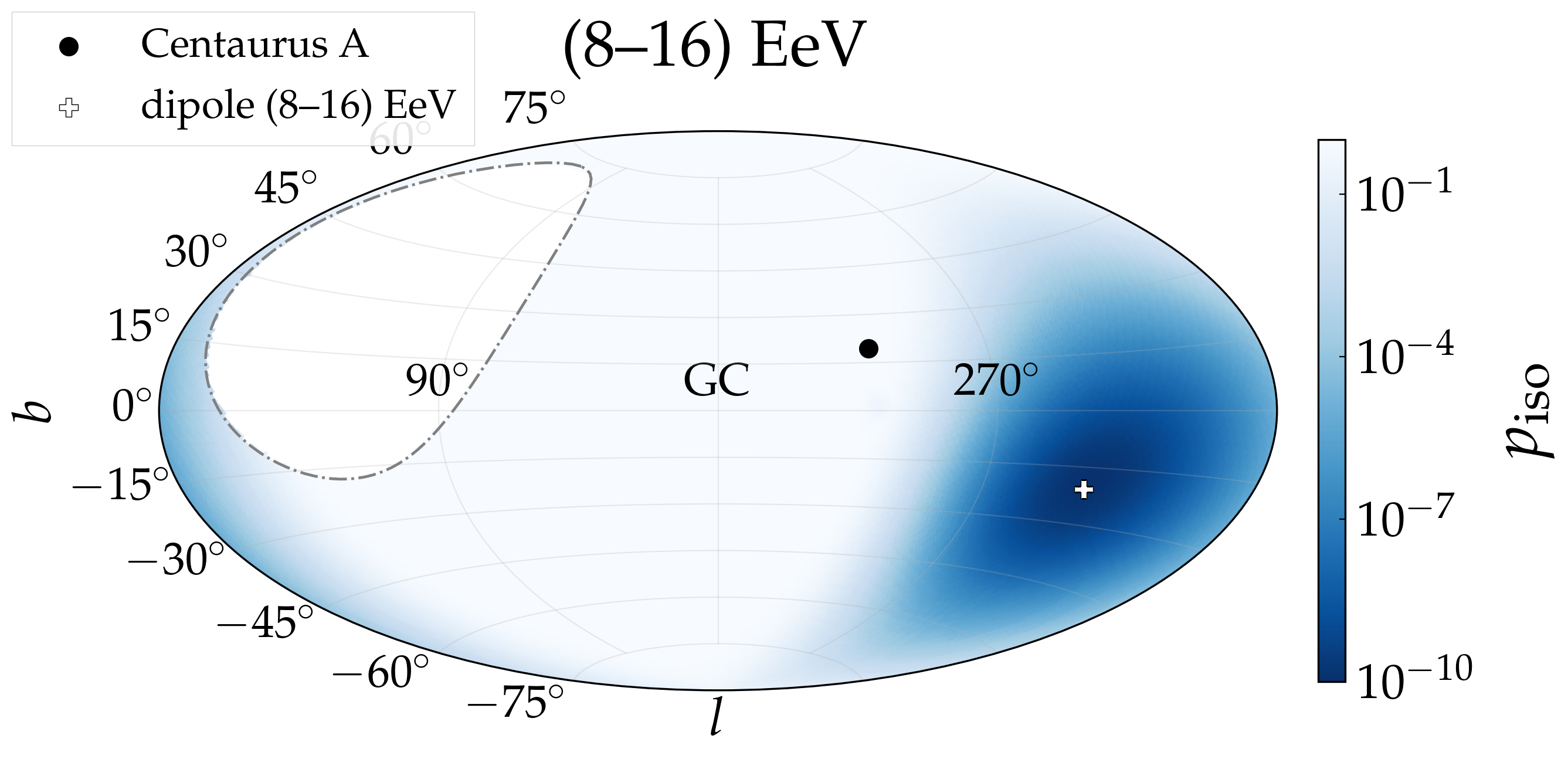}
        \caption{}
    \end{subfigure}
    \begin{subfigure}{0.47\textwidth}
        \includegraphics[width=\linewidth]{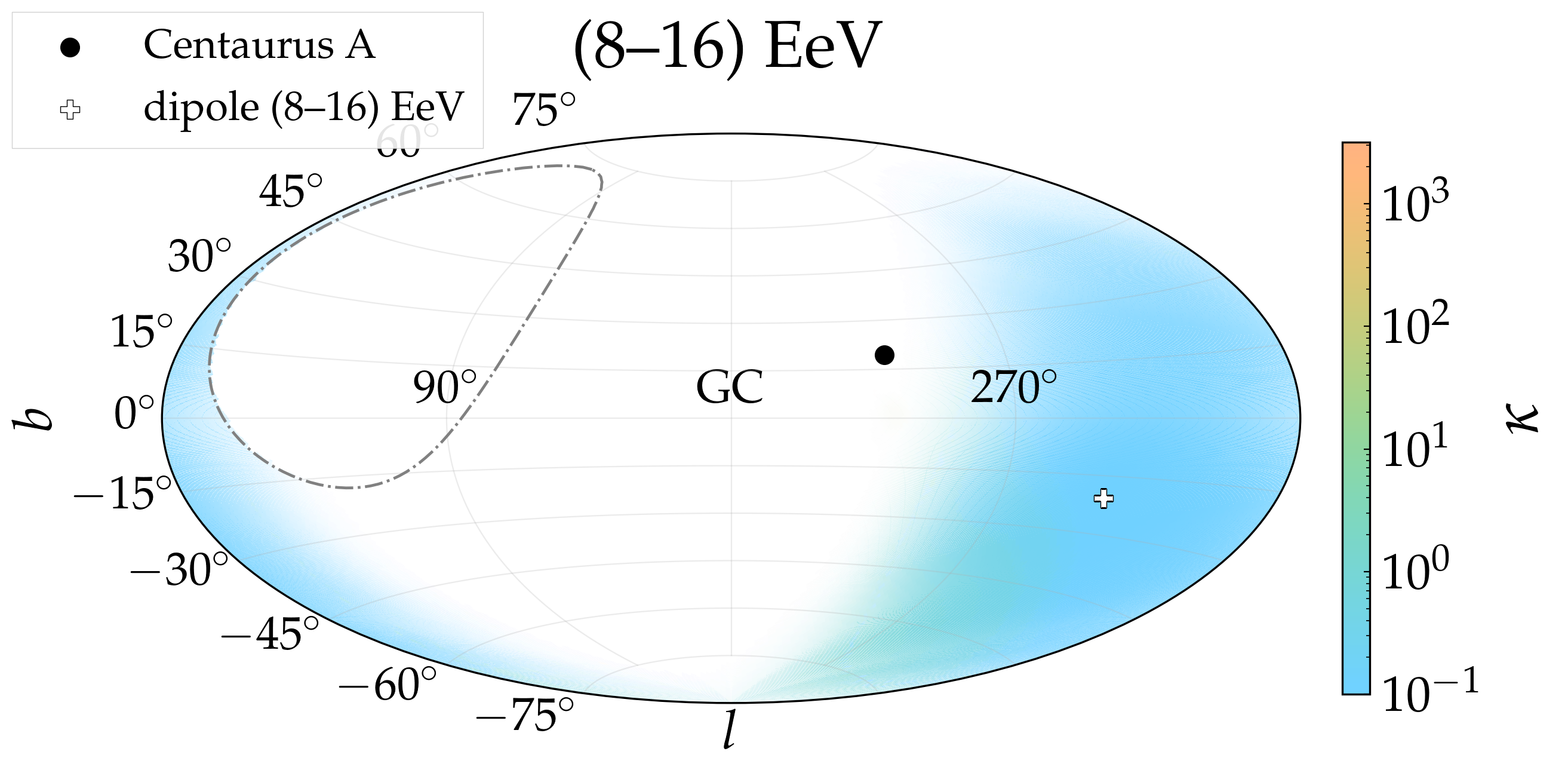}
        \caption{}
    \end{subfigure}
    \\[0.01cm]
    \begin{subfigure}{0.47\textwidth}
        \includegraphics[width=\linewidth]{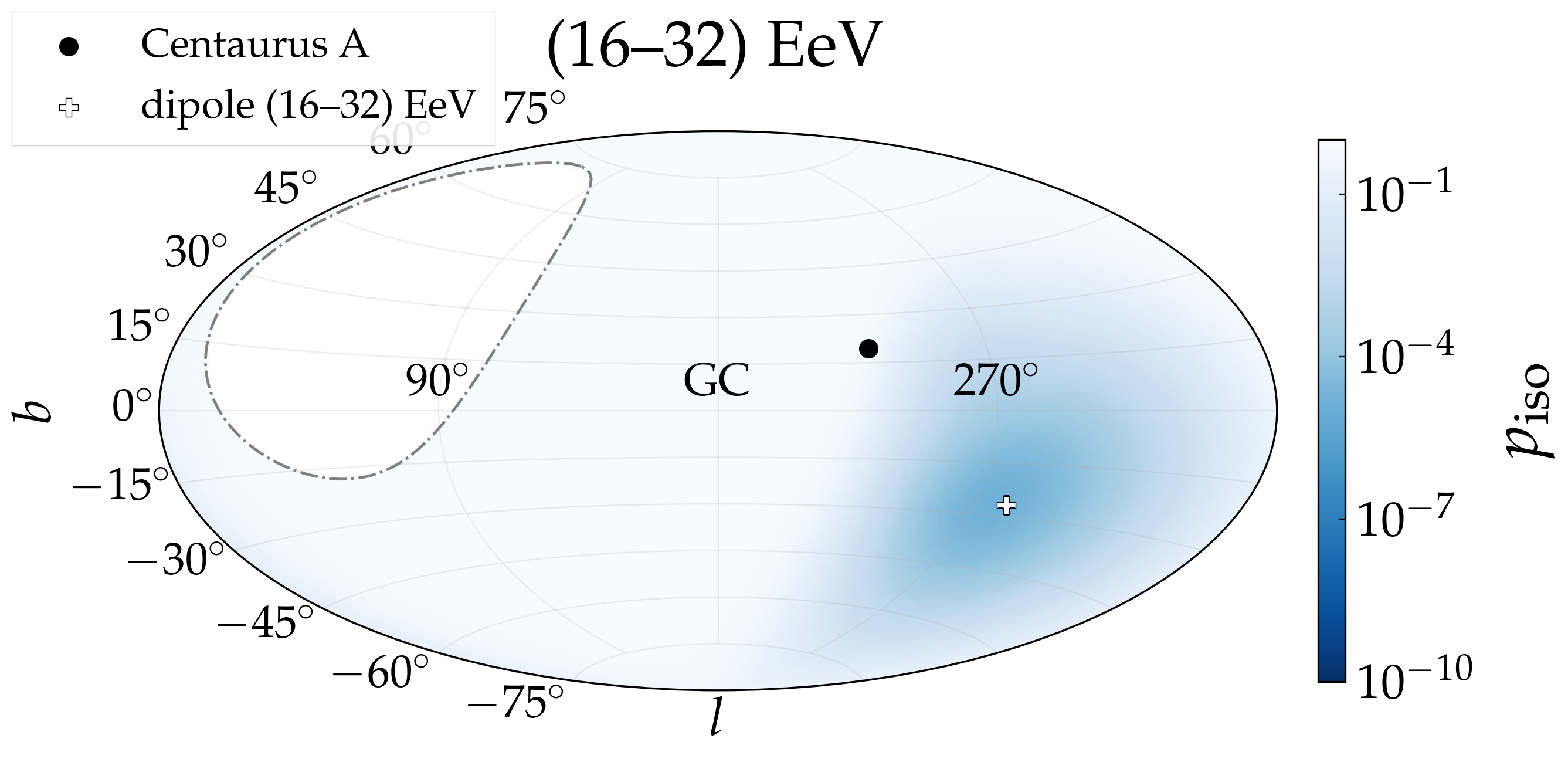}
        \caption{}
    \end{subfigure}
    \begin{subfigure}{0.47\textwidth}
        \includegraphics[width=\linewidth]{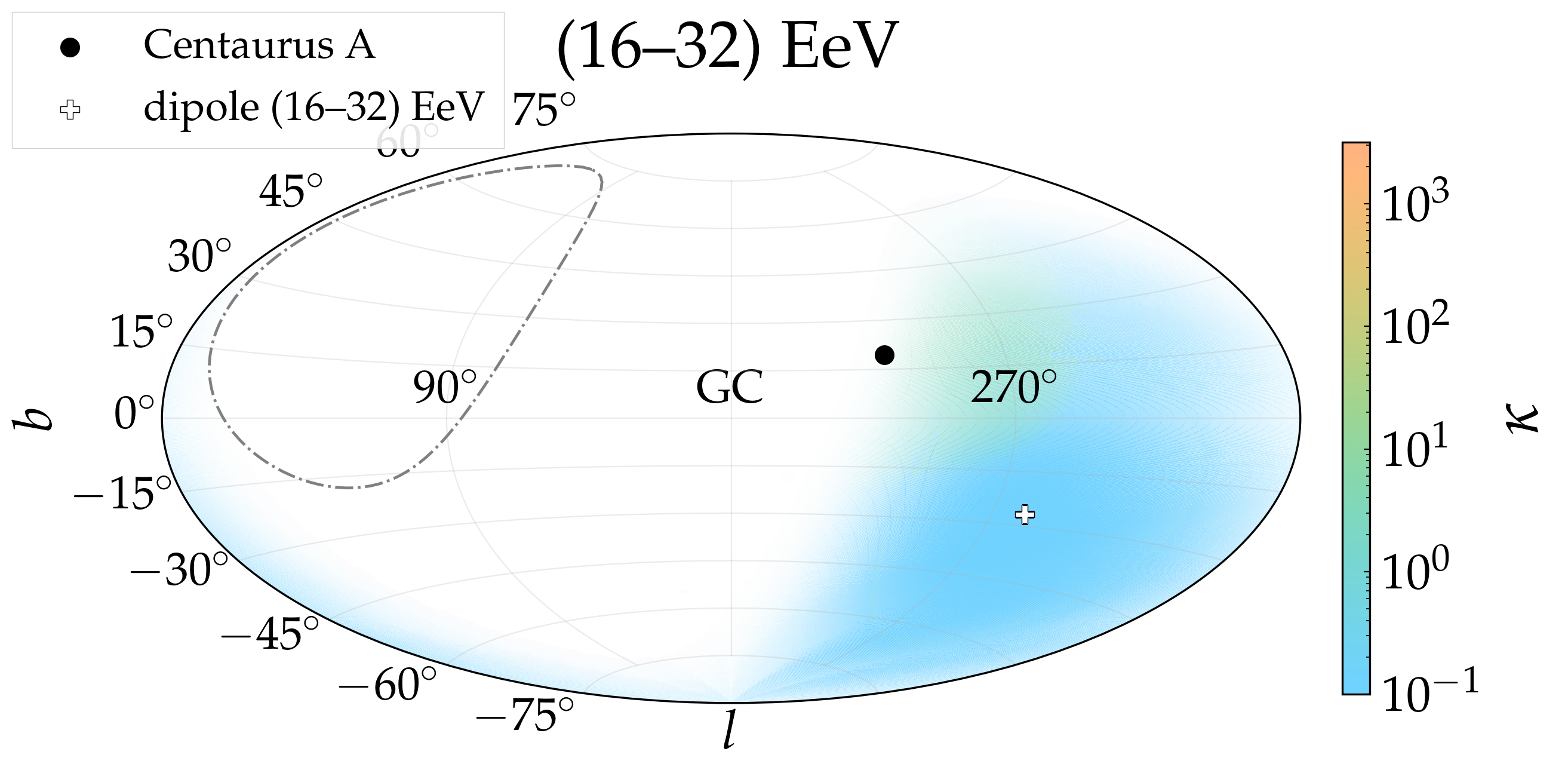}
        \caption{}
    \end{subfigure}
    \\[0.01cm]
    \begin{subfigure}{0.47\textwidth}
        \includegraphics[width=\linewidth]{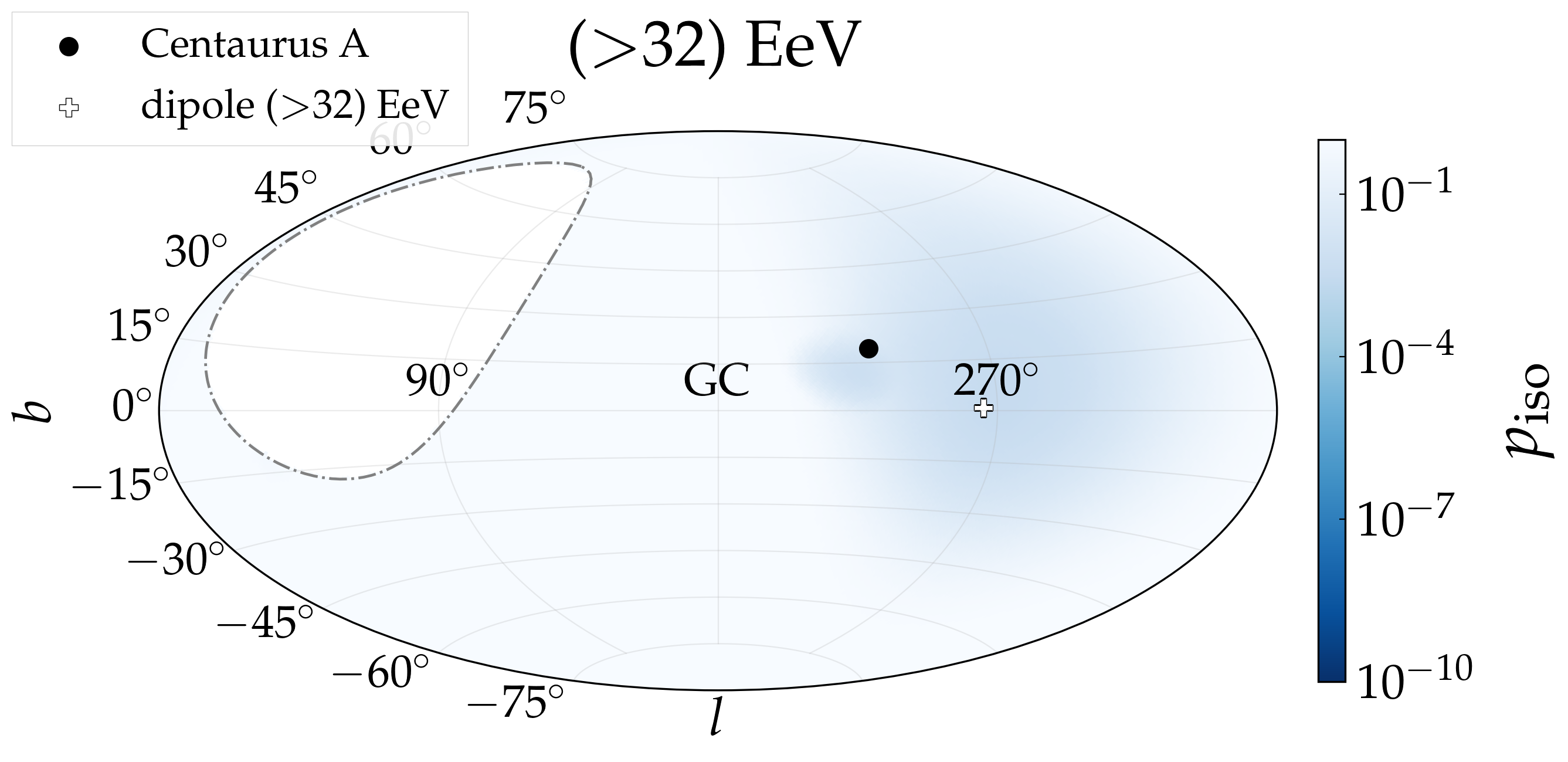}
        \caption{}
    \end{subfigure}
    \begin{subfigure}{0.47\textwidth}
        \includegraphics[width=\linewidth]{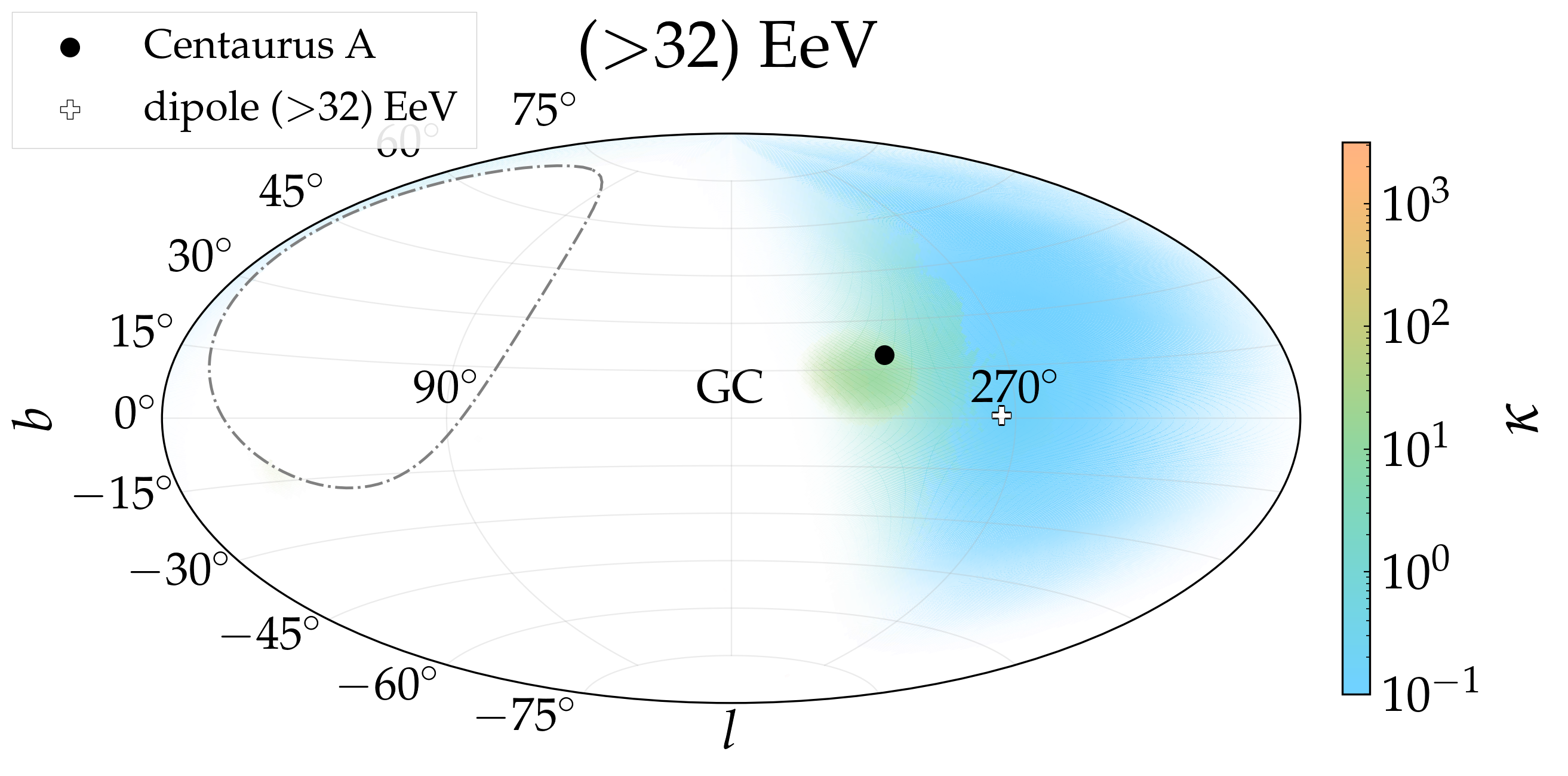}
        \caption{}
    \end{subfigure}
    \caption{
    Maps in Galactic coordinates of the lowest $p_\mathrm{iso}$ (left column) and $\kappa$ (right column) obtained by maximizing over $\alpha$ while profiling over $\kappa$.
    The maps are derived by profiling over the reference position using a \texttt{HEALPix} grid with $\texttt{nSide}=64$.
    In the $\kappa$ maps, the opacity of each point is scaled according to the ratio between the local $\log_{10} p_\mathrm{iso}$ value and the minimum $\log_{10} p_\mathrm{iso}$ in the map.
    The maps are depicted for the energy intervals (a,b) (4--8)\,EeV, (c,d) (8--16)\,EeV,  (e,f) (16--32)\,EeV, and (g,h) (>32)\,EeV. 
    In each plot, the position of Centaurus~A and the position of the center of the dipole modulation obtained for the corresponding energy interval in ref.\ \cite{ref:lsa} are indicated with a black dot and a white cross, respectively.
    }
    \label{fig:maps_ts}
\end{figure*}

In the left column of figure~\ref{fig:maps_ts}, we show maps in Galactic coordinates of the lowest $p_\mathrm{iso}$ obtained by maximizing equation~\eqref{eq:ts} while profiling over $\kappa$ and the reference direction $\bm{u}_0$.
In these maps, we compute the profile likelihood as a function of $\kappa$ and $\bm{u}_0$.
For each reference direction, using the center positions of a \texttt{HEALPix} grid~\cite{ref:healpix} defined with the parameter $\texttt{nSide}=64$ and for each $\kappa$, we find the $\alpha$ that minimizes $-2\ln(\mathcal{L}(\alpha;\bm{u}_0,\kappa))$.
We then evaluate $p_\mathrm{iso}$ associated to each scanned value of $\kappa$.
In each pixel of the maps, we show the lowest $p_\mathrm{iso}$ derived.

In the right column of the same figure, the corresponding maps with the best-fit $\kappa$ are depicted.
In this second set of maps, the opacity of each point is scaled according to the ratio between the local $\log_{10} p_\mathrm{iso}$ value and the minimum $\log_{10} p_\mathrm{iso}$ in the map.
This procedure suppresses large-$p_\mathrm{iso}$ areas, with white areas corresponding to values of $p_\mathrm{iso}\simeq1$.
Dashed curves represent the boundaries of the field of view of the Pierre Auger Observatory.
For the (8--16)\,EeV and (16--32)\,EeV intervals, the minimum $p_\mathrm{iso}$ is associated with low values of $\kappa$.
For (>32)\,EeV, two distinct regions appear: one at low $\kappa$, around the dipole direction, and a second at intermediate $\kappa$, located within the $27^\circ$ circular region around Centaurus~A (we note that the position of Centaurus~A in Galactic coordinates is ($l,b$)=(309.5$^\circ$, 19.4$^\circ$)).

\newpage

\subsection{Results for additional energy bins}
\label{sec:res_add}

\begin{table}[tbp]
    \renewcommand{\arraystretch}{1.3}
    \begin{center}
    \begin{tabular}{|c|c|c|c|c|c|c|c|c|}
        \hline
         $\bm{E/}$\,\textbf{EeV} & $\bm{\kappa}$ & $\bm{\alpha\,(\%)}$ &  $\bm{\textbf{R.A.}/{}^\circ}$ & $\bm{{\textbf{dec}/{}^\circ}}$ & $\bm{\mathrm{TS}}$ & $\bm{d\,(\%)}$ & $\bm{\sigma/{}^\circ}$ & $\bm{p_\textbf{iso}}$ \\
        \hline
        \phantom{0}4--8\phantom{0} & $2.5\phantom{0}$ & $\phantom{0}0.7^{+0.2}_{-0.2}$ & $112^{+26}_{-33}$ & $-65^{+16}_{-13}$ & $10.1$ & --- & $36.2$ & $0.023$ \\
        \phantom{0}5--10 & $2.0\phantom{0}$ & $\phantom{0}1.4^{+0.3}_{-0.3}$ & $129^{+12}_{-13}$ & $-40^{+13}_{-12}$ & $18.5$ & --- & $40.6$ & $4.6\times10^{-4\phantom{0*}}$ \\
        \phantom{0}6--13 & $0.10$ & $\phantom{0.}39^{+8\phantom{.0}}_{-8\phantom{.0}}$ & $\phantom{0}98^{+12}_{-12}$ & $-28^{+14}_{-13}$ & $27.4$ & $3.8^{+0.8}_{-0.8}$ & --- & $4.9\times10^{-6*\phantom{0}}$ \\
        \phantom{0}8--16 & $0.10$ & $\phantom{0}67^{+10\phantom{.}}_{-10\phantom{.}}$ & $\phantom{0}94^{+9\phantom{0}}_{-9\phantom{0}}$ & $-28^{+10}_{-10}$ & $49.2$ & $6.7^{+1.0}_{-1.0}$ & --- & $1.1\times10^{-10}{}^*$ \\
        10--20 & $0.10$ & $\phantom{0}69^{+12\phantom{.}}_{-12\phantom{.}}$ & $\phantom{0}95^{+11}_{-11}$ & $-30^{+12}_{-11}$ & $34.1$ & $6.9^{+1.2}_{-1.2}$ & --- & $1.9\times10^{-7*\phantom{0}}$ \\
        13--25 & $0.10$ & $\phantom{0}87^{+13\phantom{.}}_{-18\phantom{.}}$ & $\phantom{0}87^{+15}_{-15}$ & $-50^{+12}_{-10}$ & $26.9$ & $8.7^{+1.8}_{-1.7}$ & --- & $5.9\times10^{-6*\phantom{0}}$ \\
        16--32 & $0.10$ & $\phantom{.}100_{-17\phantom{.}}$ & $\phantom{0}95^{+17}_{-16}$ & $-53^{+11}_{-10}$ & $25.2$ & $\phantom{.}10^{+2\phantom{.0}}_{-2}$ & --- & $1.3\times10^{-5*\phantom{0}}$ \\
        20--40 & $0.13$ & $\phantom{0,}95^{+5\phantom{.0}}_{-22\phantom{.}}$ & $100^{+18}_{-18}$ & $-54^{+12}_{-12}$ & $21.2$ & $\phantom{.}12^{+3\phantom{.0}}_{-3}$ & --- & $8.9\times10^{-5*\phantom{0}}$  \\
        25--51 & $0.13$ & $\phantom{.}100_{-19\phantom{.}}$ & $108^{+17}_{-17}$ & $-45^{+14}_{-13}$ & $19.4$ & $\phantom{.}13^{+4\phantom{.0}}_{-3}$ & --- & $2.3\times10^{-4\phantom{0*}}$ \\
        32--64 & $0.16$ & $\phantom{0.}95^{+5}_{-26\phantom{.}}$ & $143^{+18}_{-18}$ & $-39^{+18}_{-15}$ & $14.4$ & $\phantom{.}15^{+5\phantom{.0}}_{-4}$ & --- & $2.3\times10^{-3\phantom{0*}}$ \\
        40--$\infty$ & $16\phantom{.0}$ & $\phantom{0}2.7^{+0.7}_{-0.6}$ & $210^{+7\phantom{0}}_{-7\phantom{0}}$ & $-48^{+4\phantom{0}}_{-4\phantom{0}}$ & $22.9$ & --- & $14.4$ & $2.4\times10^{-4\phantom{0*}}$ \\
        \hline
    \end{tabular} 
    \end{center}
    \caption{\label{tab:add} Same as table~\ref{tab:st} with the additional energy bins defined by the sliding window in energy.}
\end{table}

In this section, we analyze the data using a sliding window spanning a factor of 2 in energy.
The $i$-th energy interval ($E_{\min,i}$--$E_{\max,i}$)\,EeV is defined as 
\begin{align}
    E_{\min,i} &= 2^{2+i/3}\,\mathrm{EeV}, \quad i = 0, 1, \dots, 10; \\
    E_{\max,i} &= 2^{3+i/3}\,\mathrm{EeV}, \quad i = 0, 1, \dots, 9,
\end{align}
both rounded to the nearest $\mathrm{EeV}$; and $E_{\max,10} = +\infty$.
A total of 11 energy bins are selected with the sliding window procedure, the last one being a cumulative bin.
The full set of intervals is (4--8, 5--10, 6--13, 8--16, 10--20, 13--25, 16--32, 20--40, 25--51, 32--64, >40)\,EeV.
We underline that this sliding window in energy includes the three non-open energy intervals adopted in previous anisotropy studies on large angular scales, i.e.\ (4--8, 8--16, 16--32)\,EeV.

The results derived by performing the scan in $\kappa$ yielding the lowest $p_\mathrm{iso}$ for each of the energy bins defined with the sliding window procedure are listed in table~\ref{tab:add}.
As in table~\ref{tab:st}, we report the value of $\kappa$, the best-fit estimators, and the corresponding $\mathrm{TS}$ for the outcome yielding the lowest $p_\mathrm{iso}$ in each energy interval.
The equivalent dipole amplitudes or Gaussian widths are quoted depending on whether $\kappa<1$ or $\kappa\geq1$.

Dipole amplitudes are derived for the energy intervals (6--13, 8--16, 10--20, 13--25, 16--32, 20--40, 25--51, 32--64)\,EeV.
We note that the dipole amplitude increases with energy, in agreement with the trend reported in~\cite{ref:lsa}.
The corresponding $p_\mathrm{iso}$ decreases rapidly with energy, reaching the lowest value for (8–16)\,EeV, and then it gradually increases at higher energies.

For (>40)\,EeV, the value of $\kappa\simeq16$ corresponding at the lowest $p_\mathrm{iso}$ describes an excess of intermediate angular scale ($\sigma=14.4^\circ$).
The central position of the excess is located within the $27^\circ$ region around Centaurus~A. 

\begin{figure*}[!h]
    \centering
    \begin{subfigure}{0.47\textwidth}
        \includegraphics[width=\linewidth]{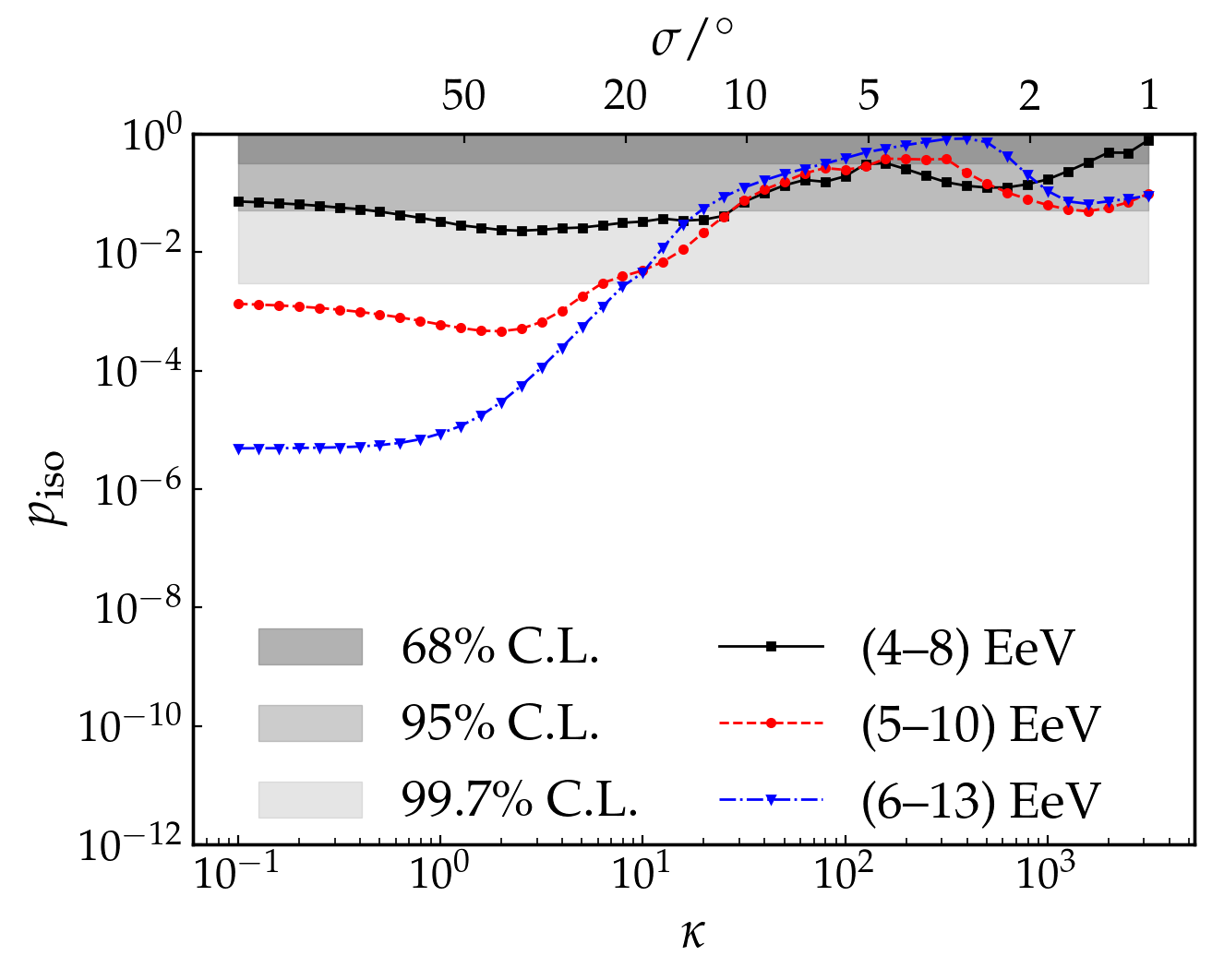}
        \caption{}
    \end{subfigure}\hspace{0.1cm}
    \begin{subfigure}{0.47\textwidth}
        \includegraphics[width=\linewidth]{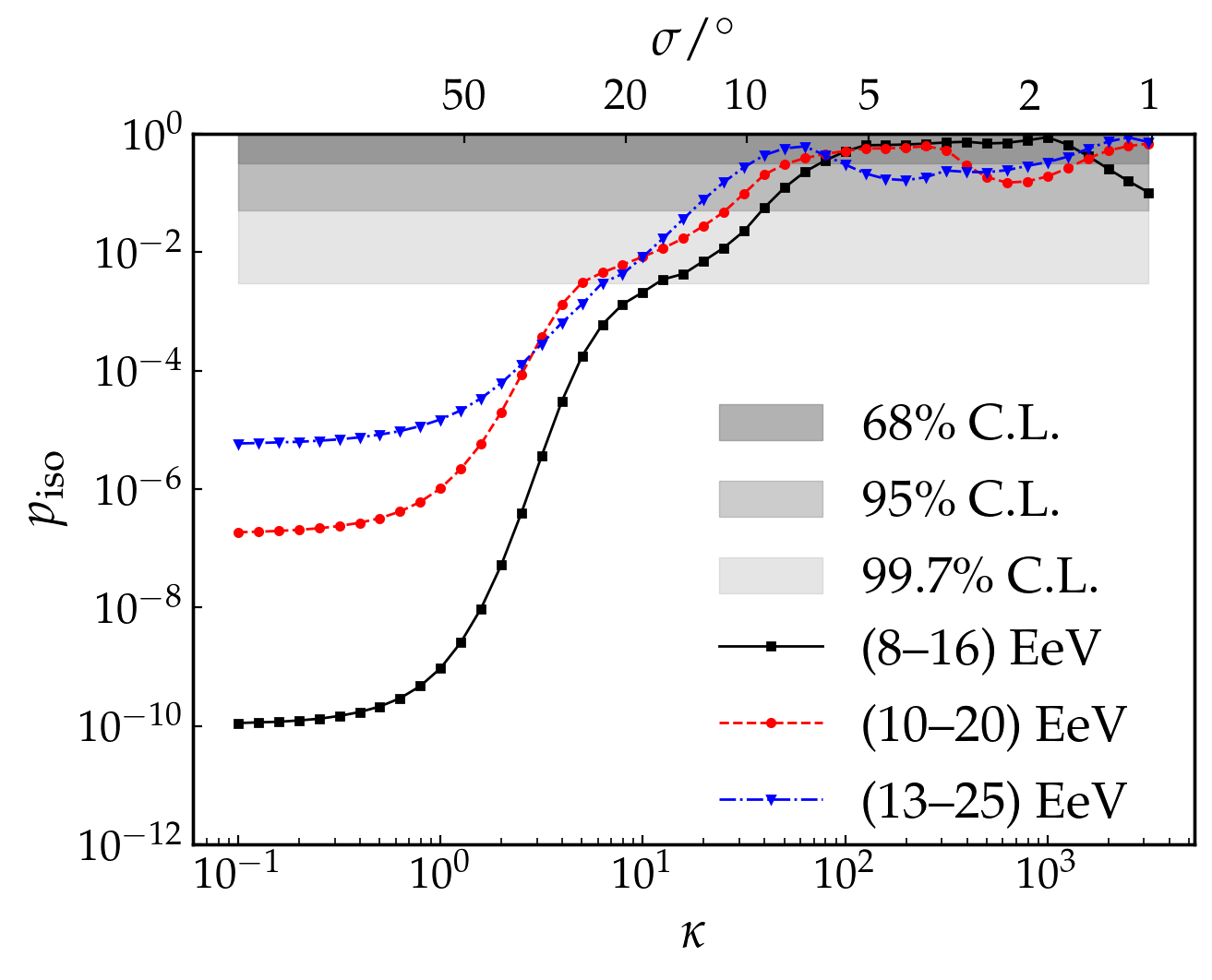}
        \caption{}
    \end{subfigure}
    \\[0.1cm]
    \begin{subfigure}{0.47\textwidth}
        \includegraphics[width=\linewidth]{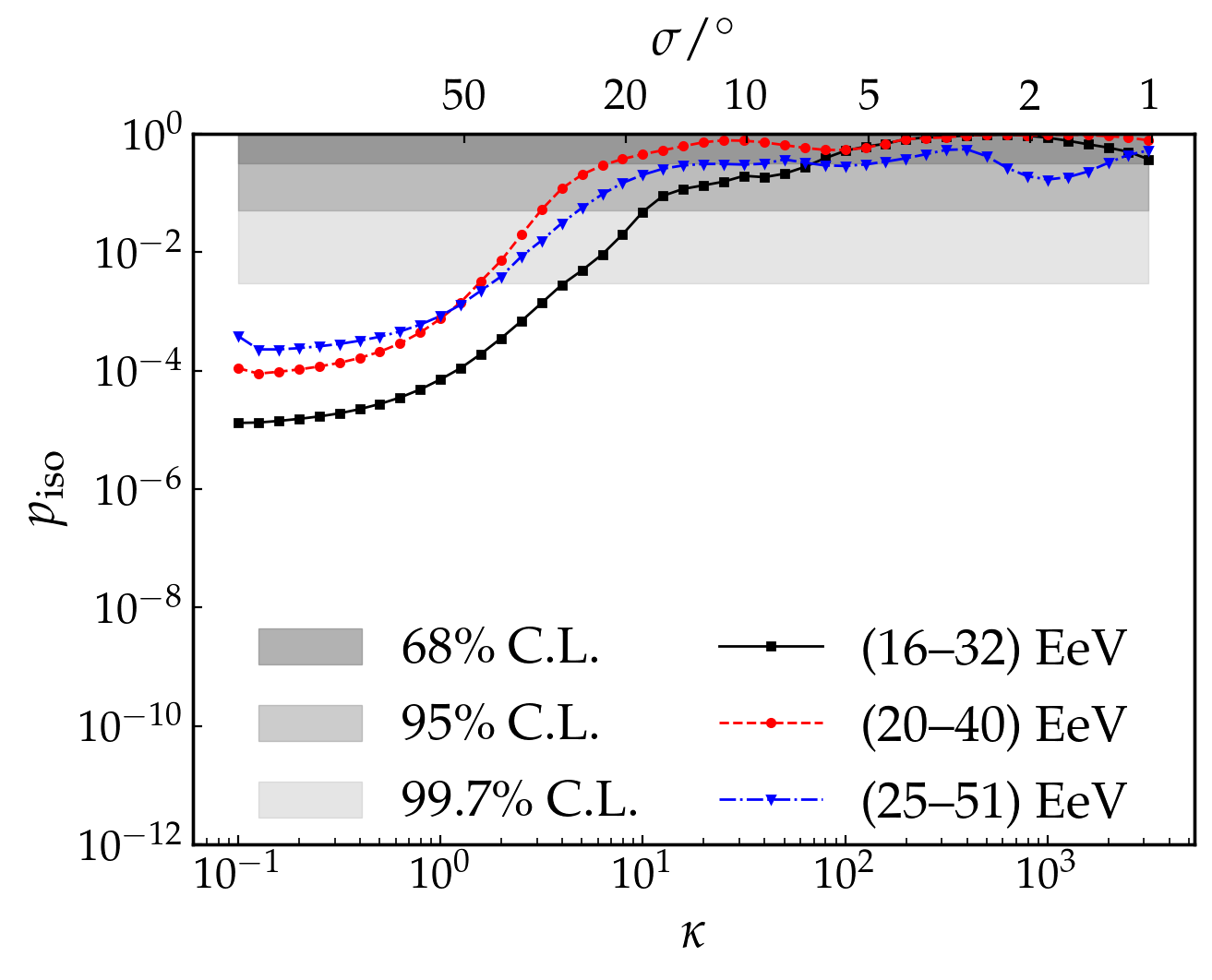}
        \caption{}
    \end{subfigure}\hspace{0.1cm}
    \begin{subfigure}{0.47\textwidth}
        \includegraphics[width=\linewidth]{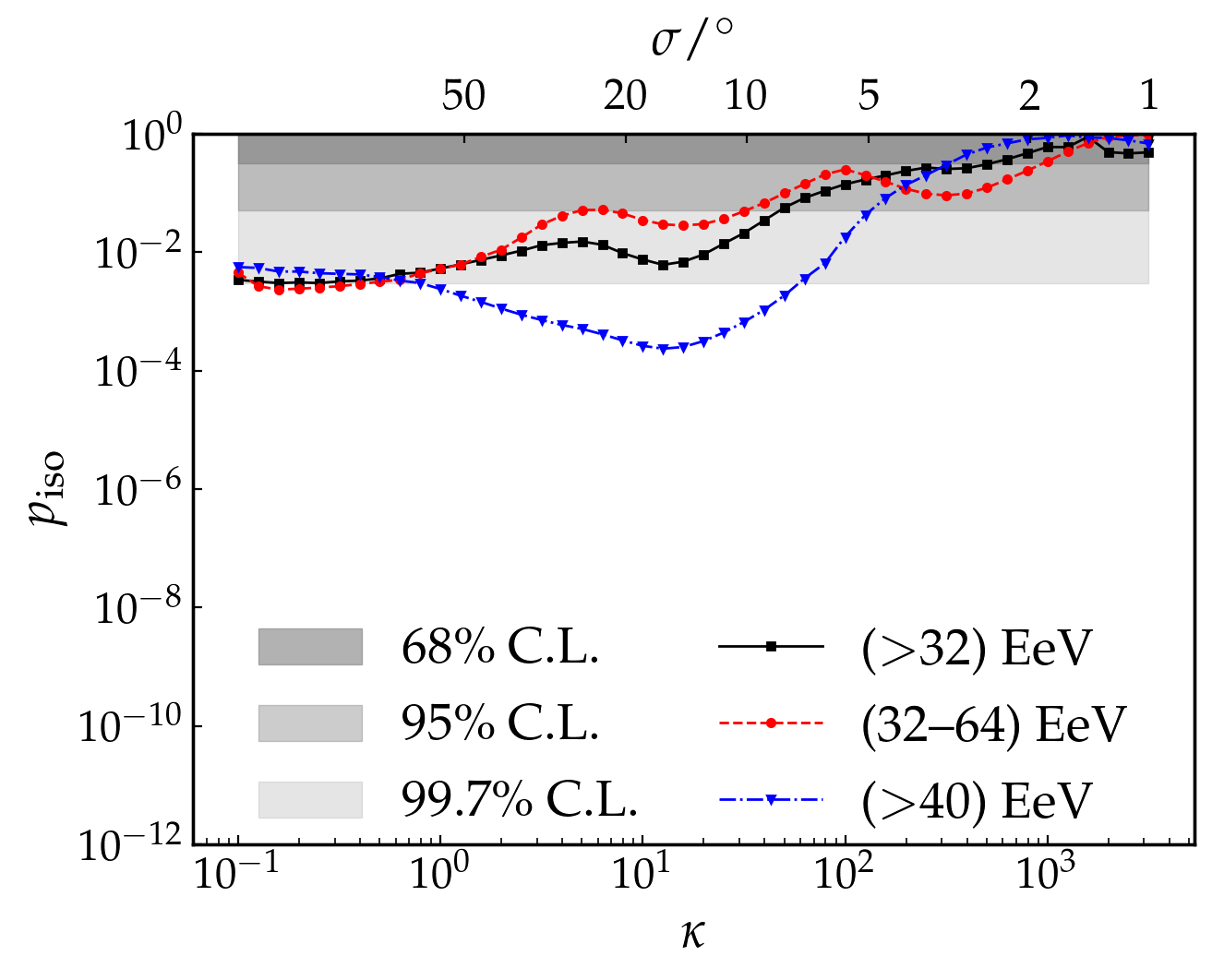}
        \caption{}
    \end{subfigure}
    \caption{Lowest $p_\mathrm{iso}$ derived by profiling over the $\kappa$ parameters.  
    The results are reported for the energy bins (a) (4--8), (5--10), and (6.3--12.6)\,EeV (b) (8--16), (10--20), and (12.6--25)\,EeV, (c) (16--32), (20--40), and (25--50)\,EeV, and (d) (>32), (32--64), and (>40)\,EeV.
    For each plot, the shadow bands correspond to the 68$\%$, 95$\%$, and 99.7$\%$ C.L.\ intervals.}
    \label{fig:plots_add}
\end{figure*}

The full sets of values of $p_\mathrm{iso}$ obtained after profiling over $\kappa$ are shown in figure~\ref{fig:plots_add}.
The energy intervals are grouped into four plots.
We note that the profiling over $\kappa$ in the (5--10)\,EeV interval yields to results with $p_\mathrm{iso}$ outside the $99.7\%$ C.L.\ starting at $\kappa=10^{-1}$, and the lowest $p_\mathrm{iso}$ is observed for $\kappa=2$.
With the exception of the (4--8), (>32), and (>40)\,EeV interval, all energy bins exhibit excesses outside of the $99.7\%$ C.L.\ band at $\kappa<1$.

In the interval (32--64)\,EeV, the most significant excess observed is at $\kappa\simeq0.16$, outside the $99.7\%$ C.L.\ band.
Moreover, a second excess outside the $95\%$\ C.L. band is present at $\kappa\simeq20$.
This excess at $\kappa\simeq20$ describes an intermediate-angular-scale overdensity located in the Centaurus region.

The combination reverses in the energy interval (>40)\,EeV.
In this bin, the most significant configuration corresponds to an overdensity at intermediate angular scales with $\kappa\simeq16$.
However, an excess outside the $99.7\%$ C.L.\ band is also present at $\kappa<1$.
The additional plots, analogous to the ones presented in figures \ref{fig:maps_ts} and  \ref{fig:maps_flux}, for the energy intervals (25--50), (32--64), and (>40)\,EeV are reported in appendix \ref{sec:appendix}.

In figure \ref{fig:plots_add}, an interesting feature appears in the highest-energy intervals, with the presence of two spatially separated excesses, one on large angular scales and another one on intermediate angular scales, outside the 95$\%$ C.L.\ band. 
We tested whether one of the excesses arises as a statistical fluctuation of the other.
To study this hypothesis, we take into consideration the energy interval (>32)\,EeV, where the most significant excess is found on large angular scales, with $\kappa=0.16$ and $p_\mathrm{iso}=3.2\times10^{-3}$ (corresponding to $d=0.14^{+0.04}_{-0.04}$) with center position $(\mathrm{R.A.}, \mathrm{dec})=(141^\circ, -44^\circ)$, and an excess on intermediate angular scales is also present with $\kappa=13$ and $p_\mathrm{iso}=6.1\times10^{-3}$ (corresponding to $\sigma=16.1^\circ$) and center position $(\mathrm{R.A.}, \mathrm{dec})=(209^\circ, -50^\circ)$.
The distance between the centers of the two excesses is $\Theta=(45^{+14}_{-14}){}^\circ$.
We perform a series of simulations of events distributed according to the likelihood present in equation~\eqref{eq:likelihood}.
The parameters ($\kappa$,$\alpha$,$\bm{u}_0$) of the likelihood are selected according to the most significant excess found in the interval (>32)\,EeV, and the number of events present in the simulations follows the number of events present in (>32)\,EeV.
We then maximize the $\mathrm{TS}$ while profiling over $\log_{10}{k}$, from $\log_{10}{k}=0.8$ to $\log_{10}{k}=2$ in steps of $0.1$.
We evaluate the frequency of obtaining an intermediate angular scale excess that is as significant as the intermediate-scale excess found in the interval (>32)\,EeV and as spatially distant from the large-scale excess as in the interval (>32)\,EeV.
In other words, we evaluate the frequency of having $p_\mathrm{iso}<6.1\times10^{-3}$ and $\Theta/\sigma_\Theta>3.3$, where $\sigma_\Theta=14^\circ$ is the uncertainty on the angular separation.
From simulations, the frequency of finding an excess on intermediate angular scales sufficiently significant and distant enough from the large-scale excess position observed on data is $2.8\times10^{-3}$.
Completing the analogous procedure on isotropic simulations, we find that the frequency of obtaining an excess on intermediate angular scales with enough separation is $2.6\times10^{-3}$.

We complete a similar procedure to study whether the large-scale excess is a statistical fluctuation of the intermediate-scale one.
We perform simulations considering the presence of an intermediate angular scale excess comparable to the ones observed on data, distributed according to the likelihood present in equation~\eqref{eq:likelihood}.
The parameters ($\kappa$,$\alpha$,$\bm{u}_0$) of the likelihood are selected according to the most significant excess of intermediate angular scale found in the interval (>32)\,EeV.
We then evaluate the frequency of obtaining an excess on large angular scales as significant as the one observed on data and enough distant from the intermediate-scale excess.
The presence of a large-scale excess is derived profiling over $\log_{10}{\kappa}$, from $\log_{10}{\kappa}=-1$ to $\log_{10}{\kappa}=0$ in steps of $0.1$.
In this case, we evaluate the frequency of having $p_\mathrm{iso}<3.2\times10^{-3}$ and $\Theta/\sigma_\Theta>3.3$.
We repeat the same procedure performing isotropic simulations.
We find that the frequency is $0.014$, while the frequency from isotropic simulations is $0.011$, therefore the probability of finding an excess at large angular scale is similar whether an intermediate-scale excess is present or not.
Based on these simulations, there is no evidence to suggest that the large-scale (intermediate-scale) excess above 32 EeV arises from a statistical fluctuation of the intermediate-scale (large-scale) excess.

\begin{figure*}[!h]
    \centering
    \begin{subfigure}{0.47\textwidth}
    \includegraphics[width=\linewidth]{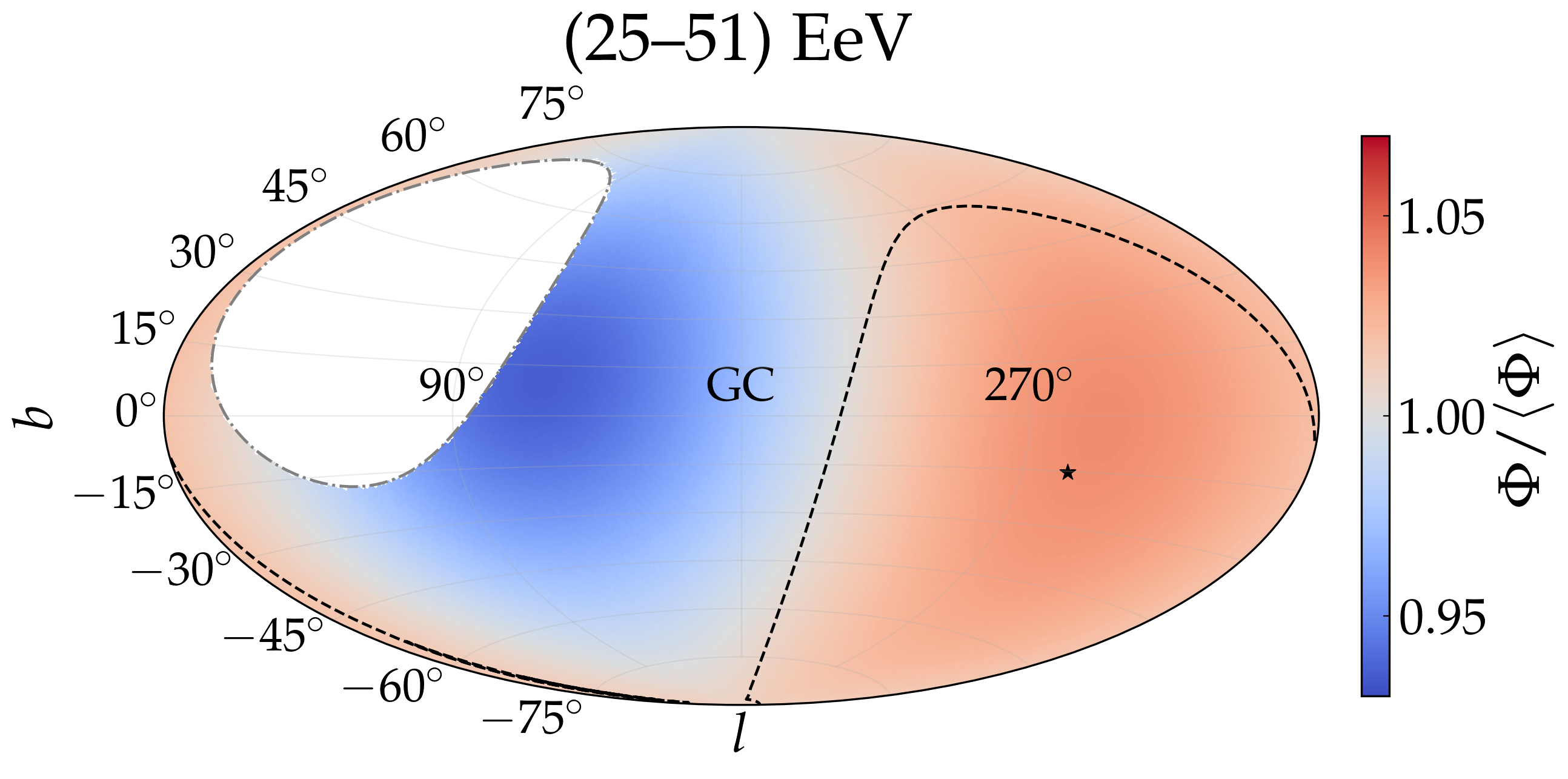}
        \caption{}
    \end{subfigure}
    \begin{subfigure}{0.47\textwidth}
    \includegraphics[width=\linewidth]{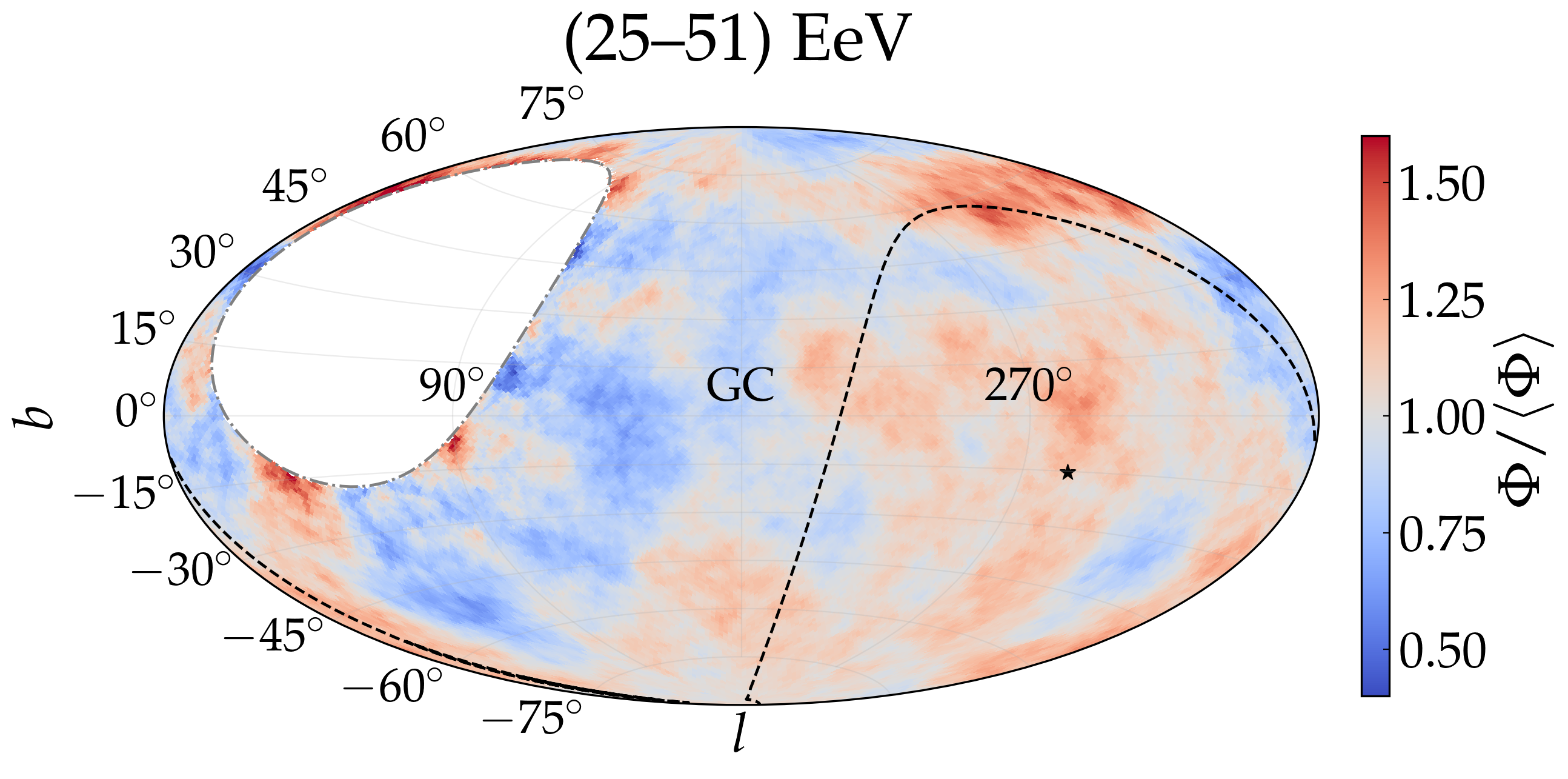}
    \caption{}
    \end{subfigure}
    \\[0.2cm]
    \begin{subfigure}{0.47\textwidth}
    \includegraphics[width=\linewidth]{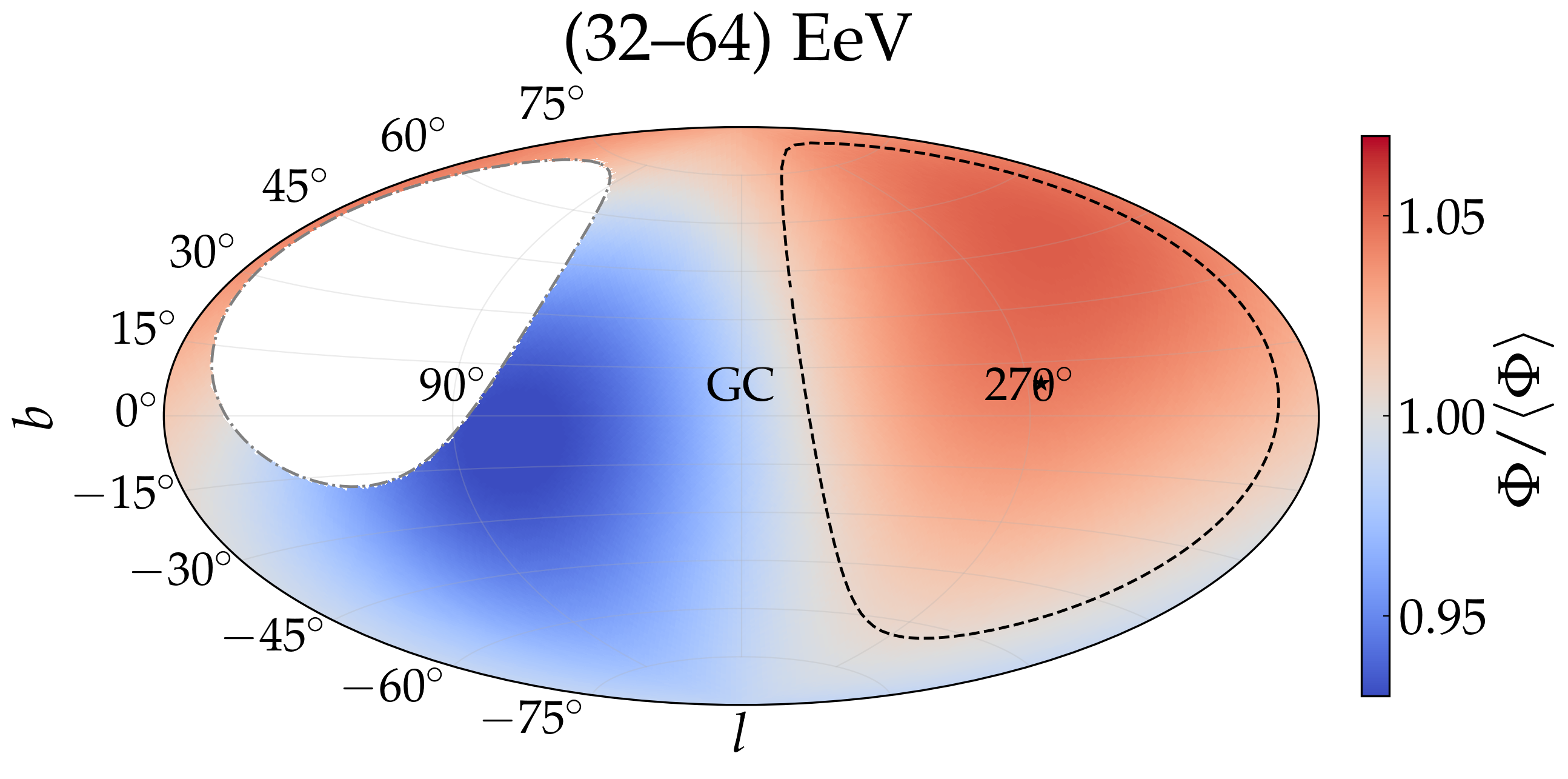}
    \caption{}
    \end{subfigure}
    \begin{subfigure}{0.47\textwidth}
    \includegraphics[width=\linewidth]{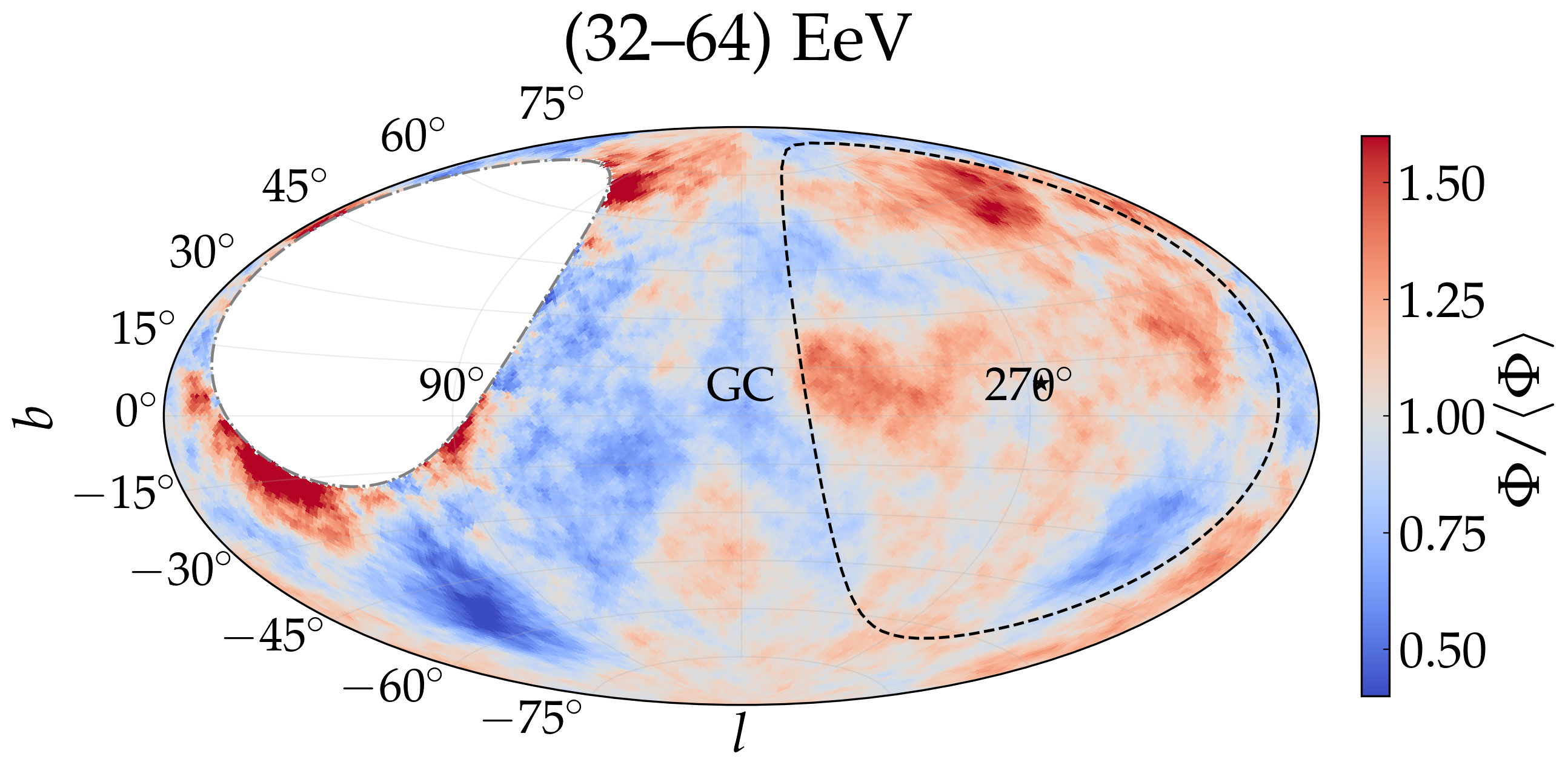}
    \caption{}
    \end{subfigure}
    \\[0.2cm]
    \begin{subfigure}{0.47\textwidth}
    \includegraphics[width=\linewidth]{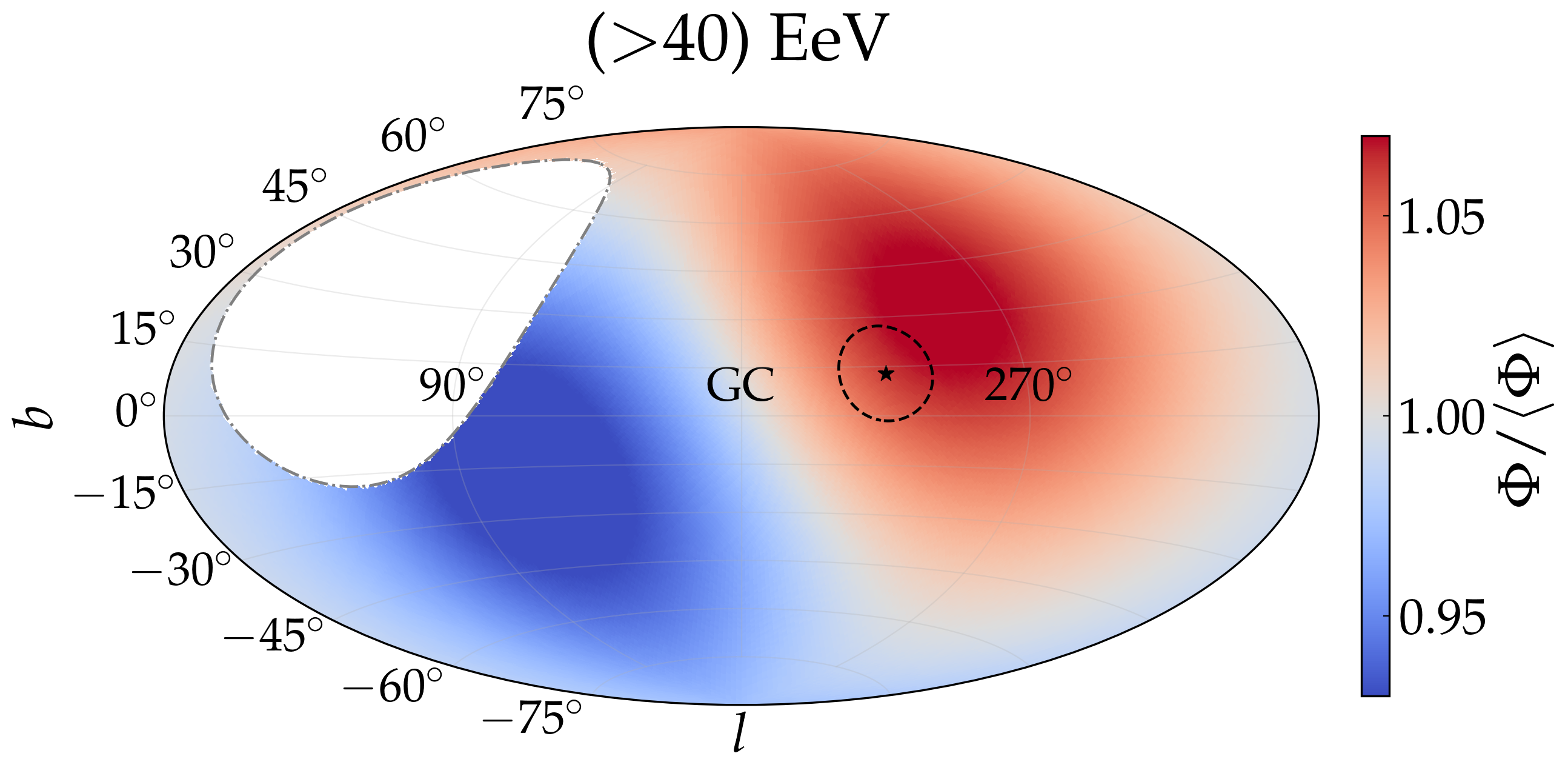}
    \caption{}
    \end{subfigure}
    \begin{subfigure}{0.47\textwidth}
    \includegraphics[width=\linewidth]{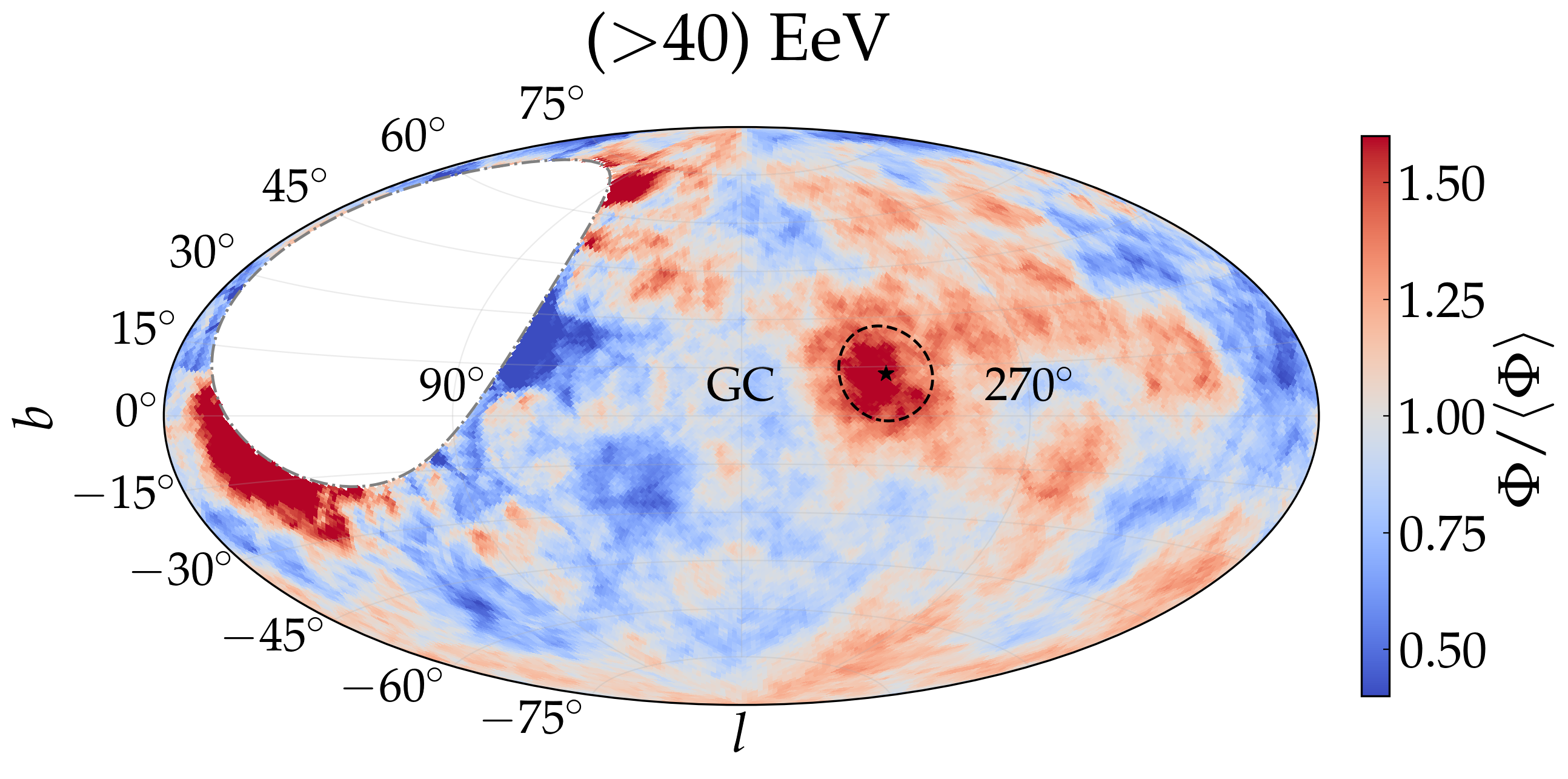}
        \caption{}
    \end{subfigure}
    \caption{Maps in Galactic coordinates of the flux for the energy intervals (a,b) (25--50)\,EeV, (c,d) (32--64) EeV, and (e,f) (>40)\,EeV.
    The maps are smoothed by a von Mises--Fisher distribution with concentration parameter $\kappa$ correspondent to $\sigma=45^\circ$ (left column) and $\sigma=15^\circ$ (right column).
    In each map, the black star indicates the position of the excess direction associated to the lowest $p_\mathrm{iso}$ obtained maximizing $\mathrm{TS}$ while profiling over $\kappa$.
    The dashed lines depict the region containing $39.3\%$ of the flux from the von Mises--Fisher distribution.
    }
    \label{fig:maps_flux}
\end{figure*}

In figure~\ref{fig:maps_flux}, maps of the flux $\Phi$ normalized to its mean value $\langle\Phi\rangle$ are depicted for the energy intervals (25--51)\,EeV, (32--64)\,EeV, and (>40)\,EeV.
To better compare the excesses observed at large and intermediate angular scales, in the left column, the flux maps are smoothed by a von Mises--Fisher distribution with concentration parameter $\kappa$ correspondent to $\sigma = 45^\circ$ ($\kappa\simeq1.5$), while in the right column we use $\sigma = 15^\circ$ ($\kappa\simeq15$). 
The two selections reproduce the smearing compatible with large-scale excesses $\sigma=45^\circ$ and intermediate-scale excesses $\sigma=15^\circ$.
In each map, the black star indicates the position of the excess direction associated to the lowest $p_\mathrm{iso}$ obtained maximizing $\mathrm{TS}$ while profiling over $\kappa$.
The dashed circles indicate the regions centered on the reference positions of the excesses for which the integral of the von Mises--Fisher distribution, using the parameters corresponding to the lowest value of $p_\mathrm{iso}$, contains 39.3$\%$ of the flux from the von Mises--Fisher distribution. 
This choice allows a direct comparison with the region enclosed by the $1\sigma$ contour of a two-dimensional Gaussian distribution.
Areas with large values of $\Phi/\langle \Phi\rangle$ are present at the edge of the field of view in all the maps with $\sigma=15^\circ$.
These are border effects given by the low statistics of events present in those regions.
In the maps with smoothing $\sigma=45^\circ$, $\Phi/\langle\Phi\rangle$ increases with energy.
This is in agreement with the increase with energy of the dipole amplitude observed on the data.
In the maps with $\sigma=15^\circ$, we observe the presence of several areas with $\Phi/\langle\Phi\rangle$ above 1.
In particular, an overdensity is present in the Centaurus region for all energy intervals considered.
The flux ratio of this region increases with energy, and has its maximum in the (>40)\,EeV interval.

\section{Conclusion and outlook}
\label{sec:concl}
With this analysis, we performed a uniform search for overdensities over different angular scales in the arrival-direction distribution of the events detected during 19 years of data acquisition of the Pierre Auger Observatory.
In particular, the scope was to thoroughly characterize the most significant overdensities present in the dataset using a von~Mises--Fisher distribution.

Using the standard energy intervals (4--8, 8--16, 16--32, >32)\,EeV defined in~\cite{ref:lsa_science, ref:lsa}, we find that the most significant overdensities correspond to large-scale excesses in all energy bins above 8\,EeV. 
Moreover, the dipolar amplitudes of these excesses in the intervals (8--16, 16--32, >32)\,EeV, as well as their increase with energy, are compatible with the results of previous analyses by the Pierre Auger Collaboration.

Using a sliding window in energy, we further find that the most significant excesses are on the large angular scale, for all energy bins starting with (6--13)\,EeV and up to (>40)\,EeV, where the most significant finding is an overdensity of intermediate angular scale ($\kappa\simeq16$, $\sigma=14.4^\circ$) located in the Centaurus region.
We note that overdensities located in the Centaurus region that exceed the $95\%$ C.L. band on intermediate angular scales are also present for the interval (32--64)\, and (>32)\,EeV.

Based on the simulations, there is no evidence that the coexistence of the two excesses at different angular scales, with characteristics similar to those observed in the data above $32\,\mathrm{EeV}$, can be explained by one excess arising as a statistical fluctuation of the other.
Different physical scenarios can give rise to a signal like this, including the superposition of the contributions from a local dominant source and from a population of farther away sources; as well as a mixed mass composition of cosmic rays at given energy, with particles of different charge dominantly contributing to the different excesses. 
Further studies, possibly including the mass composition information to be gathered from AugerPrime~\cite{ref:auger_prime}, will be very useful to shed light on these possibilities. Also the data that will be collected with the TA×4~\cite{ref:tax4} upgraded detectors or with future UHECR observatories, such as GRAND~\cite{ref:grand}, POEMMA~\cite{ref:poemma}, and GCOS~\cite{ref:gcos} will be helpful for these investigations.

\newpage
\appendix
\section{Maps for additional energy bins}
\label{sec:appendix}

\begin{figure*}[!ht]
    \centering
    \begin{subfigure}{0.47\textwidth}
        \includegraphics[width=\linewidth]{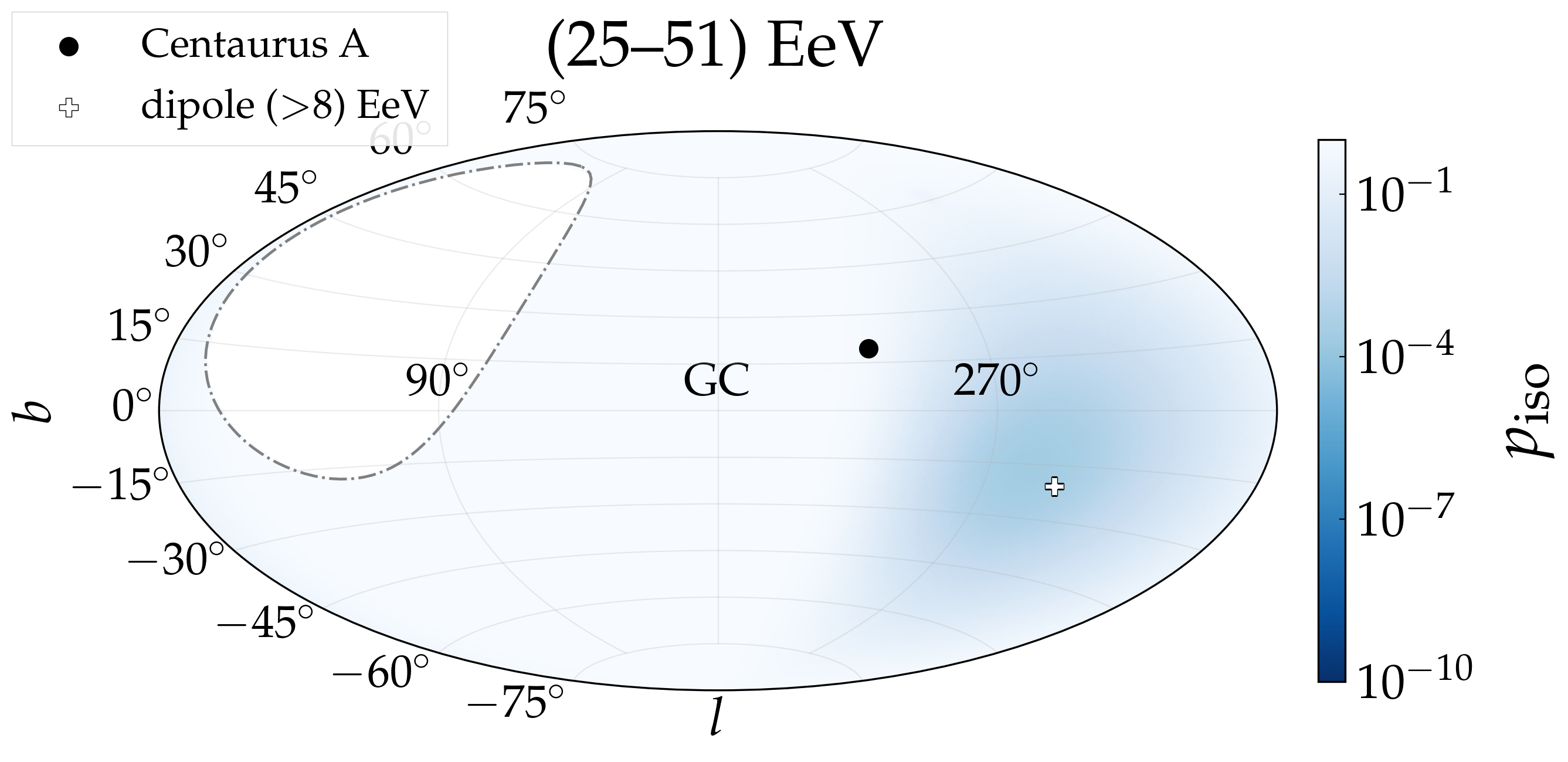}
        \caption{}
    \end{subfigure}
    \begin{subfigure}{0.47\textwidth}
        \includegraphics[width=\linewidth]{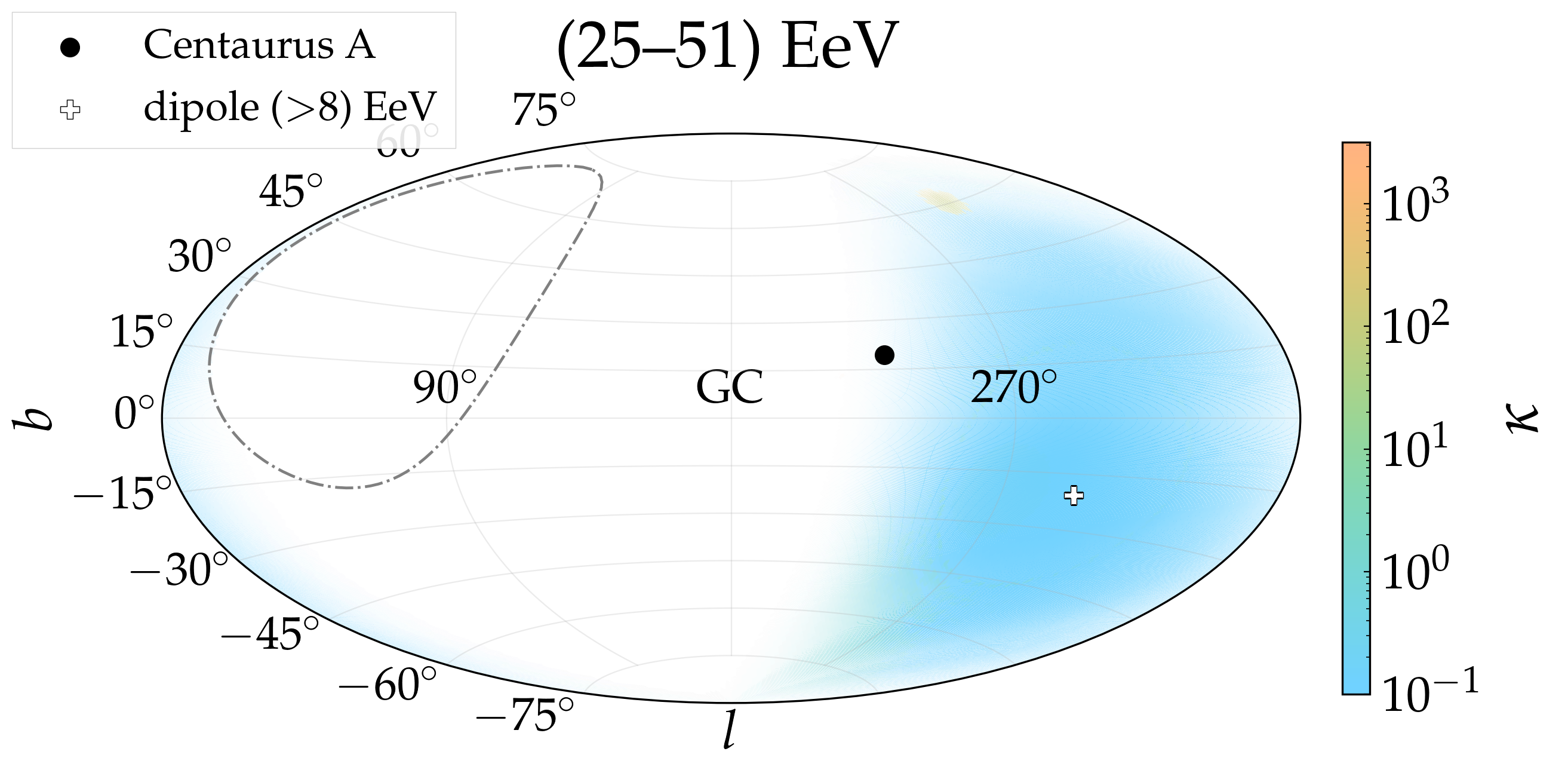}
        \caption{}
    \end{subfigure}
    \\[0.07cm]
    \begin{subfigure}{0.47\textwidth}
        \includegraphics[width=\linewidth]{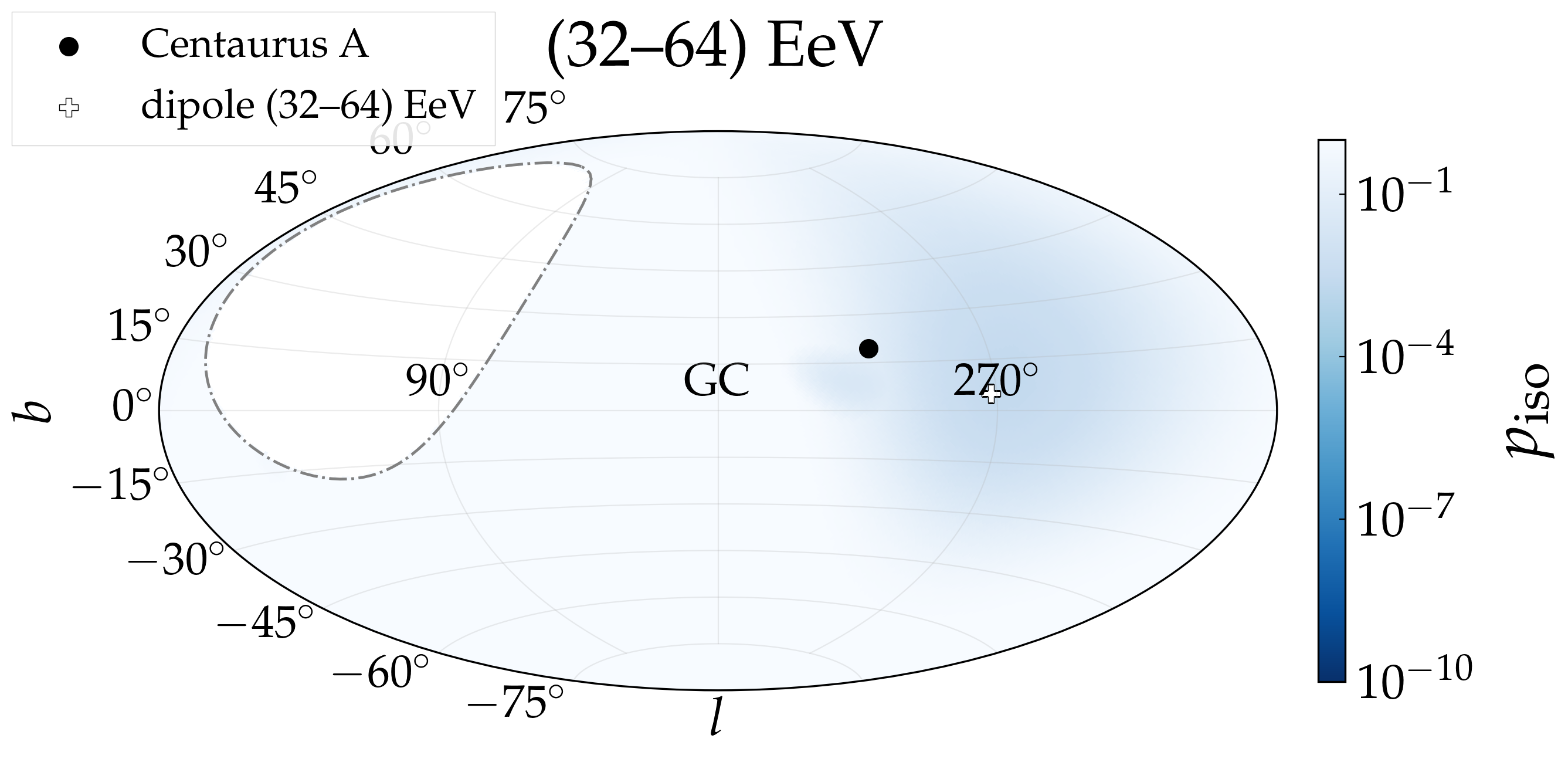}
        \caption{}
    \end{subfigure}
    \begin{subfigure}{0.47\textwidth}
        \includegraphics[width=\linewidth]{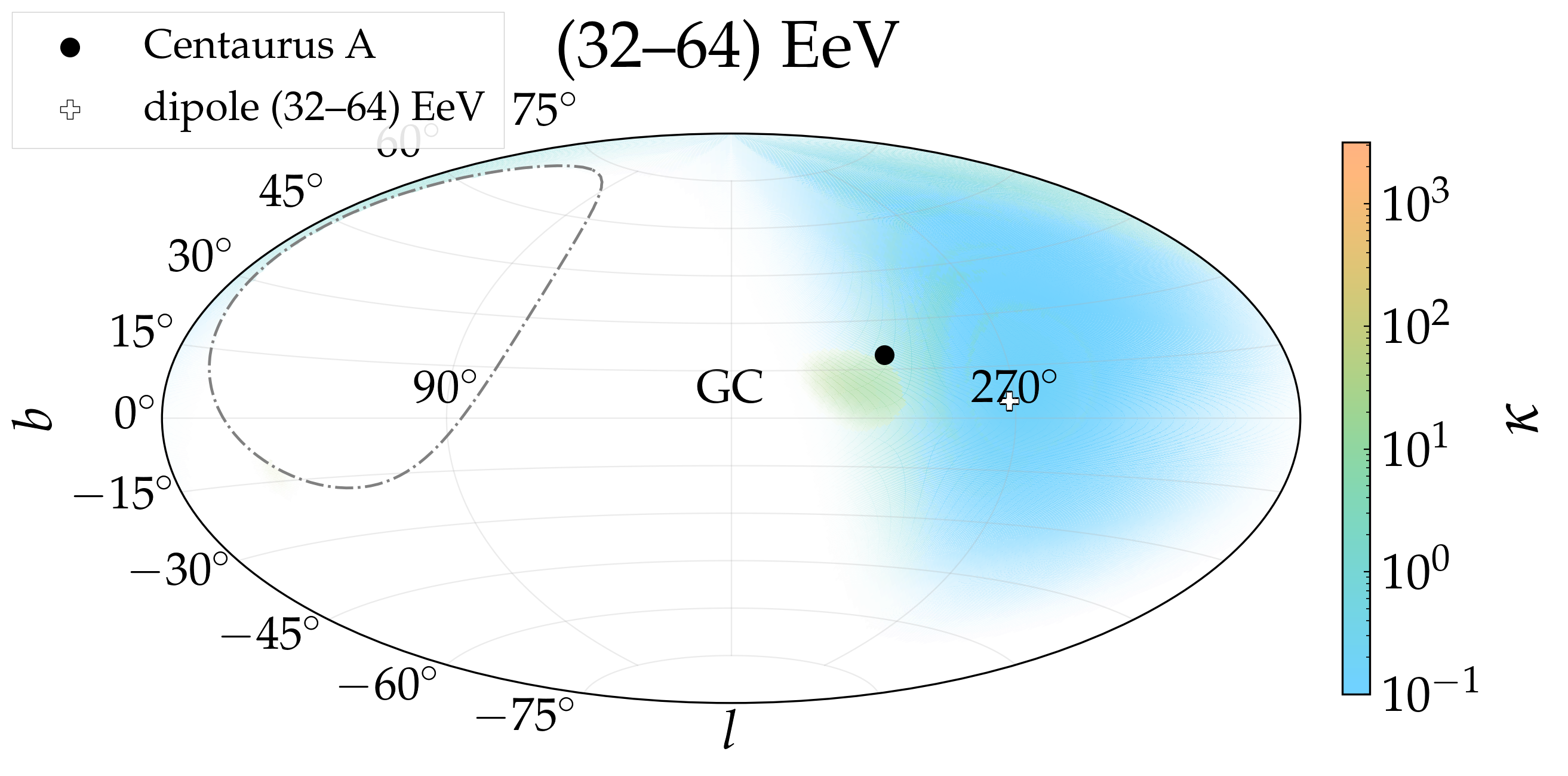}
        \caption{}
    \end{subfigure}
    \\[0.07cm]
    \begin{subfigure}{0.47\textwidth}
        \includegraphics[width=\linewidth]{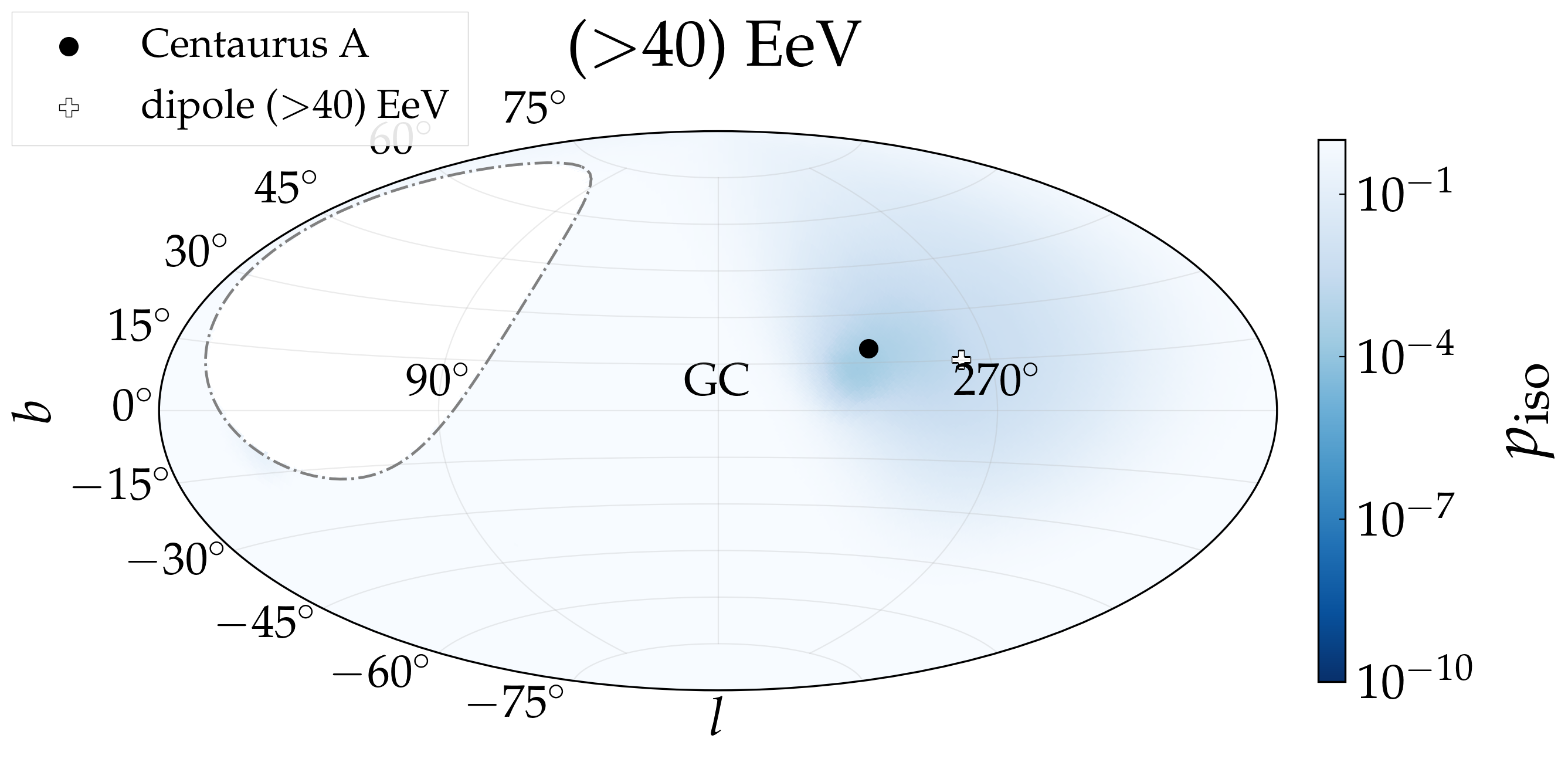}
        \caption{}
    \end{subfigure}
    \begin{subfigure}{0.47\textwidth}
        \includegraphics[width=\linewidth]{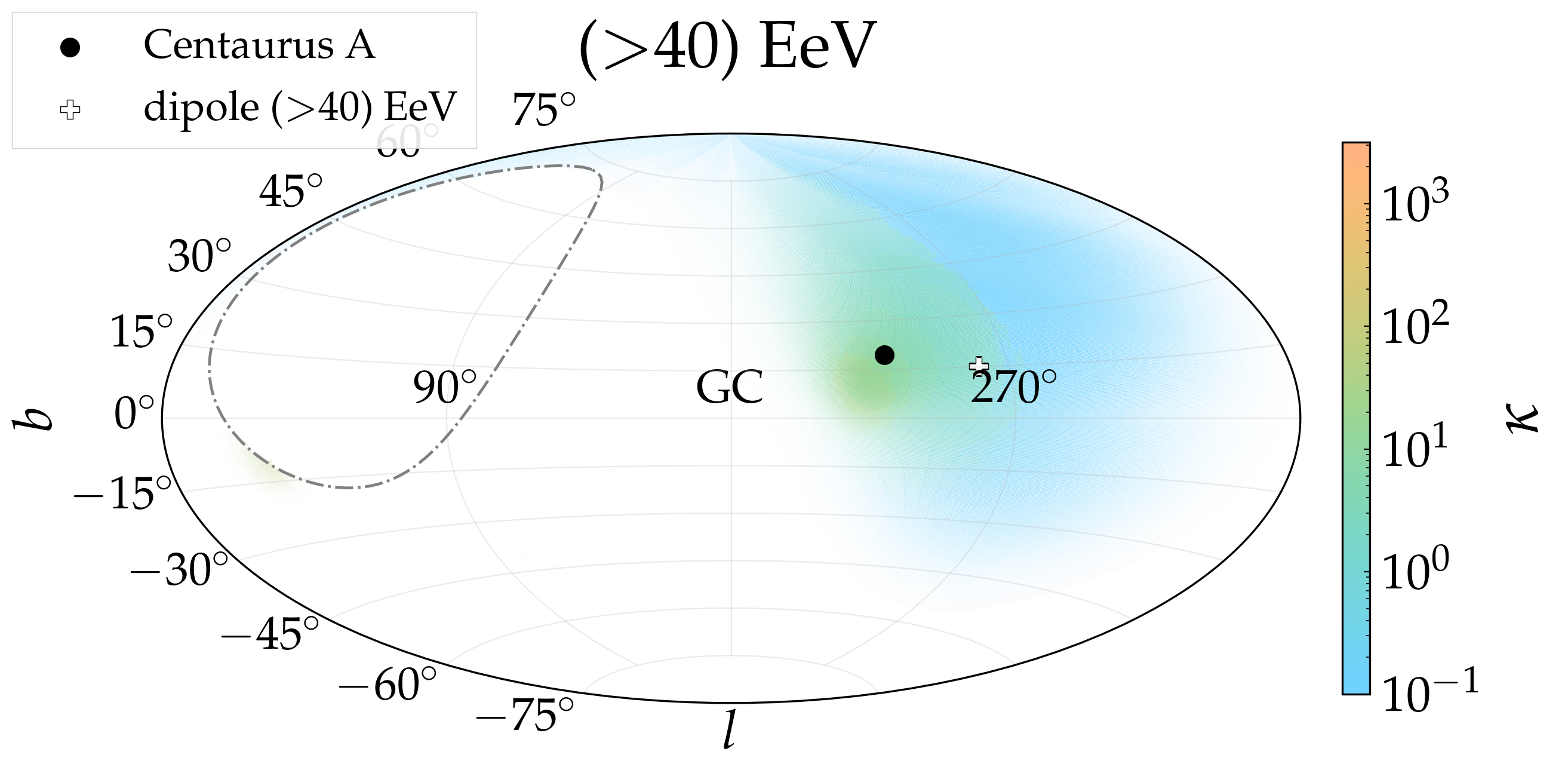}
        \caption{}
    \end{subfigure}
    \caption{
    Maps in Galactic coordinates of the lowest $p_\mathrm{iso}$ (left column) and $\kappa$ (right column) obtained by maximizing over $\alpha$ while profiling over $\kappa$.
    The maps are derived by profiling over the reference position using a \texttt{HEALPix} grid with $\texttt{nSide}=64$.
    In the $\kappa$ maps, the opacity of each point is scaled according to the ratio between the local $\log_{10}( p_\mathrm{iso})$ value and the minimum $\log_{10} (p_\mathrm{iso})$ in the map.
    The maps are depicted for the energy intervals (a,b) (25--50)\,EeV, (c,d) (32--64)\,EeV, and (e,f) (>40)\,EeV. 
    In each plot, the position of Centaurus~A and the position of the center of the dipole modulation derived for the corresponding energy intervals using the analysis presented in ref.\ ~\cite{ref:lsa} are indicated with a black dot and a white cross, respectively.
    }
    \label{fig:maps_ts_2}
\end{figure*}

\newpage
\section{Maps in equatorial coordinates}
\label{sec:appendix_eq}

\begin{figure*}[!ht]
    \centering
    \begin{subfigure}{0.47\textwidth}
        \includegraphics[width=\linewidth]{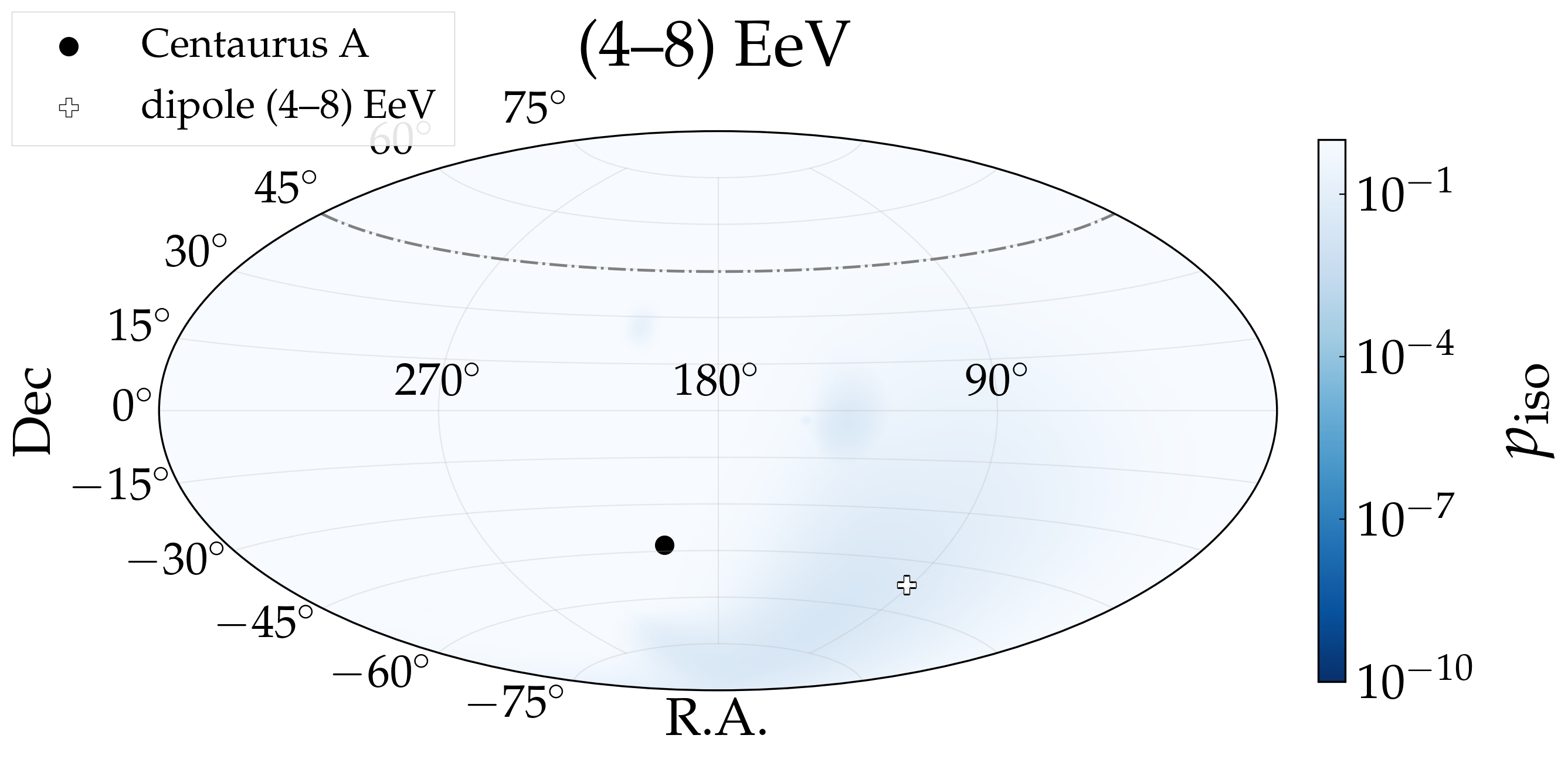}
        \caption{}
    \end{subfigure}
    \begin{subfigure}{0.47\textwidth}
        \includegraphics[width=\linewidth]{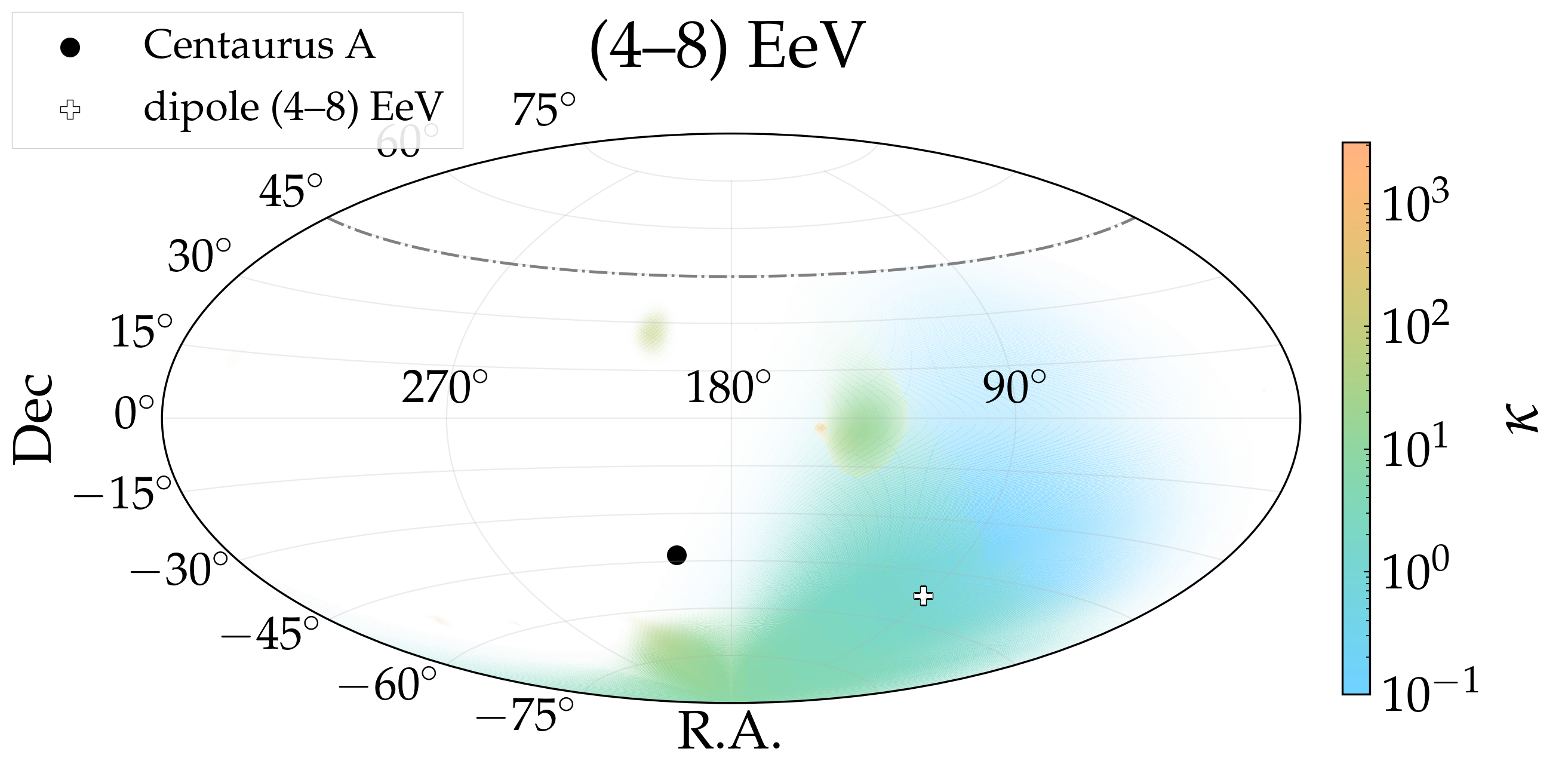}
        \caption{}
    \end{subfigure}
    \begin{subfigure}{0.47\textwidth}
        \includegraphics[width=\linewidth]{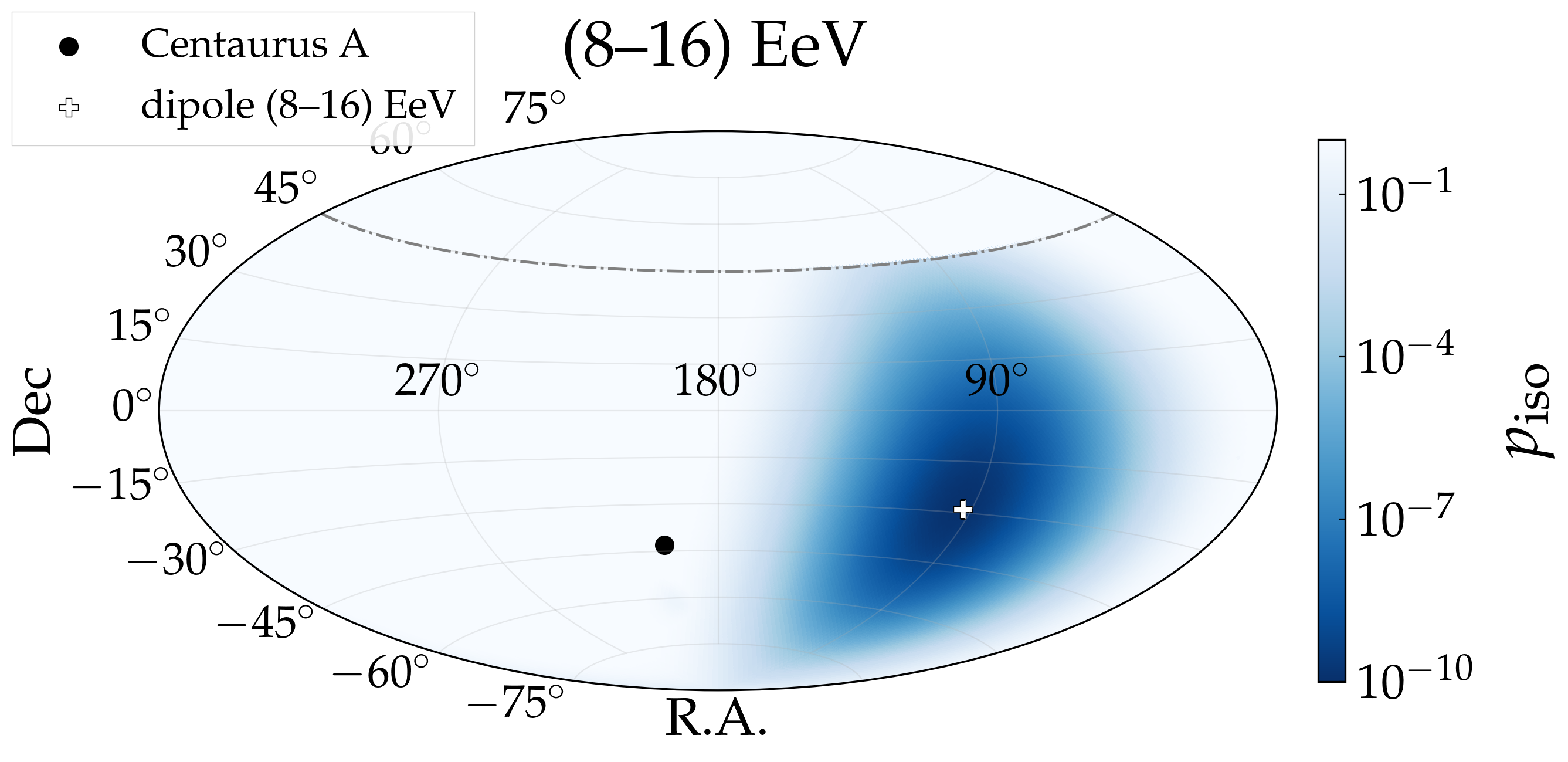}
        \caption{}
    \end{subfigure}
    \begin{subfigure}{0.47\textwidth}
        \includegraphics[width=\linewidth]{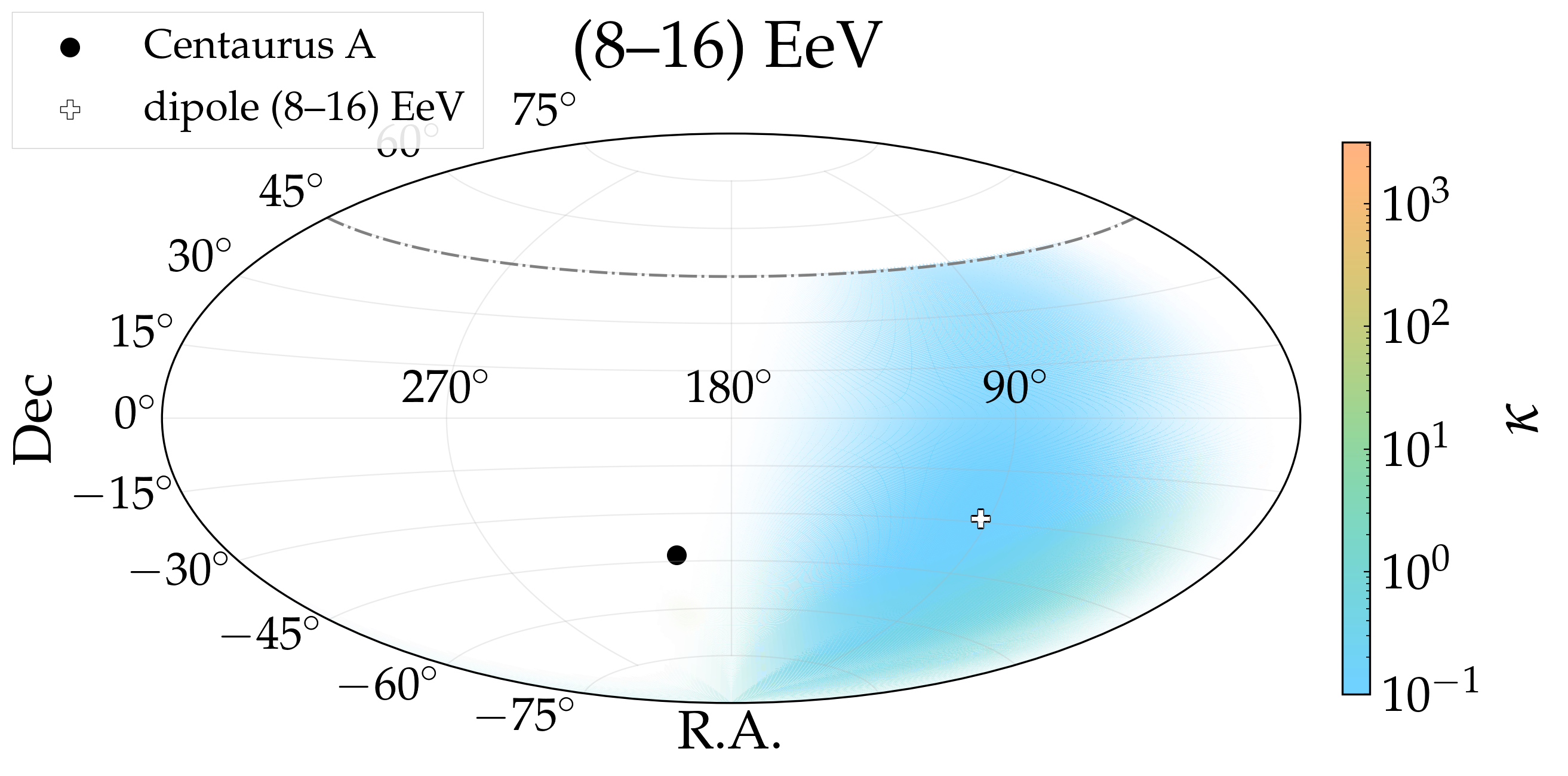}
        \caption{}
    \end{subfigure}
    \begin{subfigure}{0.47\textwidth}
        \includegraphics[width=\linewidth]{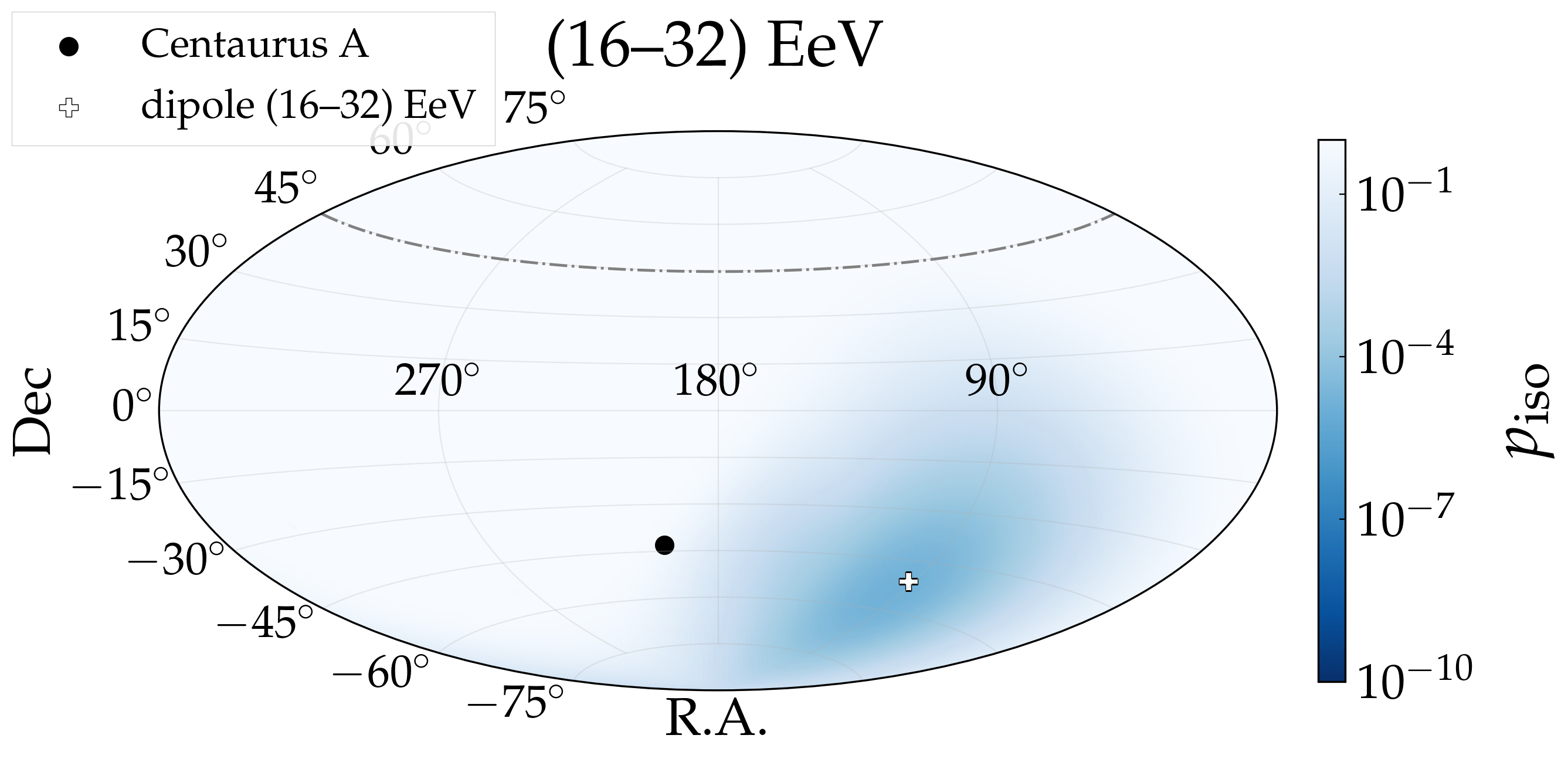}
        \caption{}
    \end{subfigure}
    \begin{subfigure}{0.47\textwidth}
        \includegraphics[width=\linewidth]{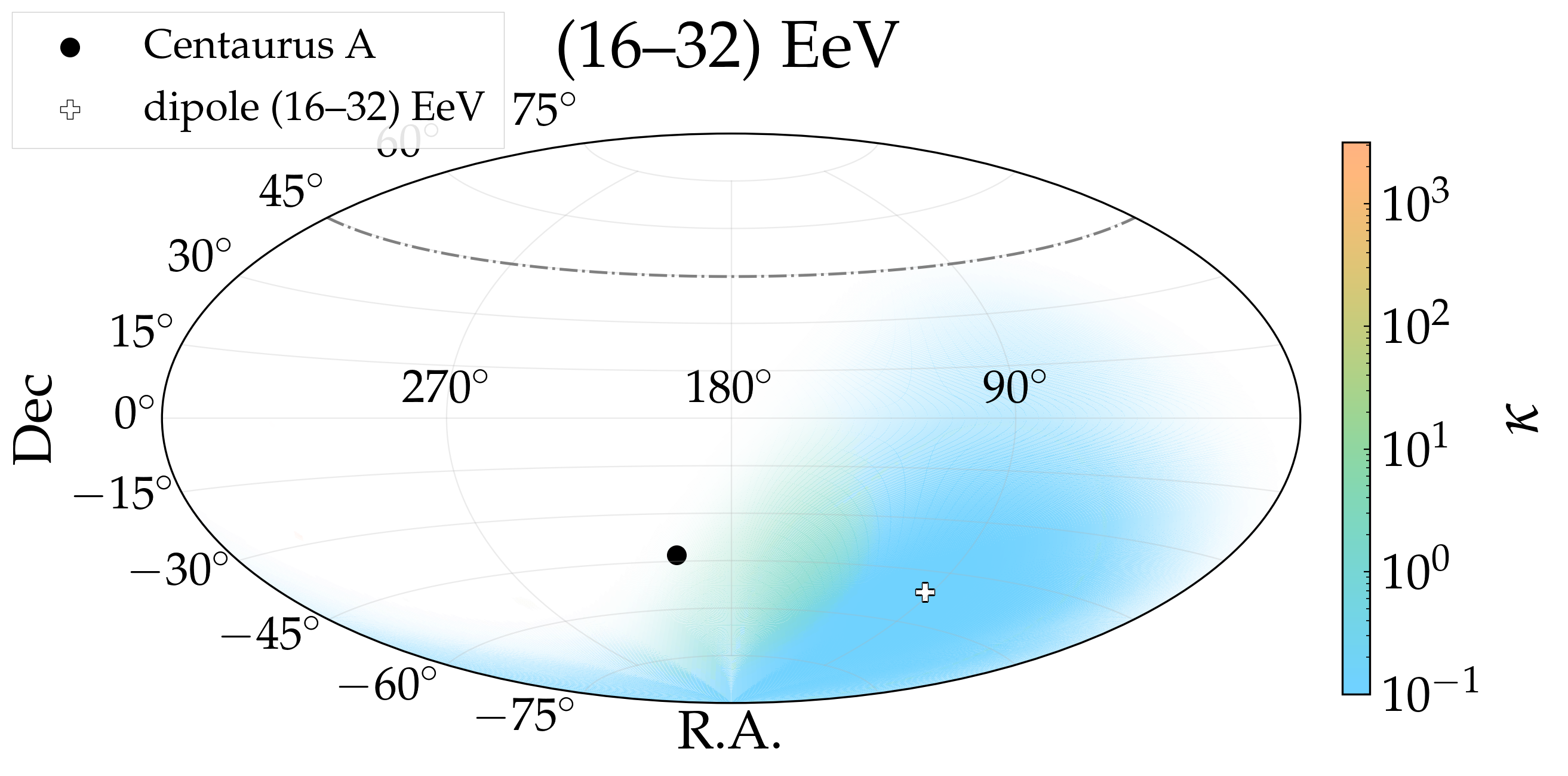}
        \caption{}
    \end{subfigure}
    \begin{subfigure}{0.47\textwidth}
        \includegraphics[width=\linewidth]{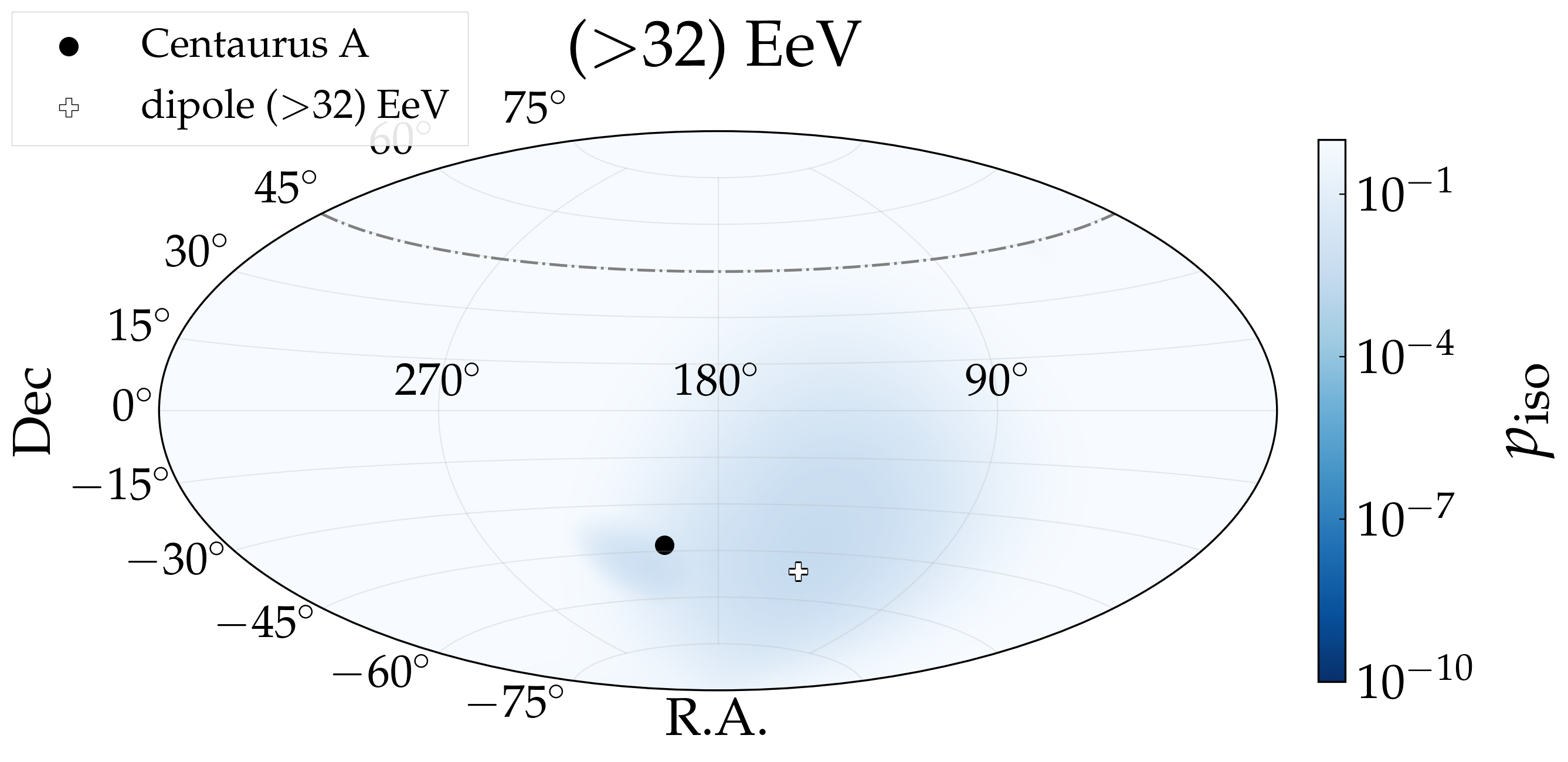}
        \caption{}
    \end{subfigure}
    \begin{subfigure}{0.47\textwidth}
        \includegraphics[width=\linewidth]{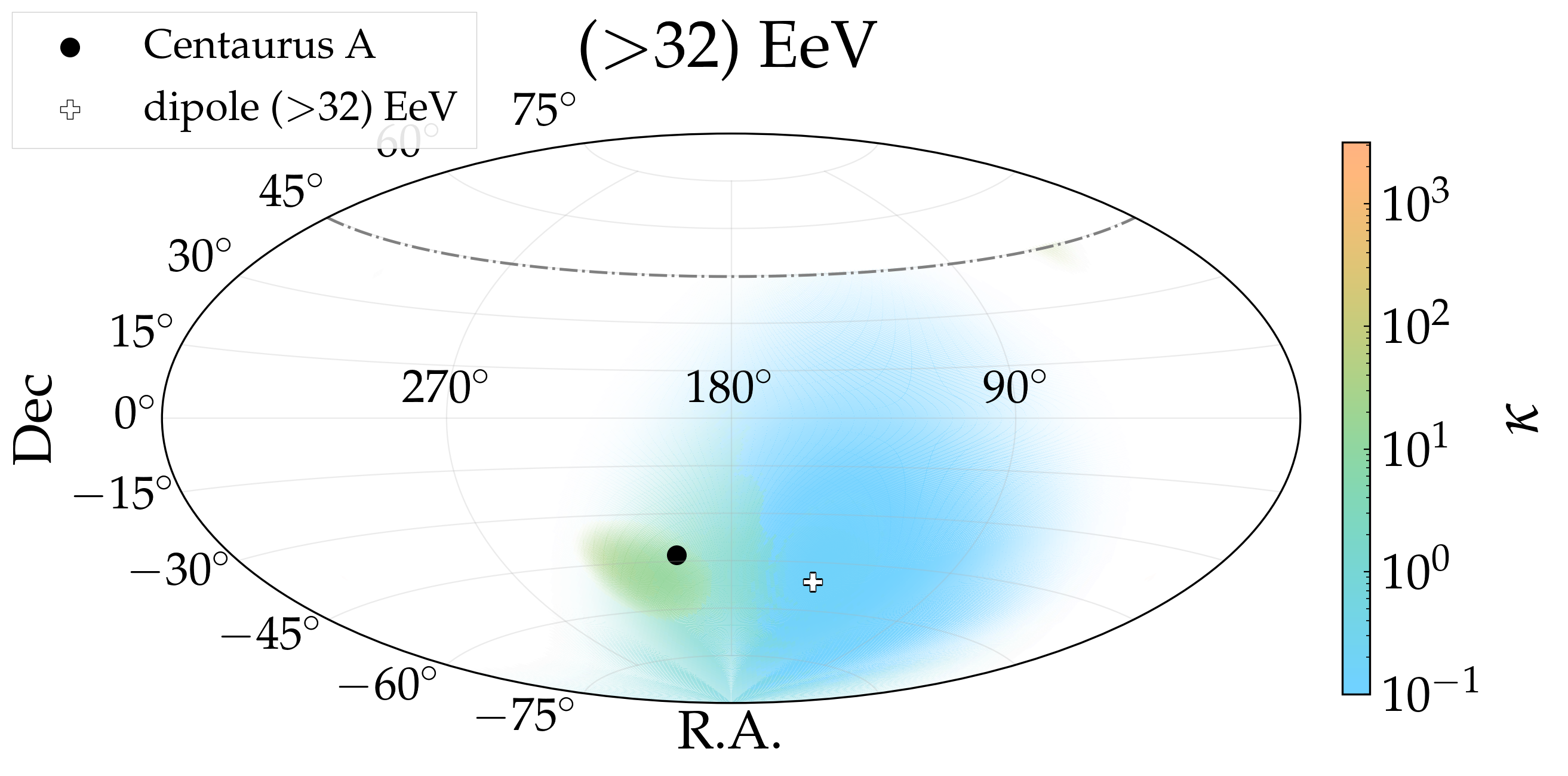}
        \caption{}
    \end{subfigure}
    \caption{
    Maps in equatorial coordinates of the lowest $p_\mathrm{iso}$ (left column) and $\kappa$ (right column) obtained by maximizing over $\alpha$ while profiling over $\kappa$.
    The maps are derived by profiling over the reference position using a \texttt{HEALPix} grid with $\texttt{nSide}=64$.
    In the $\kappa$ maps, the opacity of each point is scaled according to the ratio between the local $\log_{10} p_\mathrm{iso}$ value and the minimum $\log_{10} p_\mathrm{iso}$ in the map.
    The maps are depicted for the energy intervals (a,b) (4--8)\,EeV, (c,d) (8--16)\,EeV,  (e,f) (16--32)\,EeV, and (g,h) (>32)\,EeV. 
    In each plot, the position of Centaurus~A and the position of the center of the dipole modulation obtained for the corresponding energy interval in ref.\ \cite{ref:lsa} are indicated with a black dot and a white cross, respectively.
    }
    \label{fig:maps_ts_eq_1}
\end{figure*}

\begin{figure*}[!ht]
    \centering
    \begin{subfigure}{0.47\textwidth}
        \includegraphics[width=\linewidth]{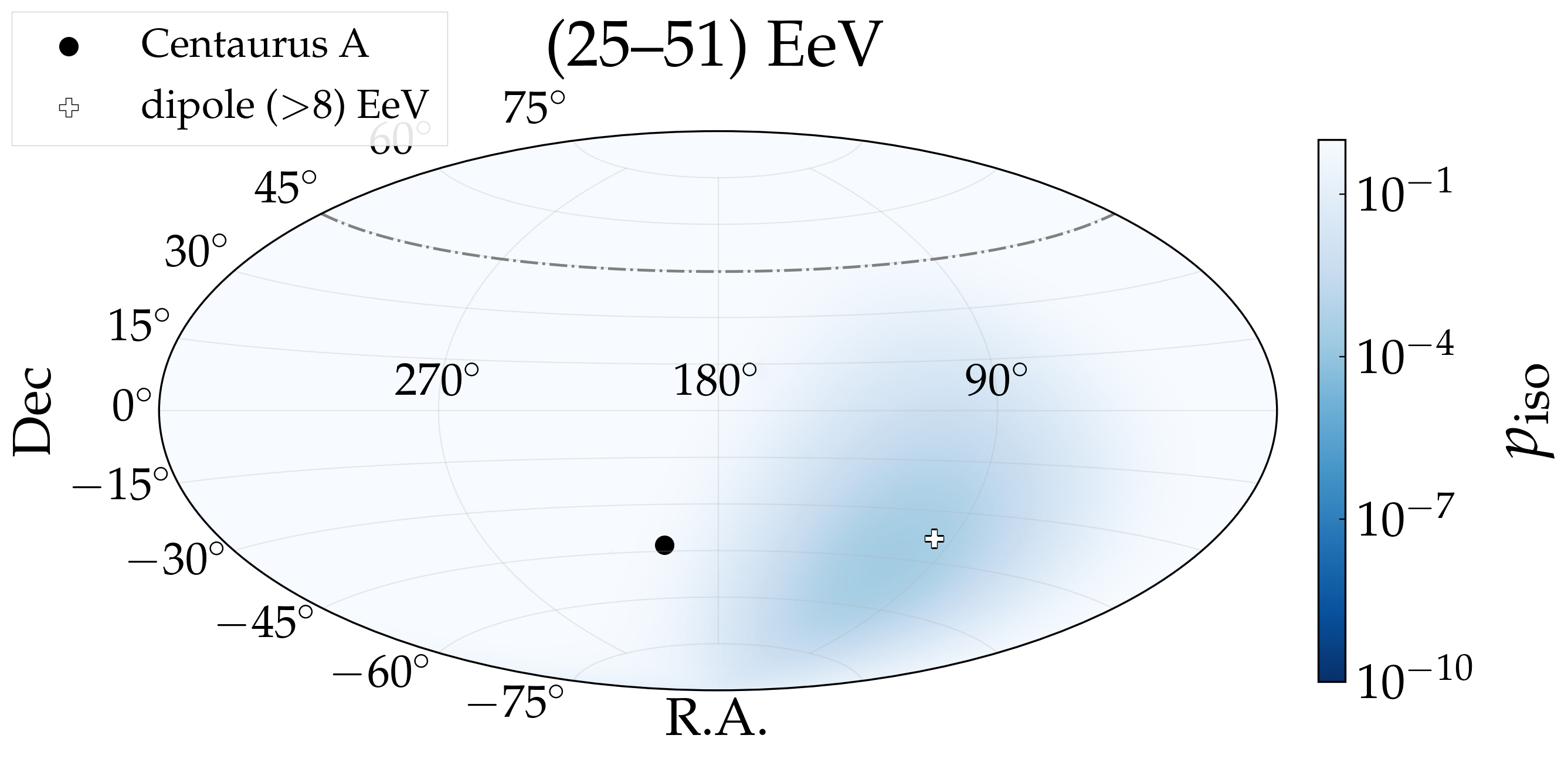}
        \caption{}
    \end{subfigure}
    \begin{subfigure}{0.47\textwidth}
        \includegraphics[width=\linewidth]{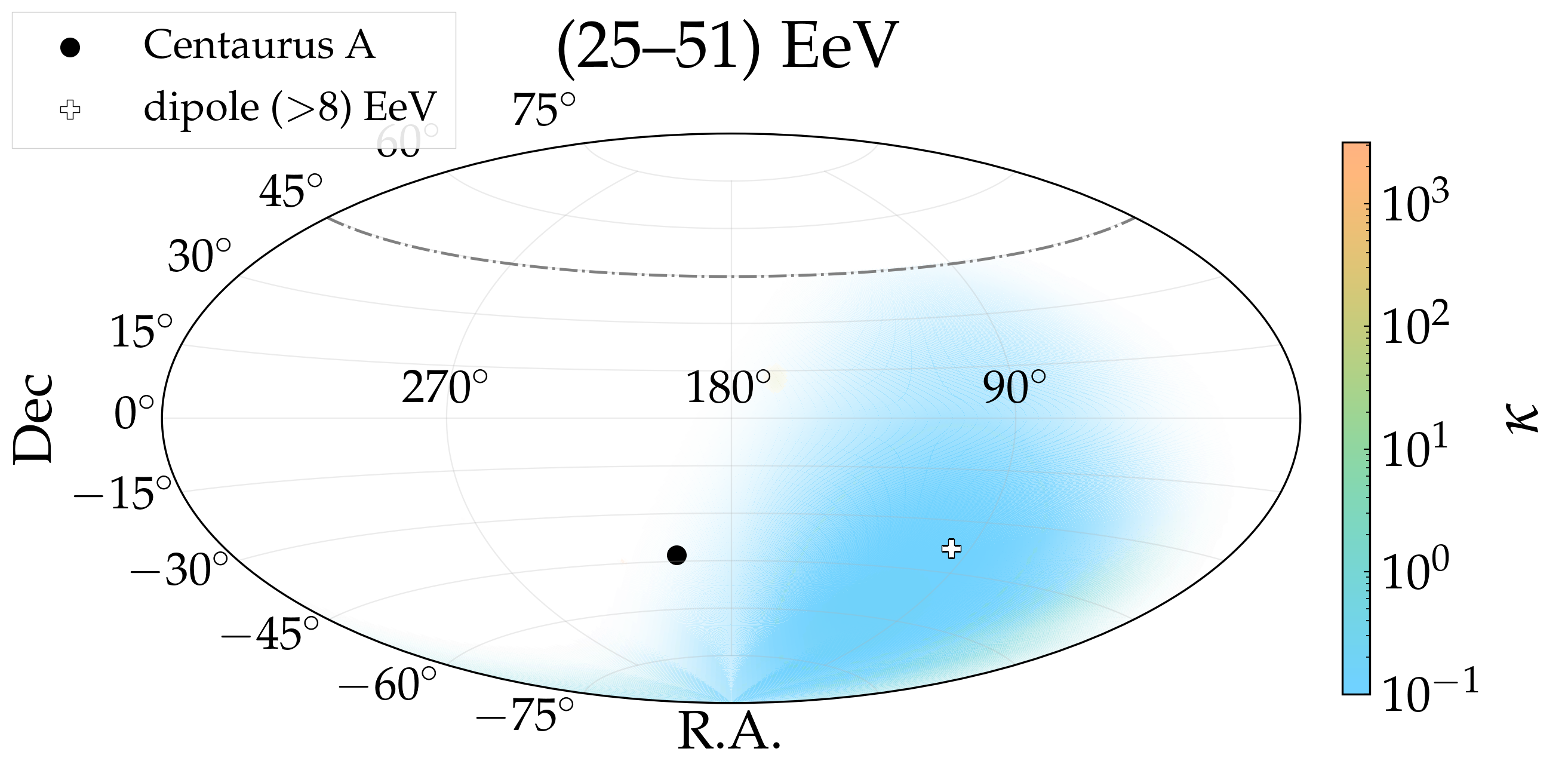}
        \caption{}
    \end{subfigure}
    \\[0.07cm]
    \begin{subfigure}{0.47\textwidth}
        \includegraphics[width=\linewidth]{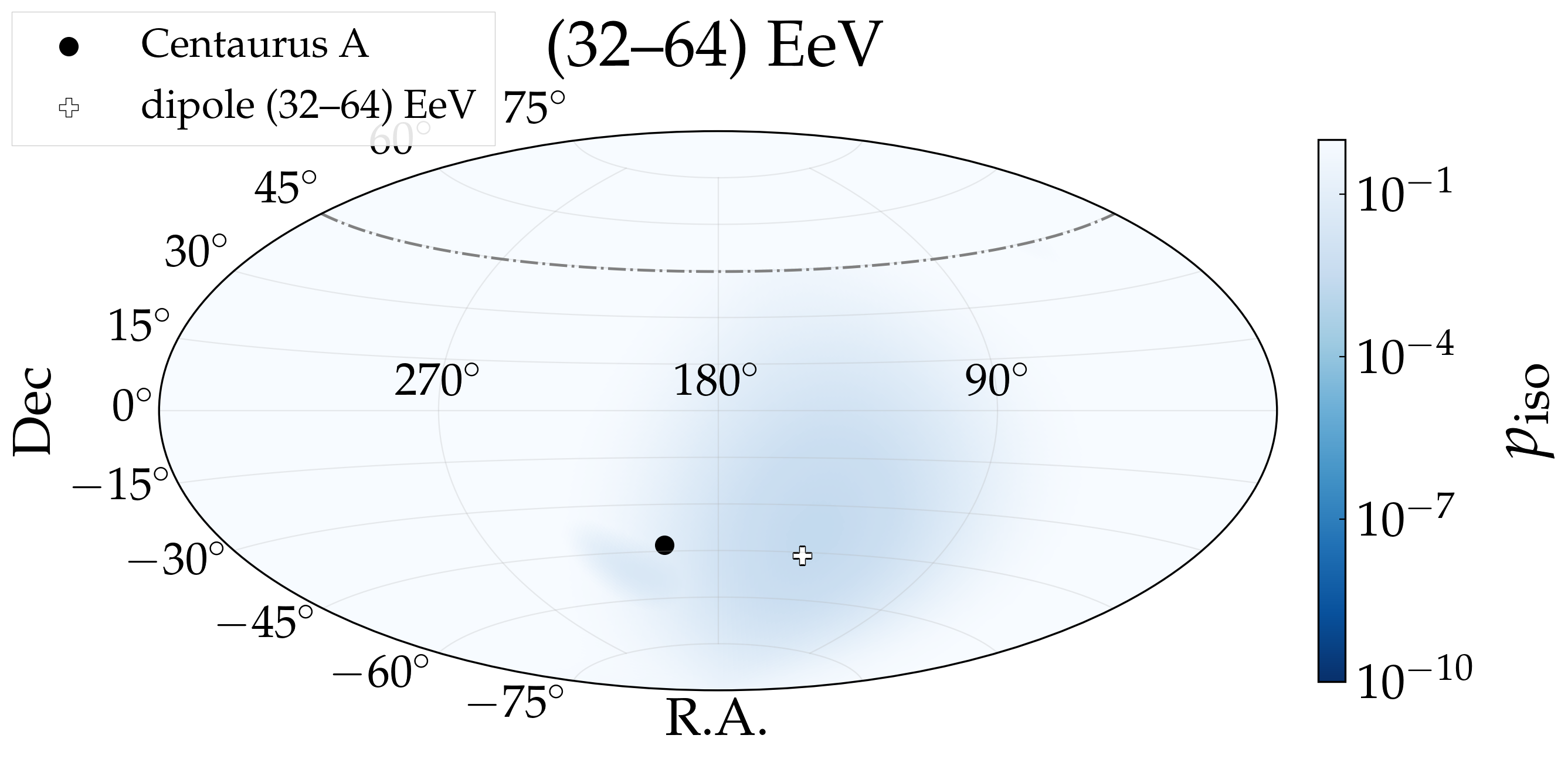}
        \caption{}
    \end{subfigure}
    \begin{subfigure}{0.47\textwidth}
        \includegraphics[width=\linewidth]{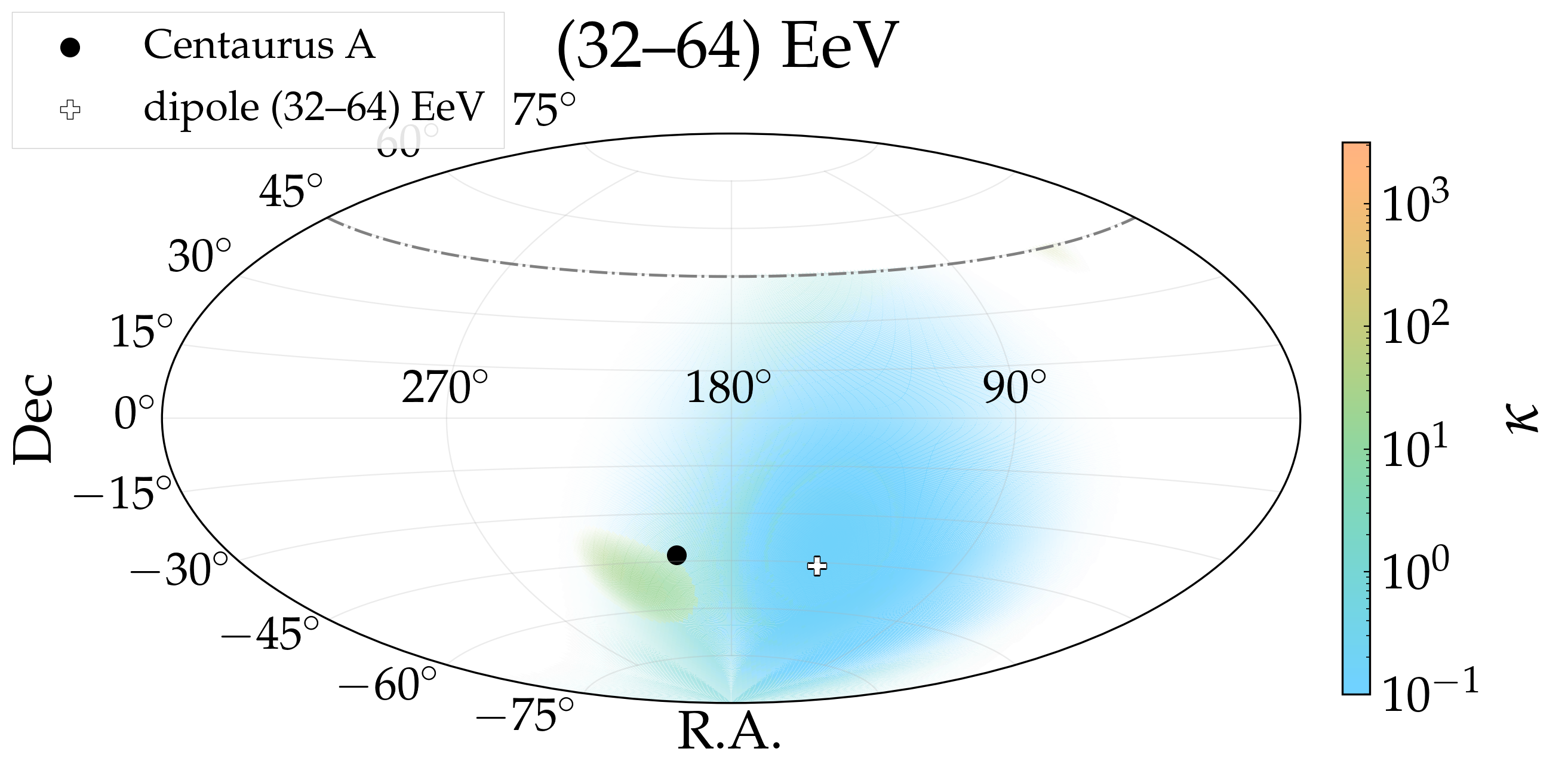}
        \caption{}
    \end{subfigure}
    \\[0.07cm]
    \begin{subfigure}{0.47\textwidth}
        \includegraphics[width=\linewidth]{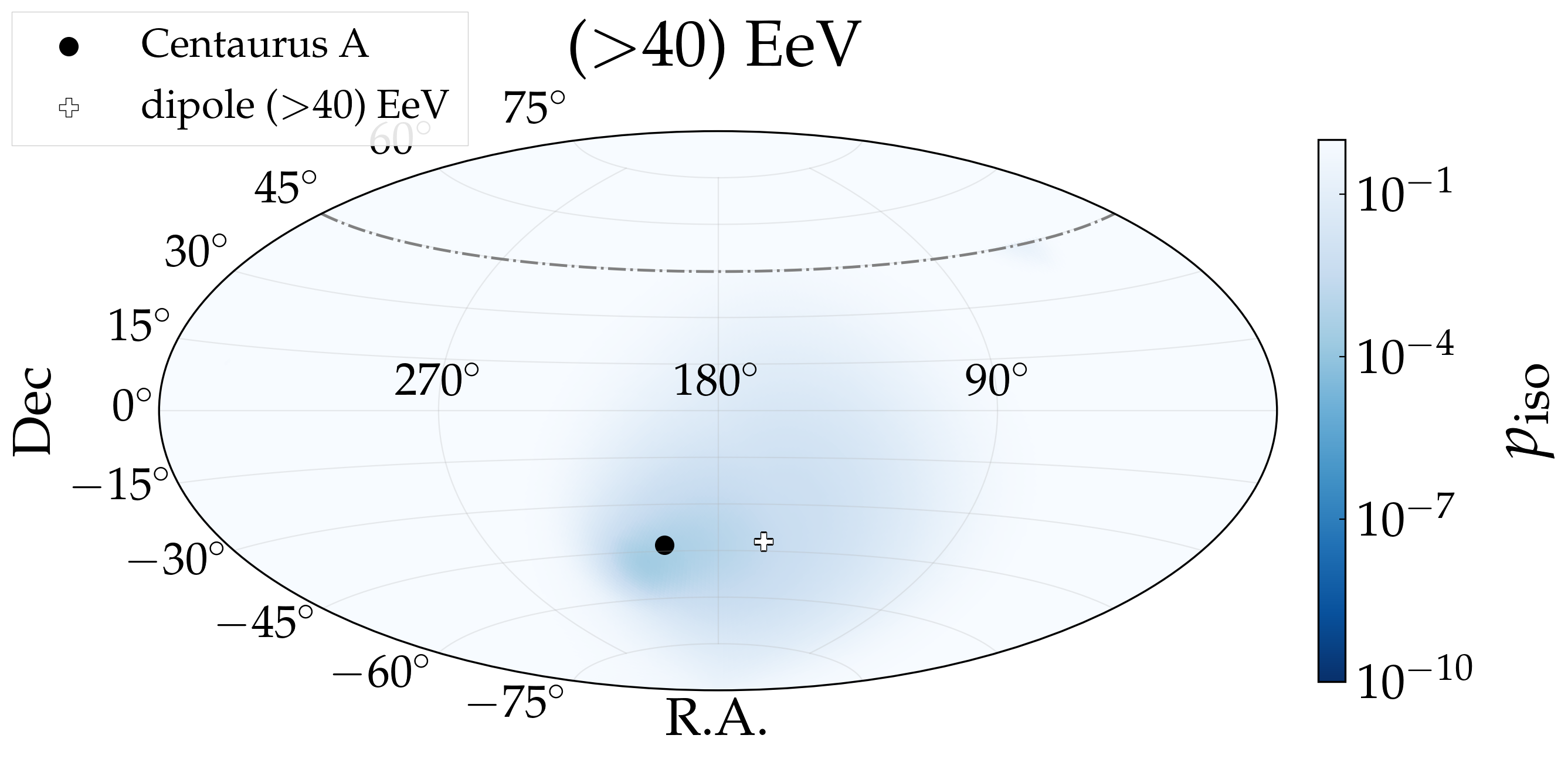}
        \caption{}
    \end{subfigure}
    \begin{subfigure}{0.47\textwidth}
        \includegraphics[width=\linewidth]{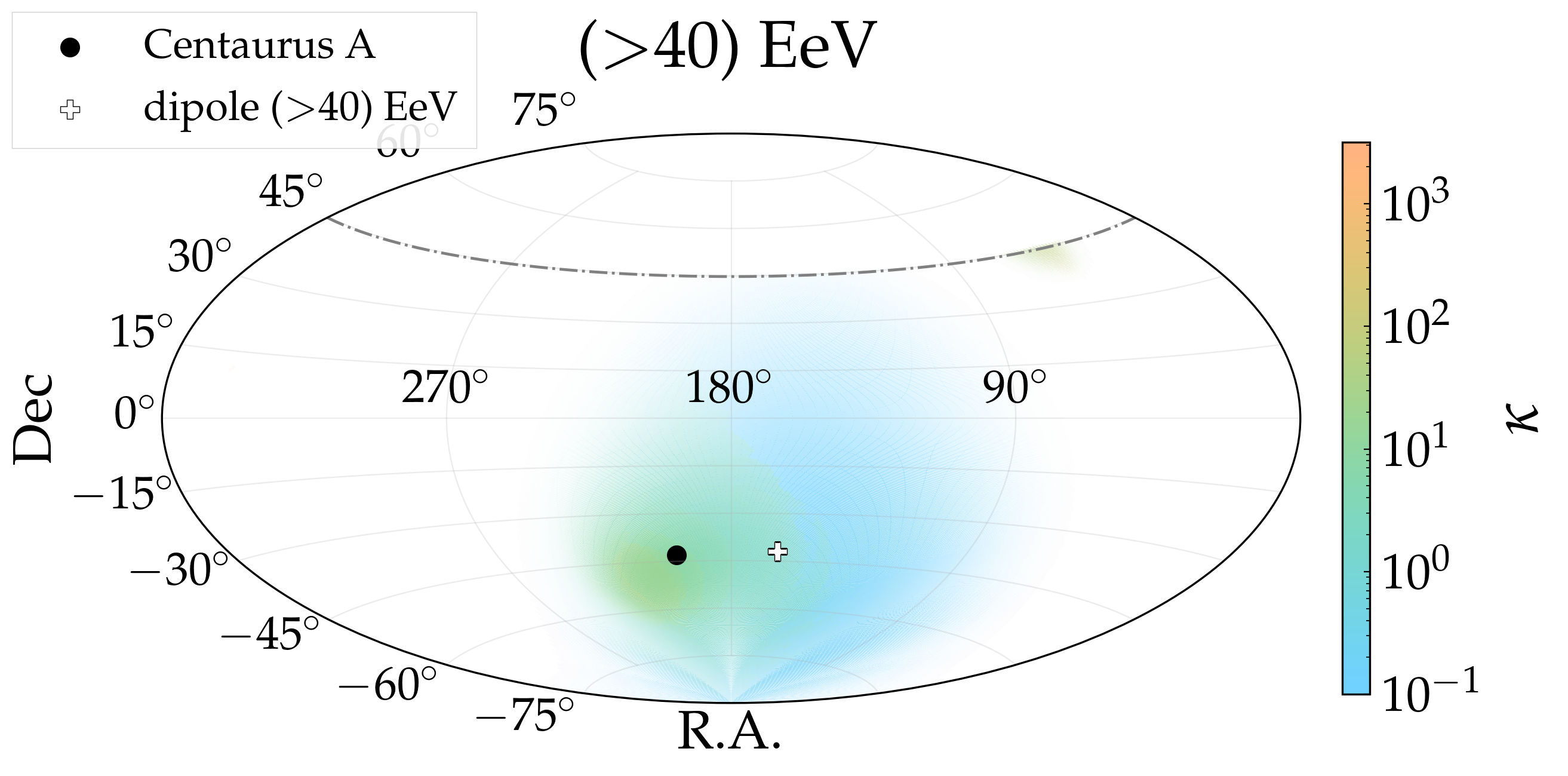}
        \caption{}
    \end{subfigure}
    \caption{
    Maps in equatorial coordinates of the lowest $p_\mathrm{iso}$ (left column) and $\kappa$ (right column) obtained by maximizing over $\alpha$ while profiling over $\kappa$.
    The maps are derived by profiling over the reference position using a \texttt{HEALPix} grid with $\texttt{nSide}=64$.
    In the $\kappa$ maps, the opacity of each point is scaled according to the ratio between the local $\log_{10}( p_\mathrm{iso})$ value and the minimum $\log_{10} (p_\mathrm{iso})$ in the map.
    The maps are depicted for the energy intervals (a,b) (25--50)\,EeV, (c,d) (32--64)\,EeV, and (e,f) (>40)\,EeV. 
    In each plot, the position of Centaurus~A and the position of the center of the dipole modulation derived for the corresponding energy intervals using the analysis presented in ref.\ ~\cite{ref:lsa} are indicated with a black dot and a white cross, respectively.
    }
    \label{fig:maps_ts_2}
\end{figure*}

\section*{Acknowledgments}

\begin{sloppypar}
The successful installation, commissioning, and operation of the Pierre
Auger Observatory would not have been possible without the strong
commitment and effort from the technical and administrative staff in
Malarg\"ue. We are very grateful to the following agencies and
organizations for financial support:
\end{sloppypar}

\begin{sloppypar}
Argentina -- Comisi\'on Nacional de Energ\'\i{}a At\'omica; Agencia Nacional de
Promoci\'on Cient\'\i{}fica y Tecnol\'ogica (ANPCyT); Consejo Nacional de
Investigaciones Cient\'\i{}ficas y T\'ecnicas (CONICET); Gobierno de la
Provincia de Mendoza; Municipalidad de Malarg\"ue; NDM Holdings and Valle
Las Le\~nas; in gratitude for their continuing cooperation over land
access; Australia -- the Australian Research Council; Belgium -- Fonds
de la Recherche Scientifique (FNRS); Research Foundation Flanders (FWO),
Marie Curie Action of the European Union Grant No.~101107047; Brazil --
Minist\'erio da Ci\^encia, Tecnologia e Inova\c{c}\~ao (MCTI); Czech Republic --
GACR 24-13049S, CAS LQ100102401, MEYS LM2023032,
CZ.02.1.01/0.0/0.0/16{\textunderscore}013/0001402, CZ.02.1.01/0.0/0.0/18{\textunderscore}046/0016010
and CZ.02.1.01/0.0/0.0/17{\textunderscore}049/0008422 and
CZ.02.01.01/00/22{\textunderscore}008/0004632; France -- Centre de Calcul IN2P3/CNRS;
Centre National de la Recherche Scientifique (CNRS); Institut National
de Physique Nucl\'eaire et de Physique des Particules (IN2P3/CNRS);
Germany -- Bundesministerium f\"ur Forschung, Technologie und Raumfahrt
(BMFTR); Deutsche Forschungsgemeinschaft (DFG); Ministerium f\"ur Finanzen
Baden-W\"urttemberg; Helmholtz Alliance for Astroparticle Physics (HAP);
Hermann von Helmholtz-Gemeinschaft Deutscher Forschungszentren e.V.;
Ministerium f\"ur Kultur und Wissenschaft des Landes Nordrhein-Westfalen;
Ministerium f\"ur Wissenschaft, Forschung und Kunst des Landes
Baden-W\"urttemberg; Italy -- Istituto Nazionale di Fisica Nucleare
(INFN); Istituto Nazionale di Astrofisica (INAF); Ministero
dell'Universit\`a e della Ricerca (MUR); CETEMPS Center of Excellence;
Ministero degli Affari Esteri (MAE), ICSC Centro Nazionale di Ricerca in
High Performance Computing, Big Data and Quantum Computing, funded by
European Union NextGenerationEU, reference code CN{\textunderscore}00000013; M\'exico --
Consejo Nacional de Ciencia y Tecnolog\'\i{}a (CONACYT-SECHTI)
No.~CB-A1-S-46703, Universidad Nacional Aut\'onoma de M\'exico (UNAM)
PAPIIT-IN114924; Benem\'erita Universidad Aut\'onoma de Puebla (BUAP), VIEP
and Laboratorio Nacional de Superc\'omputo del Sureste de M\'exico (LNS);
and Benem\'erita Universidad Aut\'onoma de Chiapas (UNACH); The Netherlands
-- Ministry of Education, Culture and Science; Netherlands Organisation
for Scientific Research (NWO); Dutch national e-infrastructure with the
support of SURF Cooperative; Poland -- Ministry of Science and Higher
Education, grant No.~2026/WK/05; National Science Centre, grant
No.~2022/45/B/ST9/02163; Portugal -- Portuguese national funds and FEDER
funds within Programa Operacional Factores de Competitividade through
Funda\c{c}\~ao para a Ci\^encia e a Tecnologia (COMPETE); Romania -- Ministry of
Education and Research, contract no.~30N/2023 under Romanian National
Core Program LAPLAS VII, and grant no.~PN 23 21 01 02; Slovenia --
Slovenian Research and Innovation Agency, grants P1-0031, I0-0033; Spain
-- Ministerio de Ciencia, Innovaci\'on y Universidades/Agencia Estatal de
Investigaci\'on MICIU/AEI /10.13039/501100011033 (PID2022-140510NB-I00,
PCI2023-145952-2, CNS2024-154676, and Mar\'\i{}a de Maeztu CEX2023-001318-M),
Xunta de Galicia (CIGUS Network of Research Centers, Consolidaci\'on
ED431C-2025/11 and ED431F-2022/15) and European Union ERDF; USA --
Department of Energy, Contracts No.~DE-AC02-07CH11359,
No.~DE-FR02-04ER41300, No.~DE-FG02-99ER41107 and No.~DE-SC0011689;
National Science Foundation, Grant No.~0450696, and NSF-2013199; The
Grainger Foundation; Astrophysics Centre for Multi-messenger studies in
Europe (ACME) EU Grant No 101131928; and UNESCO.
\end{sloppypar}

\bibliographystyle{JHEP}
\bibliography{bibliography}

@article{ref:auger,
    author = "Aab, Alexander and others",
    collaboration = "Pierre Auger",
    title = "{The Pierre Auger Cosmic Ray Observatory}",
    eprint = "1502.01323",
    archivePrefix = "arXiv",
    primaryClass = "astro-ph.IM",
    reportNumber = "FERMILAB-PUB-15-034-AD-AE-CD-TD",
    doi = "10.1016/j.nima.2015.06.058",
    journal = "Nucl. Instrum. Meth. A",
    volume = "798",
    pages = "172--213",
    year = "2015"
}

@article{ref:ta,
    author = "Abu-Zayyad, T. and others",
    collaboration = "Telescope Array",
    title = "{The surface detector array of the Telescope Array experiment}",
    eprint = "1201.4964",
    archivePrefix = "arXiv",
    primaryClass = "astro-ph.IM",
    doi = "10.1016/j.nima.2012.05.079",
    journal = "Nucl. Instrum. Meth. A",
    volume = "689",
    pages = "87--97",
    year = "2013"
}

@article{ref:fisher,
    author = "Harari, Diego and Mollerach, Silvia and Roulet, Esteban",
    title = "{Angular distribution of cosmic rays from an individual source in a turbulent magnetic field}",
    eprint = "1512.08289",
    archivePrefix = "arXiv",
    primaryClass = "astro-ph.HE",
    doi = "10.1103/PhysRevD.93.063002",
    journal = "Phys. Rev. D",
    volume = "93",
    number = "6",
    pages = "063002",
    year = "2016"
}

@article{ref:lsa,
    author = "Halim, A. Abdul and others",
    collaboration = "Pierre Auger",
    title = "{Large-scale Cosmic-ray Anisotropies with 19 yr of Data from the Pierre Auger Observatory}",
    eprint = "2408.05292",
    archivePrefix = "arXiv",
    primaryClass = "astro-ph.HE",
    reportNumber = "FERMILAB-PUB-24-1012-V",
    doi = "10.3847/1538-4357/ad843b",
    journal = "Astrophys. J.",
    volume = "976",
    number = "1",
    pages = "48",
    year = "2024"
}

@article{ref:lsa_science,
    author = "Aab, Alexander and others",
    collaboration = "Pierre Auger",
    title = "{Observation of a Large-scale Anisotropy in the Arrival Directions of Cosmic Rays above $8 \times 10^{18}$ eV}",
    eprint = "1709.07321",
    archivePrefix = "arXiv",
    primaryClass = "astro-ph.HE",
    reportNumber = "FERMILAB-PUB-17-354",
    doi = "10.1126/science.aan4338",
    journal = "Science",
    volume = "357",
    number = "6537",
    pages = "1266--1270",
    year = "2017"
}

@article{ref:lsa_2018,
    author = "Aab, A. and others",
    collaboration = "Pierre Auger",
    title = "{Large-scale cosmic-ray anisotropies above 4 EeV measured by the Pierre Auger Observatory}",
    eprint = "1808.03579",
    archivePrefix = "arXiv",
    primaryClass = "astro-ph.HE",
    reportNumber = "FERMILAB-PUB-18-393-ND-PPD-TD",
    doi = "10.3847/1538-4357/aae689",
    journal = "Astrophys. J.",
    volume = "868",
    number = "1",
    pages = "4",
    year = "2018"
}

@article{ref:apj2022,
    author = "Abreu, Pedro and others",
    collaboration = "Pierre Auger",
    title = "{Arrival Directions of Cosmic Rays above 32 EeV from Phase One of the Pierre Auger Observatory}",
    eprint = "2206.13492",
    archivePrefix = "arXiv",
    primaryClass = "astro-ph.HE",
    reportNumber = "FERMILAB-PUB-22-491-AD-PPD-SCD-TD",
    doi = "10.3847/1538-4357/ac7d4e",
    journal = "Astrophys. J.",
    volume = "935",
    number = "2",
    pages = "170",
    year = "2022"
}

@article{ref:small_icrc23,
    author = "Abdul Halim, Adila and others",
    collaboration = "Pierre Auger",
    title = "{An update on the arrival direction studies made with data from the Pierre Auger Observatory}",
    doi = "10.22323/1.444.0252",
    journal = "PoS",
    volume = "ICRC2023",
    pages = "252",
    year = "2023"
}

@article{ref:auger_prime,
    author = "Abdul Halim, Adila and others",
    collaboration = "Pierre Auger",
    title = "{AugerPrime: Status and first results}",
    eprint = "2508.08056",
    archivePrefix = "arXiv",
    primaryClass = "astro-ph.IM",
    reportNumber = "PoS-ICRC2025-385",
    doi = "10.22323/1.501.0385",
    journal = "PoS",
    volume = "ICRC2025",
    pages = "385",
    year = "2025"
}

@article{ref:prd2020,
    author = "Aab, Alexander and others",
    collaboration = "Pierre Auger",
    title = "{Measurement of the cosmic-ray energy spectrum above $2.5{\times} 10^{18}$  eV using the Pierre Auger Observatory}",
    eprint = "2008.06486",
    archivePrefix = "arXiv",
    primaryClass = "astro-ph.HE",
    reportNumber = "FERMILAB-PUB-20-432-E-TD",
    doi = "10.1103/PhysRevD.102.062005",
    journal = "Phys. Rev. D",
    volume = "102",
    number = "6",
    pages = "062005",
    year = "2020"
}

@article{ref:prl2020,
    author = "Aab, Alexander and others",
    collaboration = "Pierre Auger",
    title = "{Features of the Energy Spectrum of Cosmic Rays above 2.5{\texttimes}10$^{18}$ eV Using the Pierre Auger Observatory}",
    eprint = "2008.06488",
    archivePrefix = "arXiv",
    primaryClass = "astro-ph.HE",
    reportNumber = "FERMILAB-PUB-20-431-E-TD",
    doi = "10.1103/PhysRevLett.125.121106",
    journal = "Phys. Rev. Lett.",
    volume = "125",
    number = "12",
    pages = "121106",
    year = "2020"
}

@article{ref:vert_rec,
    author = "Aab, A. and others",
    collaboration = "Pierre Auger",
    title = "{Reconstruction of events recorded with the surface detector of the Pierre Auger Observatory}",
    eprint = "2007.09035",
    archivePrefix = "arXiv",
    primaryClass = "astro-ph.IM",
    reportNumber = "FERMILAB-PUB-20-487-AD-E-TD",
    doi = "10.1088/1748-0221/15/10/P10021",
    journal = "JINST",
    volume = "15",
    number = "10",
    pages = "P10021",
    year = "2020"
}

@article{ref:incl_rec,
    author = "Aab, Alexander and others",
    collaboration = "Pierre Auger",
    title = "{Reconstruction of Inclined Air Showers Detected with the Pierre Auger Observatory}",
    eprint = "1407.3214",
    archivePrefix = "arXiv",
    primaryClass = "astro-ph.HE",
    reportNumber = "FERMILAB-PUB-14-234-AD-AE-CD-TD",
    doi = "10.1088/1475-7516/2014/08/019",
    journal = "JCAP",
    volume = "08",
    pages = "019",
    year = "2014"
}

@article{ref:atmo,
    author = "Aab, Alexander and others",
    collaboration = "Pierre Auger",
    title = "{Impact of atmospheric effects on the energy reconstruction of air showers observed by the surface detectors of the Pierre Auger Observatory}",
    eprint = "1702.02835",
    archivePrefix = "arXiv",
    primaryClass = "astro-ph.IM",
    reportNumber = "FERMILAB-PUB-17-031-AD-CD-TD",
    doi = "10.1088/1748-0221/12/02/P02006",
    journal = "JINST",
    volume = "12",
    number = "02",
    pages = "P02006",
    year = "2017"
}

@article{ref:geom,
    author = "Abreu, P. and others",
    collaboration = "Pierre Auger",
    title = "{The effect of the geomagnetic field on cosmic ray energy estimates and large scale anisotropy searches on data from the Pierre Auger Observatory}",
    eprint = "1111.7122",
    archivePrefix = "arXiv",
    primaryClass = "astro-ph.IM",
    reportNumber = "FERMILAB-PUB-11-590-AD-AE-CD-TD",
    doi = "10.1088/1475-7516/2011/11/022",
    journal = "JCAP",
    volume = "11",
    pages = "022",
    year = "2011"
}

@article{ref:sgp,
    author = "Abdul Halim, Adila and others",
    collaboration = "Pierre Auger",
    title = "{The Distribution of Ultrahigh-energy Cosmic Rays along the Supergalactic Plane Measured at the Pierre Auger Observatory}",
    eprint = "2407.06874",
    archivePrefix = "arXiv",
    primaryClass = "astro-ph.HE",
    reportNumber = "FERMILAB-PUB-24-1040",
    doi = "10.3847/1538-4357/adbdc5",
    journal = "Astrophys. J.",
    volume = "984",
    number = "2",
    pages = "123",
    year = "2025"
}

@article{ref:ta_icrc25,
    author = "Kim, Jihyun and Ivanov, Dmitri and Kawata, Kazumasa and Sagawa, Hiroyuki and Thomson, Gordon",
    collaboration = "Telescope Array",
    title = "{Medium-scale anisotropies measured by Telescope Array surface detectors}",
    eprint = "2510.02739",
    archivePrefix = "arXiv",
    primaryClass = "astro-ph.HE",
    doi = "10.22323/1.501.0302",
    journal = "PoS",
    volume = "ICRC2025",
    pages = "302",
    year = "2025"
}

@article{ref:ta_hot,
    author = "Abbasi, R. U. and others",
    collaboration = "Telescope Array",
    title = "{Indications of Intermediate-Scale Anisotropy of Cosmic Rays with Energy Greater Than 57 EeV in the Northern Sky Measured with the Surface Detector of the Telescope Array Experiment}",
    eprint = "1404.5890",
    archivePrefix = "arXiv",
    primaryClass = "astro-ph.HE",
    doi = "10.1088/2041-8205/790/2/L21",
    journal = "Astrophys. J. Lett.",
    volume = "790",
    pages = "L21",
    year = "2014"
}

@article{ref:healpix,
    author = "G{\'o}rski, K. M. and Hivon, E. and Banday, A. J. and Wandelt, B. D. and Hansen, F. K. and Reinecke, M. and Bartelman, M.",
    title = "{HEALPix - A Framework for high resolution discretization, and fast analysis of data distributed on the sphere}",
    eprint = "astro-ph/0409513",
    archivePrefix = "arXiv",
    doi = "10.1086/427976",
    journal = "Astrophys. J.",
    volume = "622",
    pages = "759--771",
    year = "2005"
}

@article{ref:tax4,
    author = "Abbasi, R. U. and others",
    collaboration = "Telescope Array",
    title = "{Surface detectors of the TAx4 experiment}",
    eprint = "2103.01086",
    archivePrefix = "arXiv",
    primaryClass = "astro-ph.IM",
    doi = "10.1016/j.nima.2021.165726",
    journal = "Nucl. Instrum. Meth. A",
    volume = "1019",
    pages = "165726",
    year = "2021"
}

@article{ref:gcos,
    author = "Alves Batista, Rafael and others",
    collaboration = "GCOS",
    title = "{Science with the Global Cosmic-ray Observatory (GCOS)}",
    eprint = "2309.17324",
    archivePrefix = "arXiv",
    primaryClass = "astro-ph.HE",
    doi = "10.22323/1.444.0281",
    journal = "PoS",
    volume = "ICRC2023",
    pages = "281",
    year = "2023"
}

@article{ref:grand,
    author = "Kotera, Kumiko",
    collaboration = "GRAND",
    title = "{The Giant Radio Array for Neutrino Detection (GRAND) project}",
    eprint = "2108.00032",
    archivePrefix = "arXiv",
    primaryClass = "astro-ph.HE",
    reportNumber = "PoS(ICRC2021)1181",
    doi = "10.22323/1.395.1181",
    journal = "PoS",
    volume = "ICRC2021",
    pages = "1181",
    year = "2021"
}

@article{ref:poemma,
    author = "Olinto, A. V. and others",
    collaboration = "POEMMA",
    title = "{The POEMMA (Probe of Extreme Multi-Messenger Astrophysics) observatory}",
    eprint = "2012.07945",
    archivePrefix = "arXiv",
    primaryClass = "astro-ph.IM",
    doi = "10.1088/1475-7516/2021/06/007",
    journal = "JCAP",
    volume = "06",
    pages = "007",
    year = "2021"
}

@article{ref:multipole,
    author = "Aab, Alexander and others",
    collaboration = "Pierre Auger",
    title = "{Multi-resolution anisotropy studies of ultrahigh-energy cosmic rays detected at the Pierre Auger Observatory}",
    eprint = "1611.06812",
    archivePrefix = "arXiv",
    primaryClass = "astro-ph.HE",
    reportNumber = "FERMILAB-PUB-16-556-AD-AE-CD-TD",
    doi = "10.1088/1475-7516/2017/06/026",
    journal = "JCAP",
    volume = "06",
    pages = "026",
    year = "2017"
}

@article{ref:scipy,
    author = "Virtanen, Pauli and others",
    title = "{SciPy 1.0--Fundamental Algorithms for Scientific Computing in Python}",
    eprint = "1907.10121",
    archivePrefix = "arXiv",
    primaryClass = "cs.MS",
    doi = "10.1038/s41592-019-0686-2",
    journal = "Nature Meth.",
    volume = "17",
    pages = "261",
    year = "2020"
}

\end{document}